\documentclass[acmtog,nonacm]{acmart}

\AtBeginDocument{%
  }

\usepackage{tikz}
\usetikzlibrary{calc}
\usepackage{cleveref}
\usepackage{bbm}
\usepackage{multirow}
\usepackage{makecell}
\usepackage{array}
\newcommand{\rgbSquare}[4][0.65em]{%
	\begingroup
	\ifmmode
		\mathord{\textcolor[RGB]{#2,#3,#4}{\rule{#1}{#1}}}%
	\else
		\textcolor[RGB]{#2,#3,#4}{\rule{#1}{#1}}%
	\fi
	\endgroup
}

\makeatletter
\DeclareRobustCommand{\optionalCref}[2]{%
	\@ifundefined{r@#1}{#2}{\Cref{#1}}%
}
\makeatother

\begin{document}

\title{SymTRELLIS: Symmetry-Enforced Voxel Latents for 3D Generation}
\author{Guangda Ji}
\email{guangda_ji@sfu.ca}
\orcid{0000-0002-0499-8774}
\affiliation{%
	\institution{Simon Fraser University}
	\city{Burnaby}
	\state{BC}
	\country{Canada}
}

\author{Qimin Chen}
\orcid{0009-0004-8447-0137}
\affiliation{%
	\institution{Simon Fraser University}
	\city{Burnaby}
	\state{BC}
	\country{Canada}
}
\email{qiminchen1120@gmail.com}

\author{Qinchan Li}
\orcid{0009-0004-2135-0269}
\affiliation{%
	\institution{Simon Fraser University}
	\city{Burnaby}
	\state{BC}
	\country{Canada}
}
\email{qinchan_li@sfu.ca}

\author{Mingrui Zhao}
\orcid{0009-0003-1737-2476}
\affiliation{%
	\institution{Simon Fraser University}
	\city{Burnaby}
	\state{BC}
	\country{Canada}
}
\email{mingrui_zhao@sfu.ca}

\author{Kai Wang}
\orcid{0009-0006-6332-6181}
\affiliation{%
	\institution{Simon Fraser University}
	\city{Burnaby}
	\state{BC}
	\country{Canada}
}
\affiliation{%
	\institution{ShanghaiTech University}
	\city{Shanghai}
	\country{China}
}
\email{kwang.ether@gmail.com}

\author{Hao Zhang}
\orcid{0000-0003-1991-119X}
\affiliation{%
	\institution{Simon Fraser University}
	\city{Burnaby}
	\state{BC}
	\country{Canada}
}
\affiliation{%
	\institution{Augmenta}
	\city{Toronto}
	\state{ON}
	\country{Canada}
}
\email{haoz@sfu.ca}

\renewcommand{\shortauthors}{Ji et al.}


\begin{teaserfigure}
	\centering
	\begin{tikzpicture}
		\node[anchor=south west, inner sep=0] (fig) at (0,0) {\includegraphics[width=\textwidth]{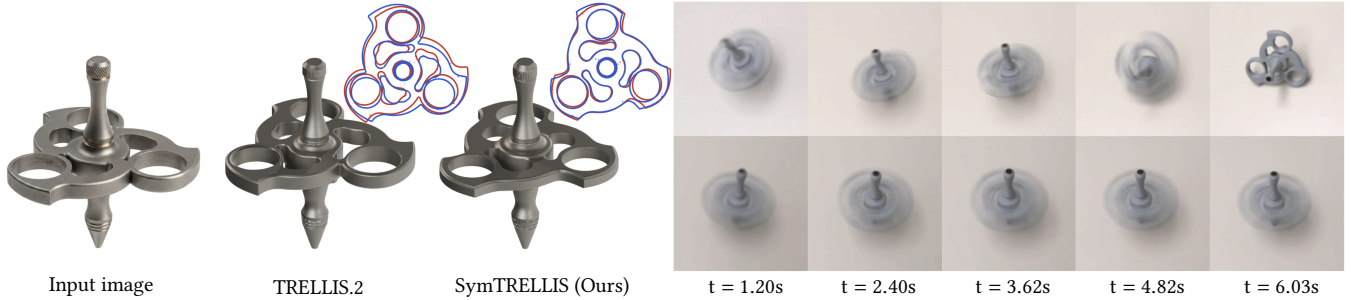}};
		\begin{scope}[x={(fig.south east)},y={(fig.north west)}]
			\node[
				anchor=center,
				align=center,
				font=\fontsize{8pt}{9pt}\selectfont,
				fill=white,
				fill opacity=0.75,
				text opacity=1,
				inner sep=2pt,
				rounded corners=1pt
			] at (0.075,0.08) {Input image};
			\node[
				anchor=center,
				align=center,
				font=\fontsize{8pt}{9pt}\selectfont,
				fill=white,
				fill opacity=0.75,
				text opacity=1,
				inner sep=2pt,
				rounded corners=1pt
			] at (0.235,0.08) {TRELLIS.2};
			\node[
				anchor=center,
				align=center,
				font=\fontsize{8pt}{9pt}\selectfont,
				fill=white,
				fill opacity=0.75,
				text opacity=1,
				inner sep=2pt,
				rounded corners=1pt
			] at (0.4,0.08) {SymTRELLIS (Ours)};
			\node[
				anchor=center,
				align=center,
				font=\fontsize{8pt}{9pt}\selectfont,
				fill=white,
				fill opacity=0.75,
				text opacity=1,
				inner sep=2pt,
				rounded corners=1pt
			] at (0.55,0.08) {$\mathrm{t}=1.20\mathrm{s}$};
			\node[
				anchor=center,
				align=center,
				font=\fontsize{8pt}{9pt}\selectfont,
				fill=white,
				fill opacity=0.75,
				text opacity=1,
				inner sep=2pt,
				rounded corners=1pt
			] at (0.648,0.08) {$\mathrm{t}=2.40\mathrm{s}$};
			\node[
				anchor=center,
				align=center,
				font=\fontsize{8pt}{9pt}\selectfont,
				fill=white,
				fill opacity=0.75,
				text opacity=1,
				inner sep=2pt,
				rounded corners=1pt
			] at (0.747,0.08) {$\mathrm{t}=3.62\mathrm{s}$};
			\node[
				anchor=center,
				align=center,
				font=\fontsize{8pt}{9pt}\selectfont,
				fill=white,
				fill opacity=0.75,
				text opacity=1,
				inner sep=2pt,
				rounded corners=1pt
			] at (0.846,0.08) {$\mathrm{t}=4.82\mathrm{s}$};
			\node[
				anchor=center,
				align=center,
				font=\fontsize{8pt}{9pt}\selectfont,
				fill=white,
				fill opacity=0.75,
				text opacity=1,
				inner sep=2pt,
				rounded corners=1pt
			] at (0.945,0.08) {$\mathrm{t}=6.03\mathrm{s}$};
		\end{scope}
	\end{tikzpicture}
	\caption{\label{fig:teaser}
		Why reconstruct a 3D spinning top without spinning it? TRELLIS.2 produces a visually plausible top with clear asymmetries, visible in the overlaid cross-section wires (red vs.~blue, in insets) after best rotation alignment. Result by our SymTRELLIS exhibits much better symmetry. Both reconstructions were 3D-printed and spun repeatedly. As shown by the time-lapse photos, exhibiting a typical case, the TRELLIS.2 top wobbles severely and stops prematurely, while the SymTRELLIS top spins stably throughout, stopping at $t=8.45s$. We quantify wobble by tracking the standard deviation of the top tip position during spin, yielding 6.55 for TRELLIS.2 vs.~0.45 for SymTRELLIS for the test shown -- a 14.5$\times$ reduction. See Supplementary for videos of the test spins.}
\end{teaserfigure}

\begin{abstract}
	Single-view 3D generative models have achieved impressive visual quality,
yet they are not designed to satisfy structural or functional requirements, and in practice, often fall short.
Symmetry is one such requirement: violations, even subtle ones, on symmetry can render a model physically unusable.
We present \emph{SymTRELLIS}, a method that enforces arbitrary finite point group symmetries (rotational, reflectional, and polyhedral) during the flow-based 3D generation of TRELLIS.2, without retraining the underlying VAE or flow model.
Our key idea is to approximate the latent-space action of spatial transformations as a \emph{learned linear operator on voxel latents},
implemented as a lightweight spatial-transform latent mapper trained on generic, non-symmetric 3D data.
At generation time, we enforce symmetry by averaging predicted flow velocities across all symmetry-equivalent transformations at each ODE step, a process we call \emph{velocity symmetrization}.
The symmetry specification can be estimated automatically from an initial TRELLIS.2 generation or supplied by the user,
enabling deliberate fold manipulation beyond what the input image suggests.
On a curated benchmark of 266 strictly symmetric objects spanning 2- to 20-fold rotations and polyhedral symmetry groups,
SymTRELLIS substantially reduces all symmetry error metrics compared to TRELLIS.2, Hunyuan3D-2.1, and TripoSG, while maintaining reconstruction accuracy comparable to the base model.
Code and data are at \href{https://symtrellis.github.io}{symtrellis.github.io}.

\end{abstract}



\maketitle

\section{Introduction}
\label{sec:intro}

Single-view 3D reconstruction and generation have seen remarkable progress in recent years,
with models such as CLAY \cite{zhang2024CLAY}, TripoSG \cite{li2025triposg}, Hunyuan3D-2 \cite{hunyuan3d2025hunyuan3d}, and the TRELLIS duologies \cite{xiang2024trellis,xiang2025trellis2} capable of producing high-quality 3D assets from a single photograph.
These models excel at visual realism, faithful alignment with the input image, and recovery of fine-grained surface detail,
but they are not designed to produce 3D models that satisfy structural or functional requirements.
An electric fan, a turbine blade, or a gear may look convincing in a 3D visualization while being geometrically malformed in ways that make it unusable in practice,
e.g., a gear with irregular teeth would not mesh.
Such a functional gap matters since the primary purpose of 3D design and creation is often not the digital model itself --- it is what the model \emph{enables in physical applications}.
A model that only looks right cannot serve this purpose.

Among the many structural and functional requirements that a 3D object must satisfy, symmetry occupies a privileged role.
The vast majority of human-made objects, from mechanical parts to daily artifacts such as furniture and vehicles, are designed to be symmetric.
This is not accidental,
since symmetry enforces balance, distributes mechanical load, and allows a single fundamental domain to be replicated in both design and production, reducing cost without sacrificing functional integrity.
Yet, despite its importance, do the most advanced single-view 3D reconstruction models of today recover symmetry reliably?
Figure~\ref{fig:teaser} offers a telling answer.
Given an image of a spinning top, an object whose very function depends on rotational symmetry,
TRELLIS.2~\cite{xiang2025trellis2} produces a visually plausible reconstruction, but one with measurable asymmetries.
We 3D-printed the result and did a test spin: the top consistently wobbles and simply does not function as it should.

\begin{figure}[t!]
	\centering
	\small
	\setlength{\tabcolsep}{2pt}
	\renewcommand{\arraystretch}{0.8}

	\newlength{\introfigresultsize}
	\setlength{\introfigresultsize}{0.185\textwidth}

	\newcommand{\placeholder}[1]{%
		\begingroup
		\setlength{\fboxsep}{0pt}%
		\fbox{%
			\begin{minipage}[c][\introfigresultsize][c]{\introfigresultsize}
				\centering
				\vspace*{\fill}
				{\small #1}
				\vspace*{\fill}
			\end{minipage}%
		}%
		\endgroup
	}

	\resizebox{\columnwidth}{!}{%
		\begin{tabular}{ccccc}
			\textbf{Input image}                                                                                                            &
			\textbf{TripoSG}                                                                                                                &
			\textbf{Hunyuan3D-2.1}                                                                                                          &
			\textbf{TRELLIS.2}                                                                                                              &
			\textbf{SymTRELLIS (Ours)}                                                                                                        \\[3pt]

			\includegraphics[width=\introfigresultsize,height=\introfigresultsize]{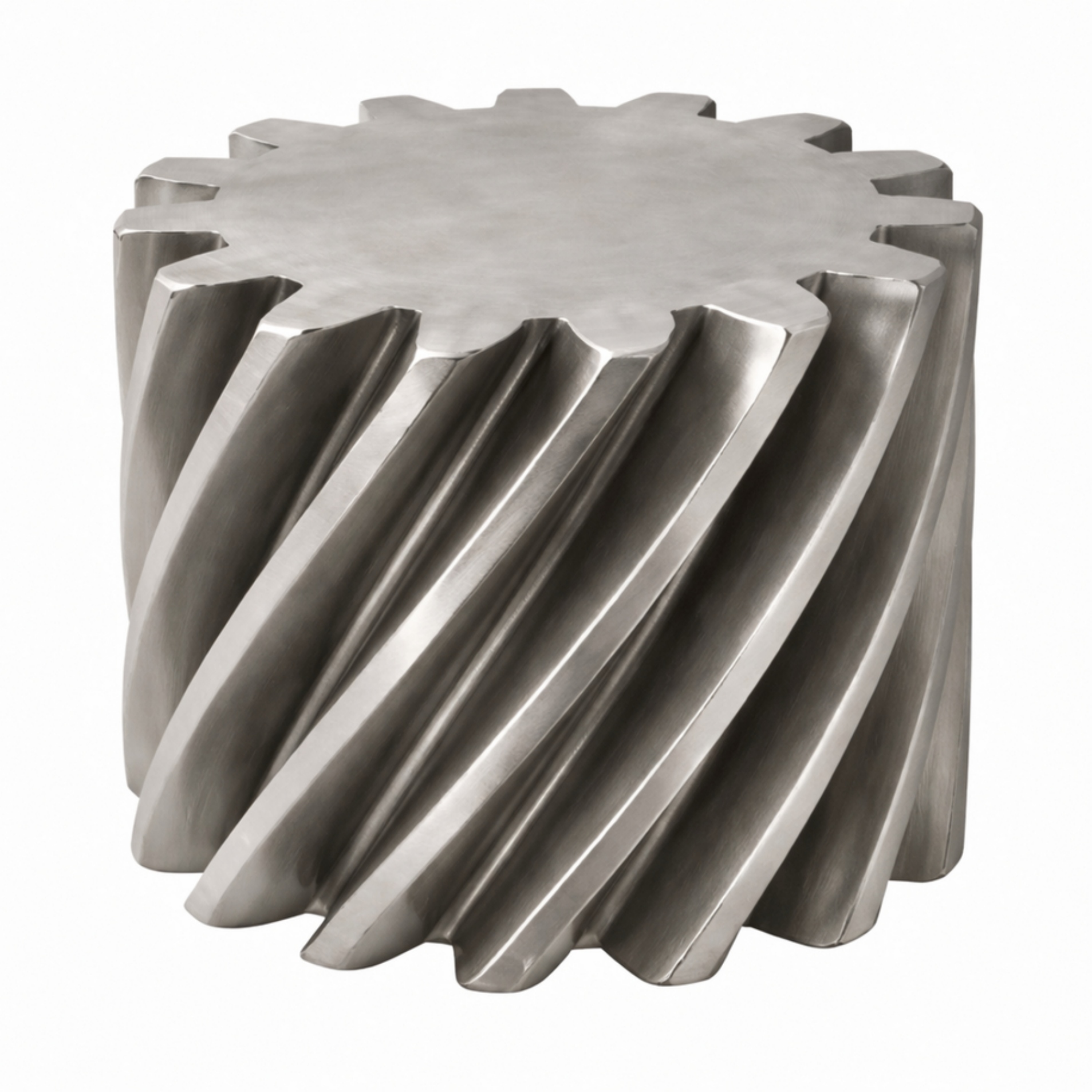}             &
			\includegraphics[width=\introfigresultsize,height=\introfigresultsize]{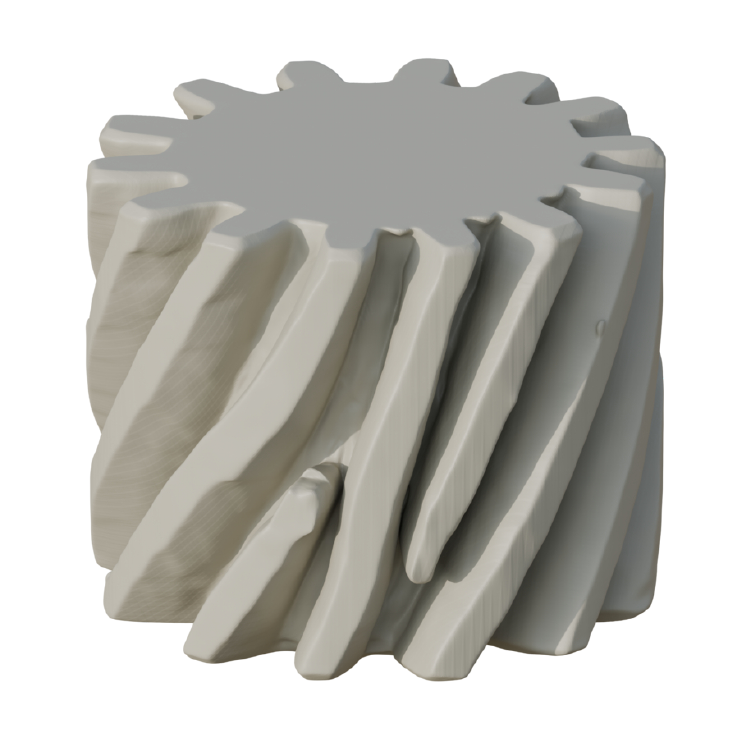}                 &
			\includegraphics[width=\introfigresultsize,height=\introfigresultsize]{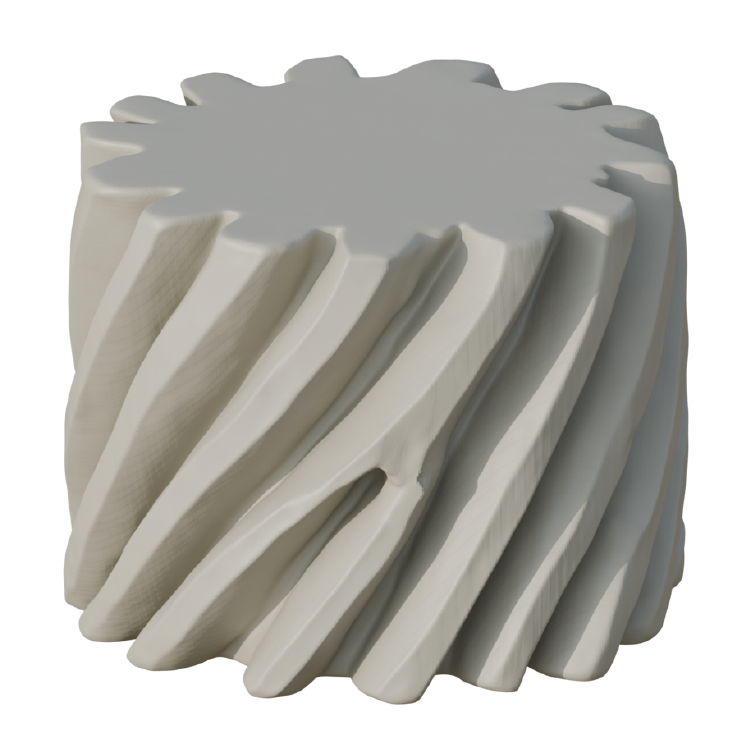}           &
			\includegraphics[width=\introfigresultsize,height=\introfigresultsize]{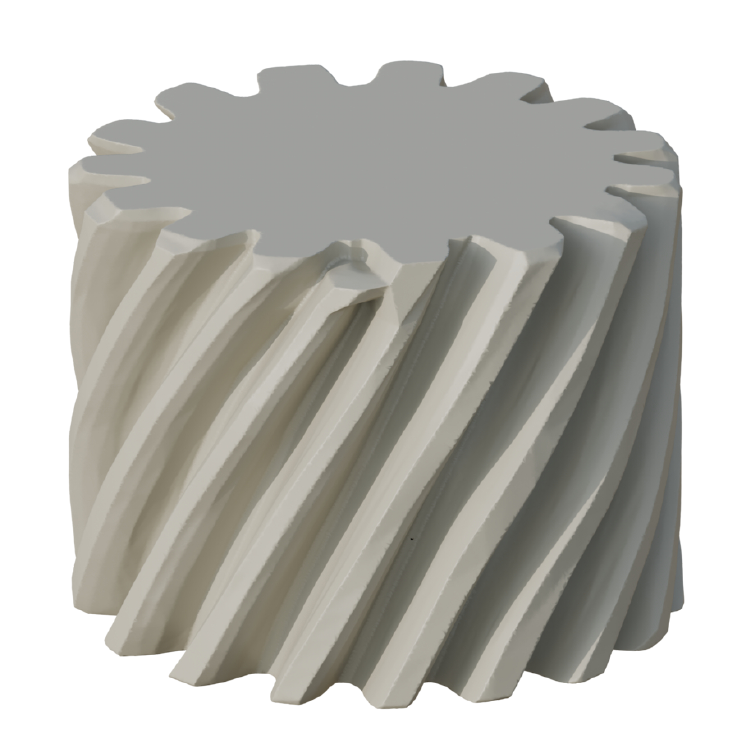}        &
			\includegraphics[width=\introfigresultsize,height=\introfigresultsize]{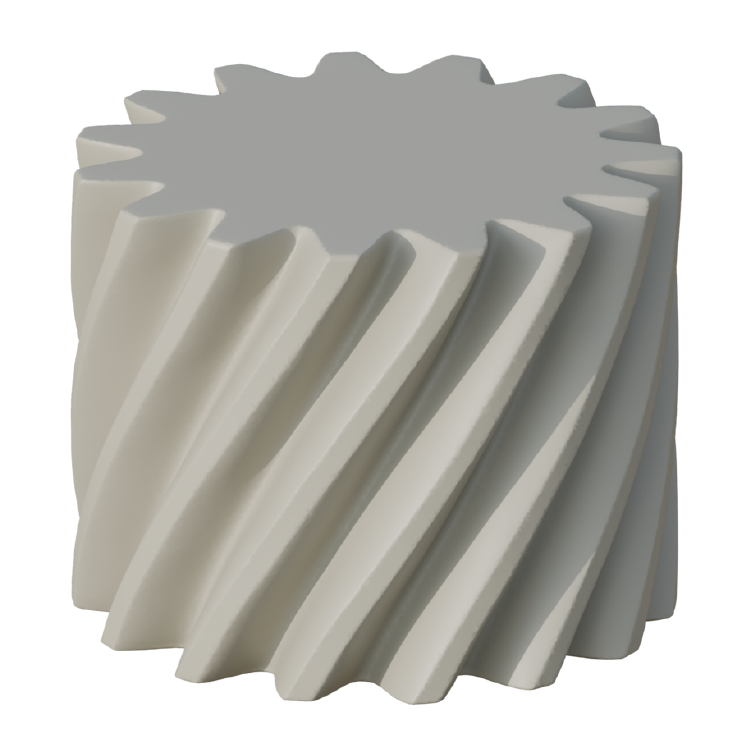}                \\[4pt]

			\includegraphics[width=\introfigresultsize,height=\introfigresultsize]{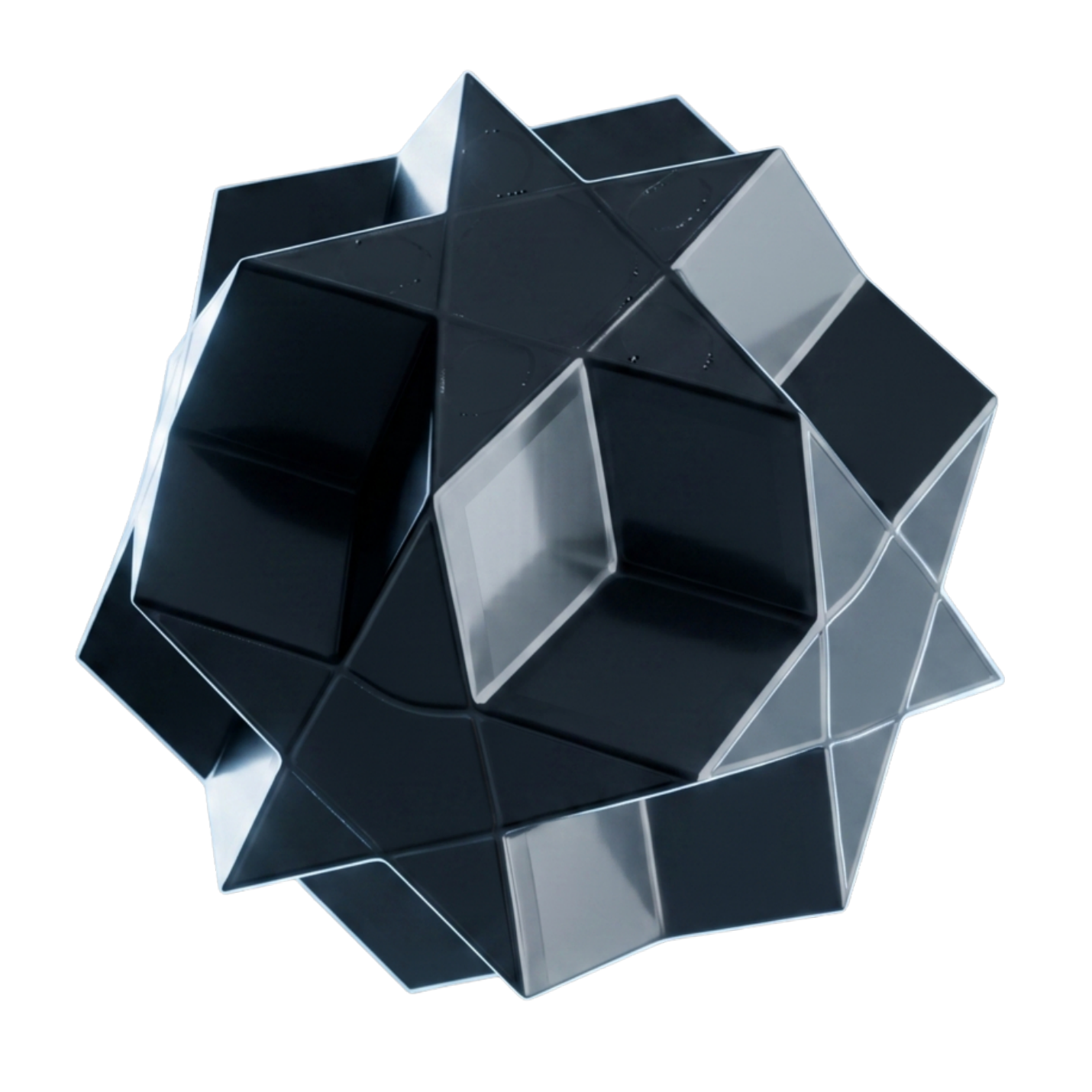}      &
			\includegraphics[width=\introfigresultsize,height=\introfigresultsize]{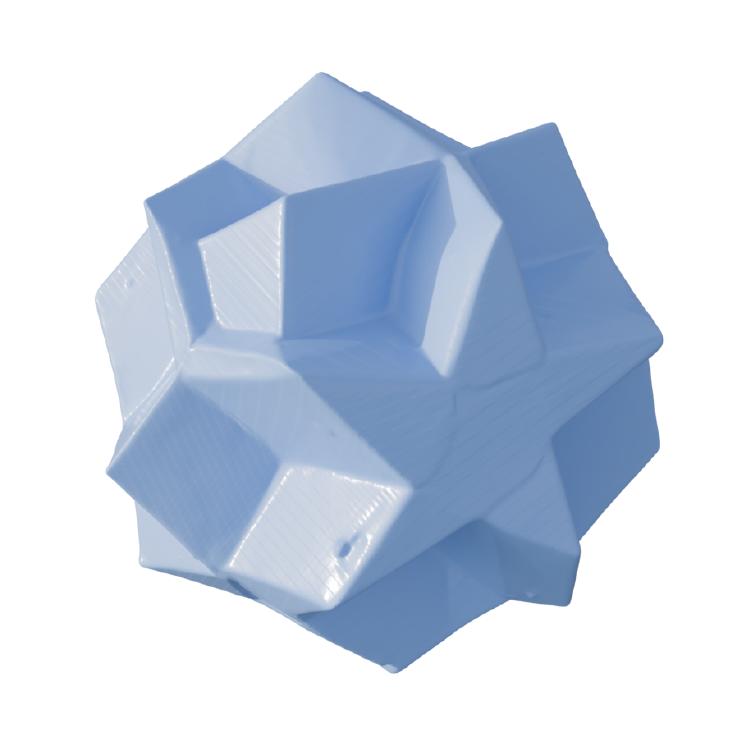}          &
			\includegraphics[width=\introfigresultsize,height=\introfigresultsize]{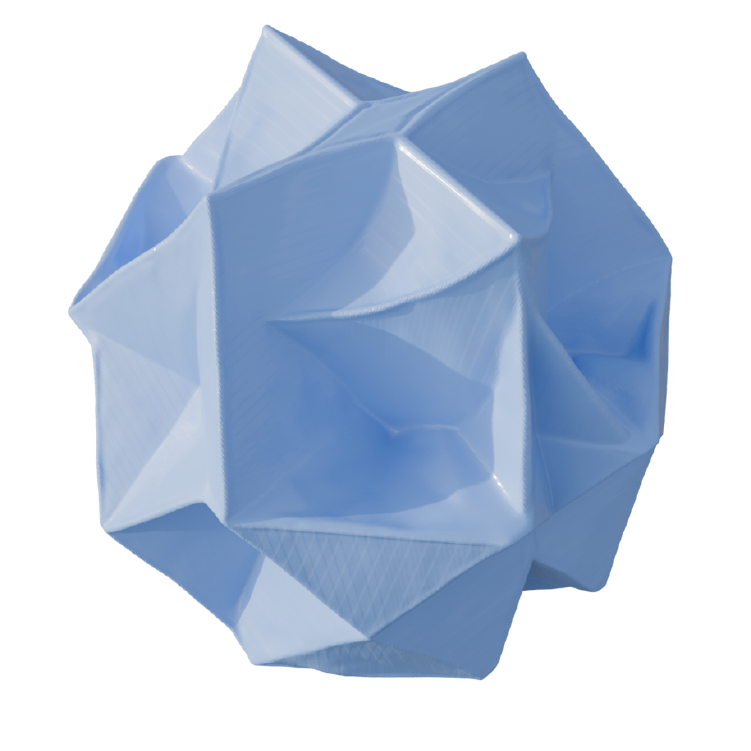}    &
			\includegraphics[width=\introfigresultsize,height=\introfigresultsize]{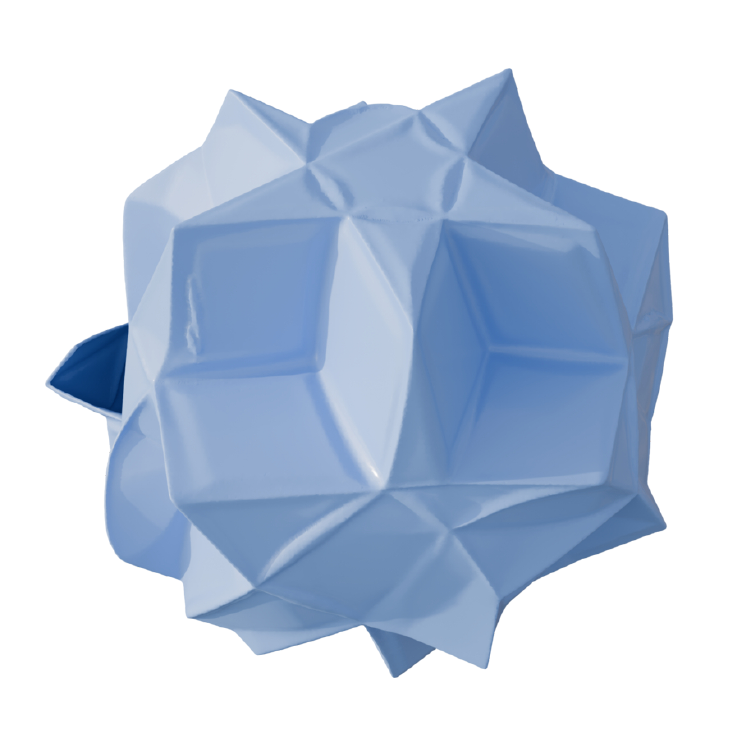} &
			\includegraphics[width=\introfigresultsize,height=\introfigresultsize]{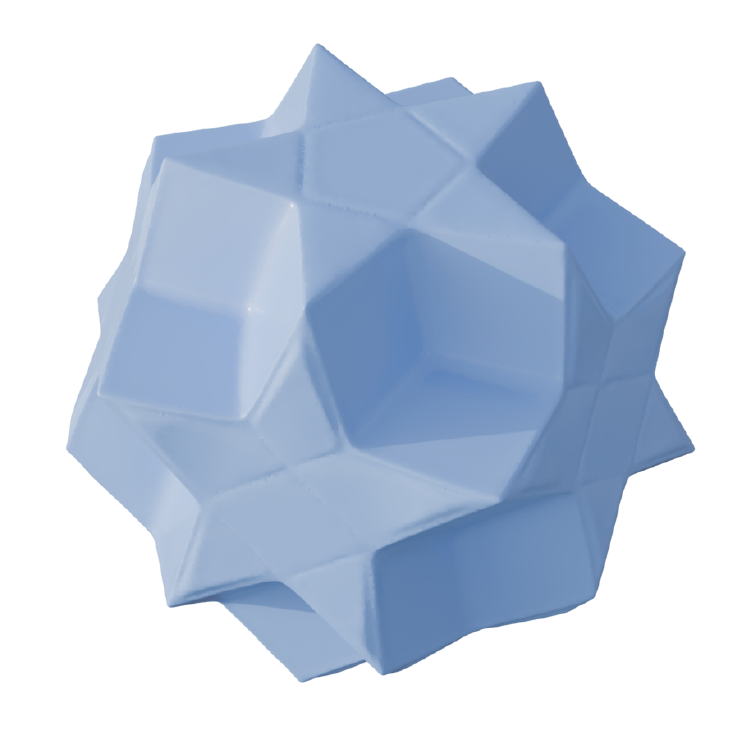}         \\[4pt]

			\includegraphics[width=\introfigresultsize,height=\introfigresultsize]{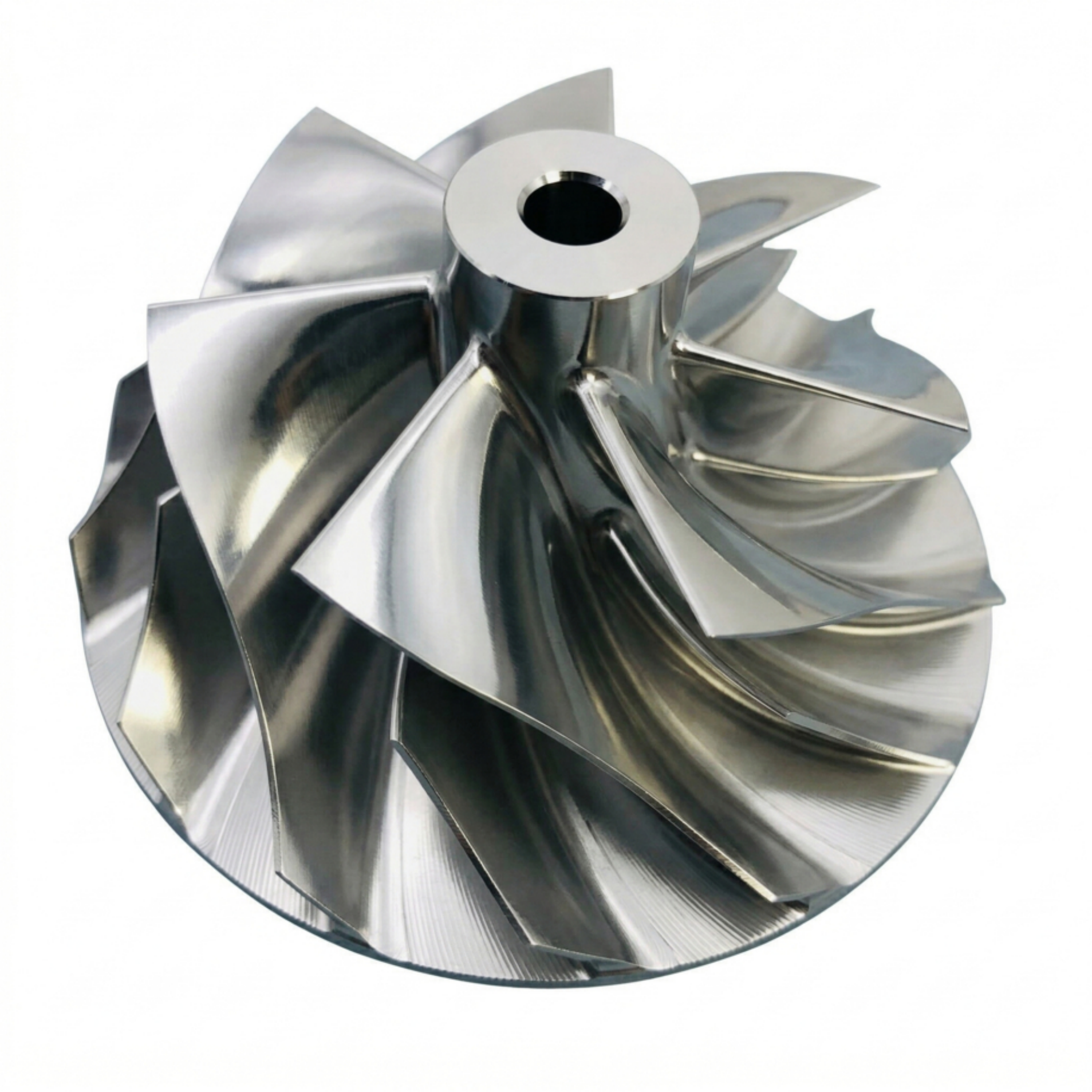}      &
			\includegraphics[width=\introfigresultsize,height=\introfigresultsize]{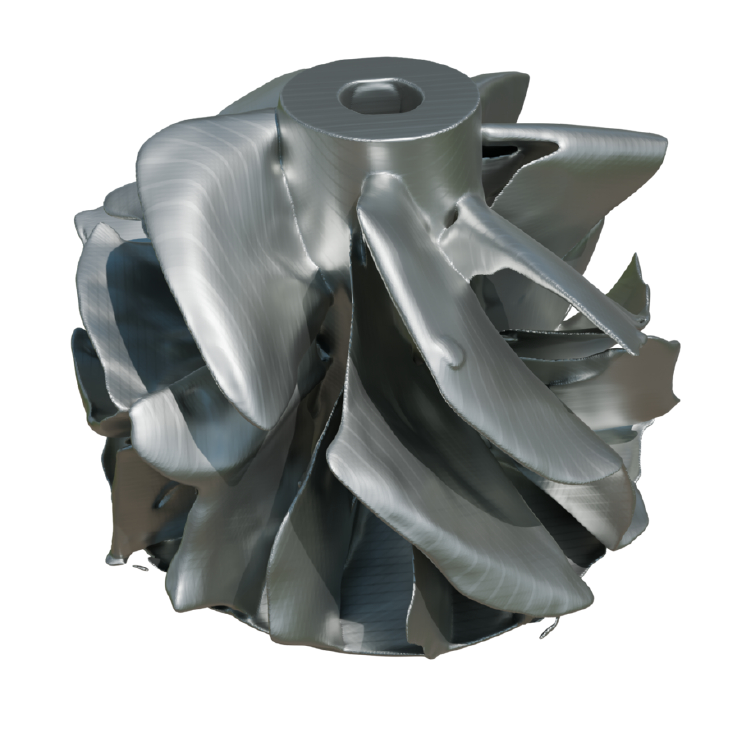}          &
			\includegraphics[width=\introfigresultsize,height=\introfigresultsize]{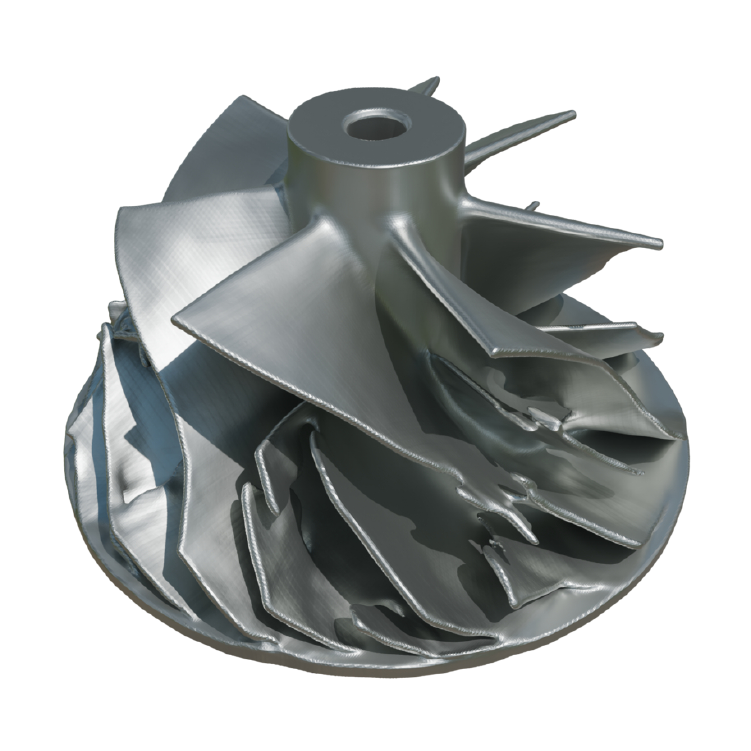}    &
			\includegraphics[width=\introfigresultsize,height=\introfigresultsize]{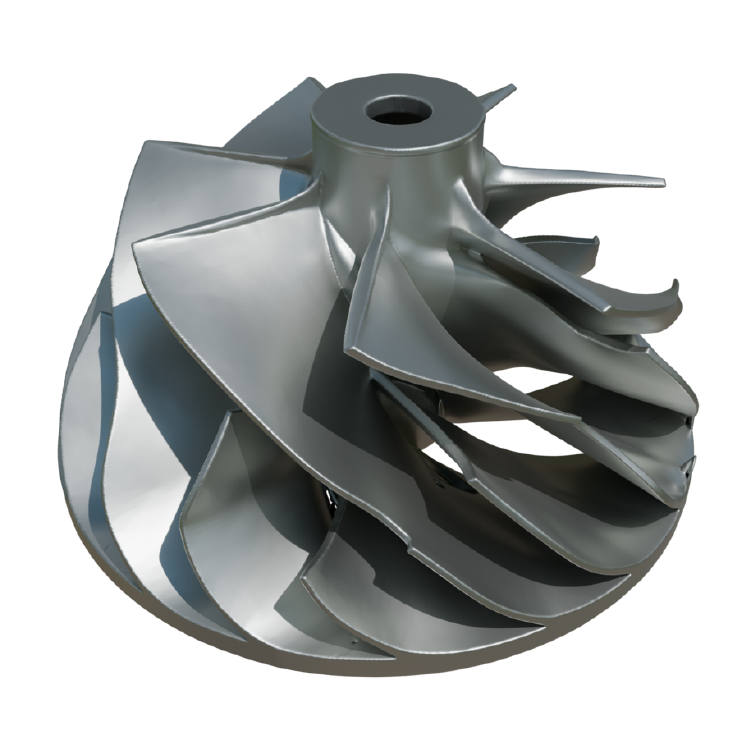} &
			\includegraphics[width=\introfigresultsize,height=\introfigresultsize]{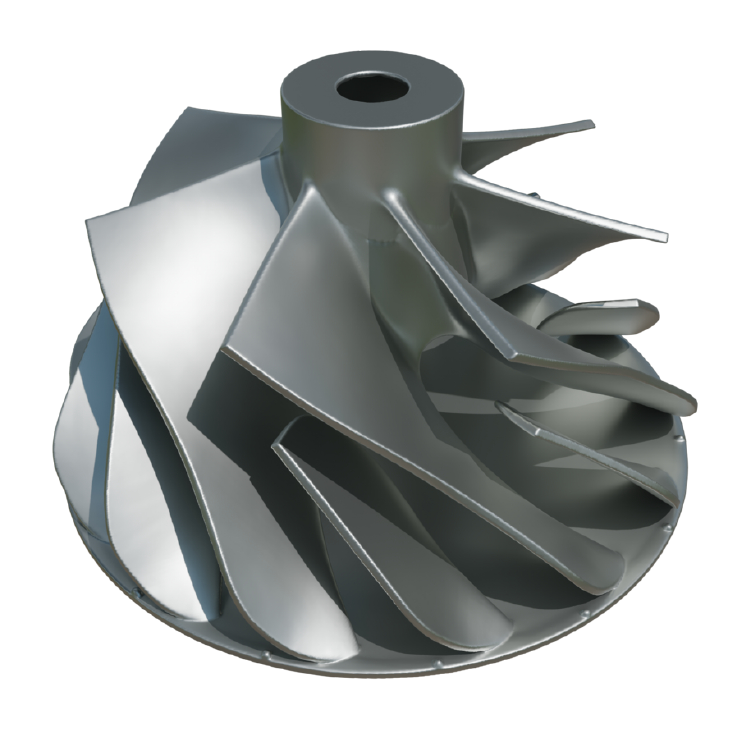}         \\[4pt]
		\end{tabular}%
	}

	\caption{
		\label{fig:sym_failure}
		Single-view 3D reconstruction results from state-of-the-art models on objects with moderately complex symmetries: a helical gear (top), a small stellated dodecahedron (middle), and a centrifugal impeller (bottom). Across all baselines, the reconstructed meshes exhibit erroneous or inconsistent fold counts, malformed sectors, and ambiguous symmetry axes, all as conditions under which post-processing symmetrization would break down. With no reliable ground-truth sector to replicate and no unambiguous symmetry group to enforce, symmetrization has no clear target. SymTRELLIS (ours) enforces symmetry during generation while remaining anchored to the input image. As a result, it is able to recover significantly more faithful and consistent symmetric structures.
	}
\end{figure}

One possible way to fix the asymmetries is to ``symmetrize"~\cite{mitra2007sym,wu2014real,yan2025symcomp} the TRELLIS.2 outputs during post-processing,
typically by warping or averaging a mesh toward a target symmetry.
However, such a symmetrization can only correct relatively small deviations from symmetries since it often requires:
a) the detected (target) symmetry is unambiguous, and b) one or more sectors in the mesh can be designated as the ``ground truth" fold to be replicated.
When averaging across sectors, the sectors cannot all be imperfect in different ways,
since otherwise averaging merely splits the differences between them, committing faithfully to none.
Figure~\ref{fig:sym_failure} shows results from state-of-the-art reconstruction models including TRELLIS.2, where these conditions are violated in one way or another.
Crucially, post-processing discards the input image, which does contain views of at least one or more ground-truth sectors.
This motivates \emph{symmetry-enforced regeneration}: rather than correcting the output after the fact, symmetries are enforced during the generation process itself,
where the image condition remains active and can anchor each symmetric sector to the available visual evidence in the input.

In this work, we propose SymTRELLIS, a novel method that enforces symmetry during the flow-based generation of TRELLIS.2, without retraining the underlying VAE or flow model.
Our key idea is to model the latent-space action of a spatial transformation, either rotation or reflection, as a \emph{learned linear operator} on the \emph{voxel latents},
implemented as a lightweight spatial-transform latent mapper.
This mapper is trained once on generic, non-symmetric 3D data and transfers to any symmetry group at inference time.
The symmetry specification, including fold count, axis, and reflection planes, can be estimated automatically from an initial TRELLIS.2 generation,
in which case SymTRELLIS performs a regeneration, or optionally supplied by the user to deliberately manipulate the symmetry structure beyond what the input image suggests.

During generation, we enforce the specified symmetry by averaging predicted flow velocities across all symmetry-equivalent transformations at each ODE step, a process we call \emph{velocity symmetrization}.
Since this coupling is applied throughout the generation process, the model can resolve fold ambiguities and recover geometry in under-represented sectors, guided at every step by the input image.
The result is a 3D model that is both faithful to the input and consistent with the detected or desired symmetry group.
Figure \ref{fig:teaser} shows our result for the spinning top.
When 3D-printed and tested, the improvement is clear: the SymTRELLIS result spins stably,
whereas the TRELLIS.2 counterpart wobbles visibly due to its residual asymmetries.
Figure~\ref{fig:sym_failure} further demonstrates our method's ability to handle more complex symmetry structures,
consistently producing results that are more complete, more symmetric, and more faithful to the input than state-of-the-art baselines.

With velocity symmetrization and the spatial-transform latent mapper providing inference-time symmetry guidance,
SymTRELLIS enforces not only simple reflectional and rotational symmetries, but also complex polyhedral symmetry groups including tetrahedral, octahedral, and icosahedral symmetries,
as well as combined rotation-reflection groups.
Besides correction, our method also supports deliberate symmetry manipulation, e.g., generating variants of an object with a user-specified fold count.
On a curated new benchmark of 250+ strictly symmetric objects spanning 2- to 20-fold rotations and polyhedral symmetries,
SymTRELLIS substantially reduces all symmetry error metrics compared to TRELLIS.2, Hunyuan3D-2.1, and TripoSG, while maintaining reconstruction accuracy comparable to the base model.
Further ablations show that structure-stage guidance is responsible for the dominant correction with additional gains from shape-stage guidance.

\section{Related work}
\label{sec:related}

\paragraph{3D generative models.}
Single-view 3D reconstruction and generation have seen rapid progress in recent years thanks to the increasing availability of compute power and 3D data, as well as the advances in latent shape representations and sampling techniques.
The majority of these models, however, are designed to prioritize visual and geometric fidelity of 3D shapes, but not to enforce structural properties such as symmetry.
Our method is complementary to these methods in that we focus on adapting such models to respect symmetry.
In this regard, one line of recent methods, such as CLAY~\cite{zhang2024CLAY}, TripoSG~\cite{li2025triposg}, and Hunyuan3D~\cite{hunyuan3d2025hunyuan3d},
represent shapes through unstructured latent tokens in which spatial and shape features are entangled, which is unsuitable for our method.
In contrast, TRELLIS~\cite{xiang2024trellis} and its successors, including TRELLIS.2~\cite{xiang2025trellis2}, take a different route, attaching latent features to a structured voxel grid.
As a result, spatial information of latent features is explicit, allowing direct transformations in this structured latent space.
Our method complements methods using such representations, specifically TRELLIS.2, and applies explicit transformation based symmetry guidance during both the sparse-structure and shape-latent generation stages.

\paragraph{Symmetry detection.}
Detecting symmetries in 3D geometry has a long history in geometry processing.
Mitra et al.~\shortcite{niloy2006partial} introduced a voting framework that pairs surface samples and clusters the induced transformations to recover partial and approximate symmetries.
Pauly et al.~\shortcite{mark2008discovering} generalized this to structural regularity, extracting repeated elements such as rotational copies and translational grids.
Wang and Huang~\shortcite{huiwang2017group} brought a group-theoretic perspective, representing symmetries through finite group representations and providing tools for validating rotational fold counts.
More recently, learning based approaches such as Reflect3D~\cite{li2025symmetry} trains models to detect reflection planes from images.
Our method operates in a simpler setting---we detect from a complete generated mesh---and accordingly uses a classical ICP-based pipeline; detection is not our contribution, and more sophisticated detectors can be substituted.

\paragraph{Symmetry enforcement.}
Making a 3D shape conform to a prescribed symmetry has been approached from several directions, each with limitations that motivate our work.
Mitra et al.~\shortcite{mitra2007sym} proposed the first general symmetrization algorithm, warping a mesh toward a target symmetry by optimizing coupled spatial and transformation-space displacements;
this fails when all symmetric sectors are imperfect in different ways, since averaging merely splits the difference.
Wu et al.~\shortcite{wu2014real} construct symmetry-preserving deformation subspaces for interactive editing, but assume the input is already symmetric---the goal is preservation, not creation.
SymmCompletion~\cite{yan2025symcomp} propagates geometry from observed to missing regions via learned local symmetry transformations,
but operates as post-processing and assumes that the partial input contains at least one reliable sector, an assumption that breaks when a generative model introduces different errors in each sector.
These methods also discard the input image, losing the visual evidence that could anchor the symmetric sectors.
The Quartet of Diffusions~\cite{zhou2026quartet} takes a different approach, building symmetry into the generative model itself through dedicated diffusion processes;
this requires training a new model and does not generalize beyond the symmetry types and object categories it was trained on.
Reflect3D~\cite{li2025symmetry} is the closest to our work: it applies symmetry guidance at inference time by adding symmetric SDS losses during DreamGaussian optimization;
however, it remains optimization-based, handles only reflection symmetry, and operates on an already-initialized 3D representation.
Our method, in contrast, enforces symmetry during the generative flow itself, before any explicit 3D representation exists,
handles arbitrary finite point groups, and keeps the input image active so that each symmetric sector remains anchored to visual evidence.

\paragraph{Visual anagrams.}
Our velocity symmetrization is inspired by visual anagrams~\cite{geng2024visual}, which average denoised predictions across pixel-space transformations at each diffusion step to produce images consistent under multiple views.
LookingGlass~\cite{chang2025lookingglass} extends this idea to Laplacian pyramid warps, showing the principle generalizes beyond simple permutations.
We adapt this mechanism to 3D spatial transformations of voxel latents, where the key difficulty is that TRELLIS.2 latent channels have no prescribed transformation law under $E(3)$---unlike pixel permutations,
the latent-space action of a rotation or reflection must be learned.

\paragraph{Equivariant generative modeling}
Respecting symmetries in the context of generative distributions and sampling dynamics is a central challenge in geometric generative modeling.
For particle systems, Equivariant Flow Matching~\cite{klein2023equivariant} aligns symmetry-equivalent configurations when training an equivariant flow.
Known transformation laws are also built directly into 2D networks.
RSF-Conv~\cite{sun2025rsf} encodes rotation and scale equivariance into its filters for retinal vessel segmentation.
Rotation Equivariant Arbitrary-Scale Image Super-Resolution~\cite{xie2025rotation} redesigns encoding and continuous reconstruction to preserve rotation equivariance throughout the super resolution process.
SymDiff~\cite{zhang2025symdiff} avoids specialized equivariant architectures by stochastically symmetrizing reverse transitions around a generic backbone.
For molecular conformers, the SO(3)-Averaged Flow~\cite{cao2025efficient} removes rotational redundancy by averaging probability flow targets during training.
Most closely related to our work,
\textit{Frame Averaging}~\cite{puny2021frame} presents a similar mathematical form to velocity symmetrization.
However, it assumes known group actions on explicit representations, such as scalars and vectors, and obtains invariance or equivariance by averaging selected transformations.
In contrast, TRELLIS.2's latent transformation laws are unknown, so that we must learn an approximate latent-space transform and average frozen-model velocities during sampling.

\section{Preliminaries}
\label{sec:prelim_flow_trellis}
We first review the flow matching model~\cite{lipman2022flow,liu2022flow} and TRELLIS.2~\cite{xiang2025trellis2}, the generative framework that our method has been built upon.

\paragraph{Flow matching}

Flow matching models generate samples through a learned velocity field.
Given a condition, the sampling trajectory of the intermediate latent $x_t$ at time step $t$ is defined by
\begin{equation}
	\frac{d x_t}{dt}=v_\theta(x_t,t;\mathrm{cond}),
	\; t\in[0,1].
\end{equation}
Starting from the initial noise \(x_0\sim\mathcal{N}(0, I)\), the model generates samples by integrating the velocity field from $t=0$ to $t=1$.
At an intermediate time \(t\), the predicted velocity can be used to estimate an \textit{endpoint} $\hat{x}_t$ via,
$\hat{x}_t := x_t + (1-t)v_\theta(x_t,t;\mathrm{cond}).$

\paragraph{TRELLIS.2}

TRELLIS.2 is a three-stage large generative model designed for high-quality image-to-3D generation.
The first stage uses a rectified flow model conditioned on the input image to predict a sparse structure latent, which is decoded into active voxels.
The second stage employs another flow model to generate shape latents conditioned on the active voxels and input image, and decodes them into mesh-producing sparse O-Voxel features.
The final stage uses a third flow model to generate texture features conditioned on the input image and the generated shape latents.

\section{Method}
\label{sec:3_methods}

In this section, we introduce a novel method for enforcing symmetries during the flow-based generation of TRELLIS.2.
We denote TRELLIS.2 latents as sparse feature fields \(x=(\mathcal{C},X)\),
where \(\mathcal{C}=\{\mathbf{c}_i\}_{i=1}^{m}\) denotes the sparse active voxels coordinate set and \(X\in\mathbb{R}^{m\times d}\) stores a \(d\)-dimensional feature at each coordinate.
The dense-grid latent in the sparse-structure stage is written in the same notation by taking \(\mathcal{C}\) to be the full grid.

Our method, SymTRELLIS, takes an input image and a symmetry specification, such as a rotational fold, a symmetry axis, and/or a reflection plane, and generates a symmetry-consistent 3D shape.
In~\Cref{sec:3_1_vsymm}, we introduce a novel velocity symmetrization process that enforces the prescribed symmetry during sampling by averaging flow predictions over symmetry-equivalent transformations of the object.
In~\Cref{sec:3_2_mapper}, we design a spatial-transform latent mapper that predicts these symmetry-equivalent transformations.
Finally, in~\Cref{sec:3_3_symm_detect}, we introduce a symmetry detection algorithm that automatically predicts the symmetry specification when no manual symmetry specification is provided.

\begin{figure*}[t]
	\centering
	\begin{tikzpicture}
		\node[anchor=south west, inner sep=0] (fig) at (0,0)
		{\includegraphics[width=\textwidth]{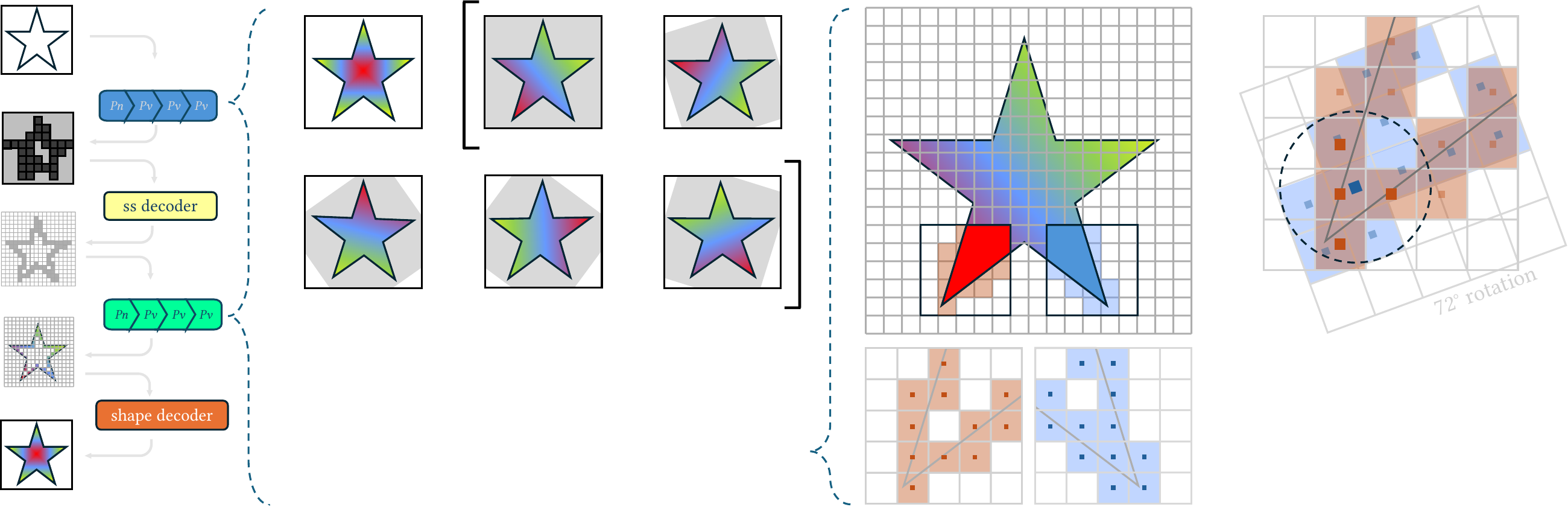}};

		\begin{scope}[
				x={(fig.south east)},
				y={(fig.north west)}
			]

			\node[
				anchor=center,
				align=center,
				font=\fontsize{5pt}{6pt}\selectfont,
				fill=white,
				fill opacity=0,
				text opacity=1,
				inner sep=1.5pt,
				rounded corners=1pt
			] at (0.025,0.83)
			{image cond};

			\node[
				anchor=center,
				align=center,
				font=\fontsize{5pt}{6pt}\selectfont,
				fill=white,
				fill opacity=0,
				text opacity=1,
				inner sep=1.5pt,
				rounded corners=1pt
			] at (0.025,0.615)
			{ss latent};

			\node[
				anchor=center,
				align=center,
				font=\fontsize{5pt}{6pt}\selectfont,
				fill=white,
				fill opacity=0,
				text opacity=1,
				inner sep=1.5pt,
				rounded corners=1pt
			] at (0.025,0.41)
			{active vox};

			\node[
				anchor=center,
				align=center,
				font=\fontsize{5pt}{6pt}\selectfont,
				fill=white,
				fill opacity=0,
				text opacity=1,
				inner sep=1.5pt,
				rounded corners=1pt
			] at (0.0275,0.21)
			{shape latent};

			\node[
				anchor=center,
				align=center,
				font=\fontsize{5pt}{6pt}\selectfont,
				fill=white,
				fill opacity=0,
				text opacity=1,
				inner sep=1.5pt,
				rounded corners=1pt
			] at (0.026,0.01)
			{shape};

			\node[
				anchor=center,
				align=center,
				font=\fontsize{5pt}{6pt}\selectfont,
				fill=white,
				fill opacity=0,
				text opacity=1,
				inner sep=1.5pt,
				rounded corners=1pt
			] at (0.1,0.843)
			{Symmetrized flow};

			\node[
				anchor=center,
				align=center,
				font=\fontsize{5pt}{6pt}\selectfont,
				fill=white,
				fill opacity=0,
				text opacity=1,
				inner sep=1.5pt,
				rounded corners=1pt
			] at (0.105,0.43)
			{Symmetrized flow};

			\node[
				anchor=center,
				align=center,
				font=\fontsize{9pt}{10pt}\selectfont,
				fill=white,
				fill opacity=0,
				text opacity=1,
				inner sep=1.5pt,
				rounded corners=1pt
			] at (0.235,0.71)
			{$x_{\text{symm}}$};

			\node[
				anchor=center,
				align=center,
				font=\fontsize{9pt}{10pt}\selectfont,
				fill=white,
				fill opacity=0,
				text opacity=1,
				inner sep=1.5pt,
				rounded corners=1pt
			] at (0.35,0.71)
			{$x$};

			\node[
				anchor=center,
				align=center,
				font=\fontsize{9pt}{10pt}\selectfont,
				fill=white,
				fill opacity=0,
				text opacity=1,
				inner sep=1.5pt,
				rounded corners=1pt
			] at (0.465,0.71)
			{$M_{72^\circ}x$};

			\node[
				anchor=center,
				align=center,
				font=\fontsize{9pt}{10pt}\selectfont,
				fill=white,
				fill opacity=0,
				text opacity=1,
				inner sep=1.5pt,
				rounded corners=1pt
			] at (0.235,0.39)
			{$M_{144^\circ}x$};

			\node[
				anchor=center,
				align=center,
				font=\fontsize{9pt}{10pt}\selectfont,
				fill=white,
				fill opacity=0,
				text opacity=1,
				inner sep=1.5pt,
				rounded corners=1pt
			] at (0.35,0.39)
			{$M_{216^\circ}x$};

			\node[
				anchor=center,
				align=center,
				font=\fontsize{9pt}{10pt}\selectfont,
				fill=white,
				fill opacity=0,
				text opacity=1,
				inner sep=1.5pt,
				rounded corners=1pt
			] at (0.465,0.39)
			{$M_{288^\circ}x$};

			\node[
				anchor=center,
				align=center,
				font=\fontsize{9pt}{10pt}\selectfont,
				fill=white,
				fill opacity=0,
				text opacity=1,
				inner sep=1.5pt,
				rounded corners=1pt
			] at (0.2832,0.86)
			{$=\hspace{-0.2em}\dfrac{1}{5}$};

			\node[
				anchor=center,
				align=center,
				font=\fontsize{9pt}{10pt}\selectfont,
				fill=white,
				fill opacity=0,
				text opacity=1,
				inner sep=1.5pt,
				rounded corners=1pt
			] at (0.405,0.86)
			{$+$};

			\node[
				anchor=center,
				align=center,
				font=\fontsize{9pt}{10pt}\selectfont,
				fill=white,
				fill opacity=0,
				text opacity=1,
				inner sep=1.5pt,
				rounded corners=1pt
			] at (0.405,0.54)
			{$+$};

			\node[
				anchor=center,
				align=center,
				font=\fontsize{9pt}{10pt}\selectfont,
				fill=white,
				fill opacity=0,
				text opacity=1,
				inner sep=1.5pt,
				rounded corners=1pt
			] at (0.29,0.54)
			{$+$};

			\node[
				anchor=center,
				align=center,
				font=\fontsize{9pt}{10pt}\selectfont,
				fill=white,
				fill opacity=0,
				text opacity=1,
				inner sep=1.5pt,
				rounded corners=1pt
			] at (0.18,0.54)
			{$+$};

			\node[
				anchor=north west,
				text width=0.33\textwidth,
				align=left,
				font=\fontsize{7.5pt}{9pt}\selectfont,
				inner sep=2pt,
				fill=white,
				fill opacity=0,
				text opacity=1,
				rounded corners=1pt
			] at (0.18,0.32) {%
				Symmetrization via averaging mapped latents from all $(R,t)$ relations induced by the given symmetry:
				\[
					\begin{aligned}
						Px & := |\mathcal{G}|^{-1}\sum_{g_n\in\mathcal{G}}
						M_\phi(g_n,\mathcal{C})\,x,                        \\
						Pv & :=\left[P( x_t + (1-t)v_t)-x_t\right]/(1-t).
					\end{aligned}
				\]
			};

			\node[
				anchor=north west,
				text width=0.23\textwidth,
				align=left,
				font=\fontsize{7.5pt}{9pt}\selectfont,
				inner sep=2pt,
				fill=white,
				fill opacity=0,
				text opacity=1,
				rounded corners=1pt
			] at (0.77,0.32) {%
				\rgbSquare{234}{57}{34}  source pose latent \\
				\rgbSquare{56}{94}{149}  target pose latent \\
				\vspace{0.7em}
				Spatial-Transform Mapping on Grid Latent:
				\[
					X_{i,\rgbSquare[0.4em]{56}{94}{149}} = \sum_{j \in \mathrm{Ball}_i} \alpha_{ji}W_{ji}(g,\mathcal{C}_{\rgbSquare[0.4em]{234}{57}{34}/\rgbSquare[0.4em]{56}{94}{149}};\theta) X_{j,\rgbSquare[0.4em]{234}{57}{34}}
				\]
			};

		\end{scope}
	\end{tikzpicture}
	\caption{
		\label{fig:02_method}
		\textbf{Symmetry-guided latent-flow sampling with learned spatial-transform mapping}
		At each sampling step, SymTRELLIS averages the flow prediction across spatial transformations.
		The star example illustrates this operation for five-fold rotational symmetry.
		The same averaging applies to any specified finite symmetry group.
		To make this averaging possible, the learned mapper predicts each spatial transform on the sparse TRELLIS.2 latents.
		It transports features from nearby transformed voxel locations and "rotate" the latent channels under the transform. Refer to the video for the method explanation.
	}
\end{figure*}
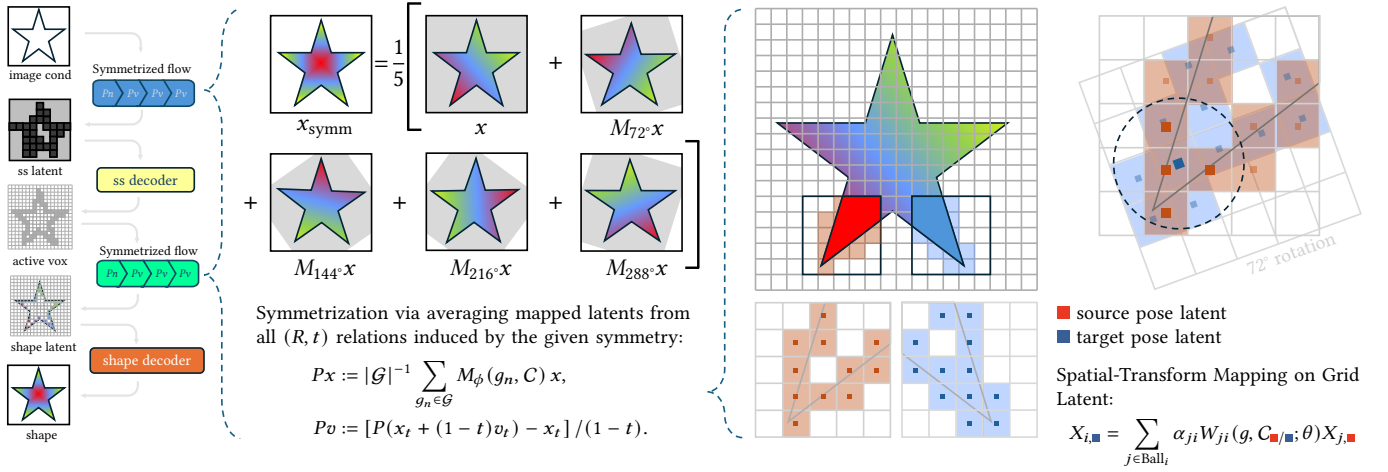

\subsection{Velocity Symmetrization}
\label{sec:3_1_vsymm}

Velocity symmetrization enforces symmetries by averaging flow predictions across symmetry-equivalent transformations at each sampling step.
We apply it to the sparse-structure and shape latent generation stages of TRELLIS.2.

Given a symmetry specification, we construct a finite group of Euclidean transformations
\(\mathcal{G}=\{g_n\}_{n=1}^{N}\subset E(3)\) that includes all possible rotations and reflections.
For an explicit shape representation \(\mathcal{S}\), exact symmetry with respect to $\mathcal{G}$ can be written as
\(\mathcal{S} = g_n \cdot \mathcal{S}\), \(\forall g_n\in\mathcal{G}\).
This suggests symmetrizing imperfect $\mathcal{S}$ can be achieved by averaging transformed shapes or penalizing deviations between \(\mathcal{S}\) and \(g_n\cdot\mathcal{S}\)~\cite{wu2014real}.
However, TRELLIS.2 utilizes an implicit representation,
of which the action of $g\cdot x$ is not defined.
The ideal latent-space action of \(g\) on a latent \(x\) is
\begin{equation}
	\rho_g(x) := \mathrm{Enc}\bigl(g\cdot\mathrm{Dec}(x)\bigr).
\end{equation}
We approximate \(\rho_g\) with a \emph{spatial-transform latent mapper} \(M_\phi(g)\), a transform-conditioned \emph{linear} operator on \(x\),
\begin{equation}
	\rho_g(x) \approx M_\phi(g)x,
\end{equation}
where the choice of linearity in the approximation is justified at the end of \Cref{sec:3_2_mapper}.
This linear form allows us to construct a symmetrization operator by averaging transformation operators,
\begin{equation}
	P_{\mathcal{G}} x
	:=
	|\mathcal{G}|^{-1}
	\sum_{g\in\mathcal{G}} M_\phi(g)x.
\end{equation}
We overload \(P_{\mathcal{G}}\) to denote the induced operation on velocity $v$ by applying the projection to \(\hat{x}_t\),
\begin{equation}
	P_{\mathcal{G}}v_t
	:=
	\frac{P_{\mathcal{G}}\hat{x}_t - x_t}{1-t}.
\end{equation}

This averaging view is inspired by \textit{visual anagrams}~\cite{geng2024visual,chang2025lookingglass}, which combine denoising predictions across transformed image views.
In our setting, the views correspond to 3D spatial transformations of the generated object, and the averaged prediction is employed for symmetry enforcement.

Intuitively, velocity symmetrization treats symmetric parts as transformed instances of the same underlying structure.
Averaging over \(\mathcal{G}\) couples their generation, so each symmetry sector contributes to the others.
This differs from post-processing methods that modify only the final mesh.
Because the coupling is applied during structure and shape generation, the model can recover parts that vanilla TRELLIS.2 misses in one symmetry sector.

In practice, full group averaging can over-smooth the denoised prediction, so we use a subset
\(\mathcal{T}\subseteq\mathcal{G}\) of pure rotations and mirror reflections and replace
\(P_{\mathcal{G}}\) by \(P_{\mathcal{T}}\).
We apply this velocity symmetrization as guidance after the classifier-free guidance stage of TRELLIS.2,
\begin{equation}
	v_t^{\mathrm{sym}}
	=
	v_t^{\mathrm{cfg}}
	+
	\lambda
	\left(
	P_{\mathcal{T}} v_t^{\mathrm{cfg}} - v_t^{\mathrm{cfg}}
	\right),
\end{equation}
where \(\lambda\) controls the symmetry guidance strength.
We apply this symmetry guidance during  duration $t\in[0, \tau]$.

\subsection{Spatial-Transform Latent Mapper}
\label{sec:3_2_mapper}

The spatial-transform latent mapper approximates the latent-space action \(\rho_g\) of a Euclidean transformation \(g\).
Given source and target coordinates, it predicts target features as a transform-conditioned linear map on the source features,
\begin{equation}
	\widehat{X}^{\mathrm{tgt}}=    M_\phi(g;\mathcal{C}^{\mathrm{src}},\mathcal{C}^{\mathrm{tgt}})
	X^{\mathrm{src}}.
\end{equation}
The coordinates are provided as inputs to the mapper, while \(M_\phi\) only transports features and does not create or remove coordinates.
During sampling, we apply the mapper on the current coordinates
\(\mathcal{C}^{\mathrm{src}}=\mathcal{C}^{\mathrm{tgt}}=\mathcal{C}\).
For fixed $\mathcal{C}$ and $g$, \(M_\phi\) defines a linear operator on \(X\).

Constructing this map is nontrivial.
First, rotations and reflections generally move target coordinates off the source sparse grid,
hence there is no canonical one-to-one correspondence between source and target coordinates.
Second, the TRELLIS.2 latent channels are learned features with no prescribed transformation laws under \(E(3)\).
They are not known to be scalars, vectors, tensors, or a mixture of these quantities.
Thus, nearest-neighbor or coordinate interpolation alone cannot define the latent transformation.

We model the mapper as a local feature transport.
Specifically, for each target coordinate \(\mathbf{c}^{\mathrm{tgt}}_i\), we gather source coordinates within radius \(r\) after mapping it back by \(g^{-1}\),
\begin{equation}
	\mathcal{N}_i(g)
	=
	\{j:\|\mathbf{c}^{\mathrm{src}}_j-g^{-1}\mathbf{c}^{\mathrm{tgt}}_i\|\le r\}.
\end{equation}
The mapped feature at \(\mathbf{c}^{\mathrm{tgt}}_i\) is computed by a local transport:
\begin{equation}
	\widehat{X}^{\mathrm{tgt}}_i=    \sum_{j\in\mathcal{N}_i(g)}
	\alpha_{ij} W_{ij}    X^{\mathrm{src}}_j.
\end{equation}
Here \(\alpha_{ij}\) is a scalar aggregation weight-normalized over the source neighbors, and
\(W_{ij}\in\mathbb{R}^{d\times d}\) is a learned feature-space transport matrix, implemented as a residual low-rank map.

A \(g\)-conditioned sparse transformer embeds source coordinates and target coordinates mapped back to the source frame,
and an edge MLP predicts \(\alpha_{ij}\) and \(W_{ij}\) for each local edge.
Additional architectural details are provided in \optionalCref{app:mapper}{the appendix}.

We train \(M_\phi\) on pairs of TRELLIS.2 latents encoded from the same object under two known spatial transformations.
The relative transformation defines \(g\), and the mapper is supervised with an MSE loss against \(X^{\mathrm{tgt}}\).
Importantly, training does not require the underlying objects to be symmetric, allowing us to use large generic 3D datasets.
Symmetry is imposed only at inference time by averaging the mapper over the specified symmetry group.

The linearity-in-feature assumption is motivated by both robustness and structure.
First, since \(M_\phi\) is trained on encoded data latents \(x_1\) but applied during sampling to predicted endpoints \(\hat{x}_t\), a linear mapping reduces nonlinear overfitting on the training distribution.
Second, for fixed coordinates and transform, the operator can also be computed once and reused across flow steps.
Finally, it follows the standard group-representation view of equivariance, where a transformation acts on a latent vector space through a linear operator~\cite{garrido2023self,keurti2023homomorphism}.
Our mapper can be viewed as an engineering instantiation of this principle.

\subsection{Symmetry Detection}
\label{sec:3_3_symm_detect}

In practice, for many examples encountered, we can employ an automatic symmetry detection from vanilla TRELLIS.2 outputs, without the need for
manual symmetry specification, which is applied only when such a detection fails or during symmetry manipulation.
Our detector primarily focuses on global cyclic rotational symmetry \(C_n\), and falls back to reflectional symmetry detection when no reliable rotational symmetry is found.
The proposed velocity symmetrization method itself is not restricted to \(C_n\) and can be applied to any finite point group once its transformations are specified.

Given a 3D shape generated by TRELLIS.2, we detect symmetries by self-registering the shape under many randomized rotations and reflection initializations using ICP.
We then discard solutions with screw or glide components, and cluster the remaining transformations using DBSCAN~\cite{ester1996density}.
Each cluster defines a candidate symmetry-equivalent rotation, with its ICP RMSE used as the fitting error.
The rotation fold is inferred from the rotation clusters and validated using~\cite{huiwang2017group}.
The principal axis is then refined by a final ICP pass initialized from the clustered transformations.
The detector outputs a ranked list of candidate axes and folds with fitting errors.

When available, we select the best-fitting axis with fold \(n\ge 3\);
otherwise, we use the best-fitting two-fold axis.
If no reliable rotation is detected, we use the best-fitting reflection plane as the detected symmetry relation.
If no reliable reflection is found, we
fall back to manual symmetry specification, if any.

\section{Experiments}
\subsection{Training and Evaluation of the Mapper}

\begin{figure}[t!]
	\centering
	\small
	\setlength{\tabcolsep}{2pt}
	\renewcommand{\arraystretch}{1.0}

	\newlength{\mappercellsize}
	\setlength{\mappercellsize}{0.220\columnwidth}

	\newlength{\mapperlabelwidth}
	\setlength{\mapperlabelwidth}{0.045\columnwidth}

	\newlength{\mapperlabelyshift}
	\setlength{\mapperlabelyshift}{0.15\mappercellsize}

	\newlength{\mapperrowsep}
	\setlength{\mapperrowsep}{1pt}

	\newlength{\mappercaptionskip}
	\setlength{\mappercaptionskip}{0.35em}

	\newcommand{\mapperimg}[1]{%
		\raisebox{-0.5\height}{%
			\includegraphics[width=\mappercellsize]{#1}%
		}%
	}

	\newcommand{\mapperrowlabel}[1]{%
		\raisebox{\dimexpr -0.5\height + \mapperlabelyshift\relax}{%
			\makebox[\mapperlabelwidth][c]{%
				\rotatebox[origin=c]{90}{{#1}}%
			}%
		}%
	}
	\vspace{\mappercaptionskip}
	\resizebox{\columnwidth}{!}{%
		\begin{tabular}{@{}ccccc@{}}
			                                                                                &
			Source                                                                          &
			Target                                                                          &
			Nearest neighbor                                                                &
			Mapped                                                                            \\[3pt]
			\mapperrowlabel{Active voxel}                                                   &
			\mapperimg{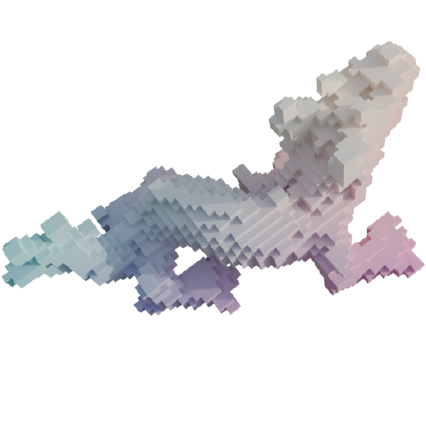}                  &
			\mapperimg{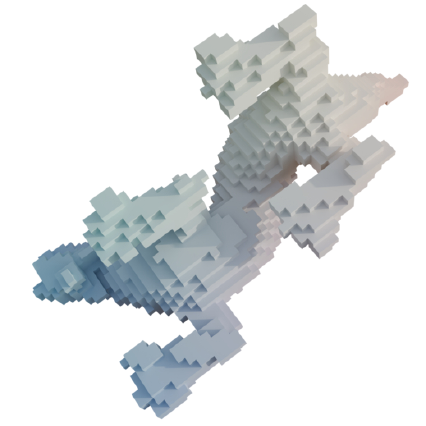}          &
			\mapperimg{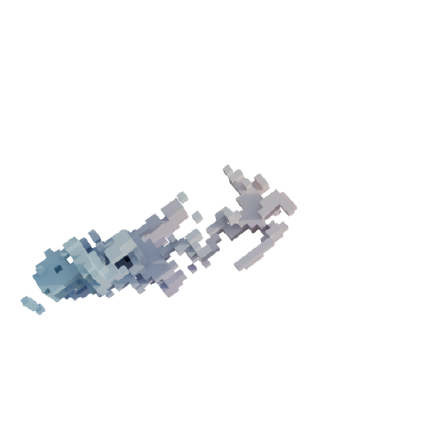} &
			\mapperimg{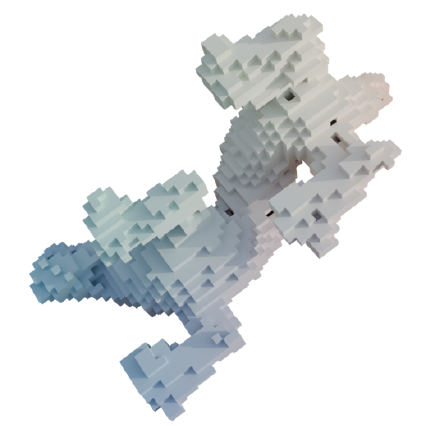}             \\[\mapperrowsep]
			\mapperrowlabel{Shape}                                                          &
			\mapperimg{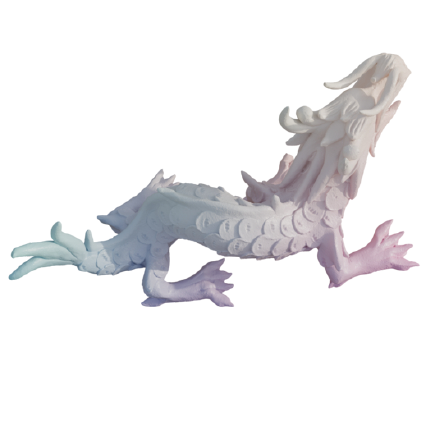}                         &
			\mapperimg{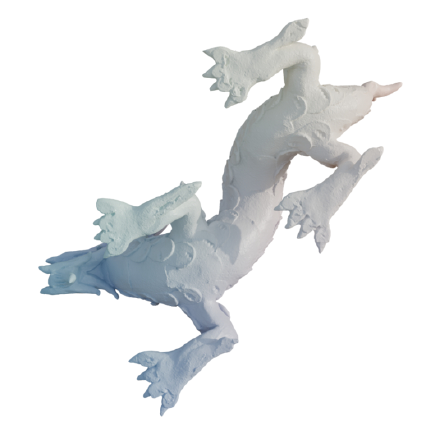}                 &
			\mapperimg{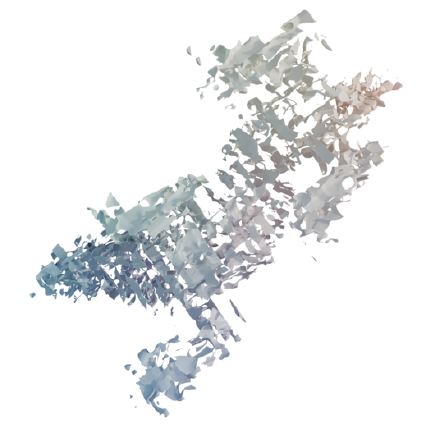}        &
			\mapperimg{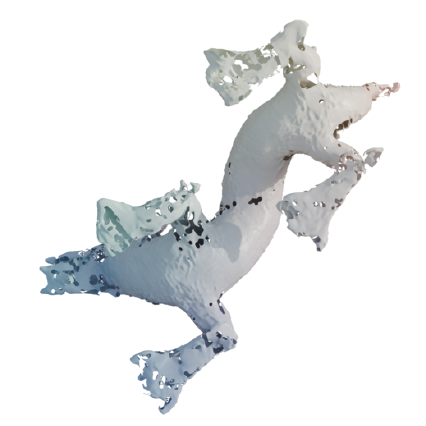}
		\end{tabular}%
	}
	\caption{
		\label{fig:03_mapper}
		Performance of the spatial-transform latent mapper.
		Rows show decoded active-voxel occupancy and decoded shape latents.
		Columns show the source latent, target latent, nearest-neighbor interpolation baseline, and our mapped latent.
		Our mapper recovers occupancy close to the target for active voxels, while partially correcting the normal-level detail for shape latents. In contrast, the nearest-neighbor baseline fails to preserve the target structure, showing that spatial transformation in latent space is non-trivial.
	}
\end{figure}

We train separate spatial-transform latent mappers for the sparse-structure and shape latents using the Sketchfab subset of Objaverse-XL~\cite{deitke2023objaverse,deitke2023objaversexl}, and evaluate them on Toys4K.
For each shape, we sample two random transformations, each consisting of rotations, translations, and reflections.
We then encode the transformed shapes using TRELLIS.2 encoders and use the relative transformation between them as $g$.

Both mappers use a 12M-parameter 3D Swin Transformer backbone.
We first train the networks with a latent-space MSE loss for $25,000$ steps,
and then fine-tune them for an additional $12,500$ steps with a decoder-space feature MSE loss weighted by 0.002.
Both training use AdamW with a learning rate of \(10^{-3}\) and batch size \(256\).
Detailed architecture can be found in the supplementary.

We evaluate the latent mappers by measuring the cosine similarity between the predicted transformed latent and the ground-truth transformed latent.
On Toys4K, the sparse-structure mapper achieves \(59\%\) mean cosine similarity and \(75\%\) IoU on decoded active voxels,
while the shape-latent mapper achieves \(88\%\) mean cosine similarity.
As shown in \Cref{fig:03_mapper}, the sparse-structure mapper is accurate enough to recover occupancy, while the shape-latent mapper preserves only part of the normal-level detail.
Therefore, We use the shape mapper only for symmetry guidance, as it is not yet sufficiently reliable as a standalone latent copy-paste tool.
\optionalCref{tab:mapper_quantitative}{Table~B1} further reports cosine similarities for identity, inverse, composition, and closure.
The similarities decrease under successive group action compositions.
This reaffirms that our mappers are not exact; they \emph{approximate} group actions in latent space.

\begin{table*}[t!]
	\centering
	\caption{Symmetry metrics and reconstruction metrics.
		$\mathrm{SD}$, $\varepsilon$-Err, and CD are reported in units of $10^{-3}$.
		Fold accuracies are reported in percentages.}
	\label{tab:main_results}

	\resizebox{1.0\textwidth}{!}{
		\begin{tabular}{lcccccccccc}
			\toprule
			 & \multicolumn{9}{c}{Symmetry}
			 & \multicolumn{1}{c}{Reconstruction}                                                                     \\
			\cmidrule(lr){2-10} \cmidrule(lr){11-11}

			\multirow{2}{*}{Method}
			 & \multirow{2}{*}{\makecell[c]{$\mathrm{SD}_{\max}\downarrow$}}
			 & \multirow{2}{*}{\makecell[c]{$\mathrm{SD}_{\mathrm{avg}}\downarrow$}}
			 & \multicolumn{3}{c}{$\varepsilon\text{-Err}_{\max}\downarrow$}
			 & \multicolumn{3}{c}{$\varepsilon\text{-Err}_{\mathrm{avg}}\downarrow$}
			 & \multirow{2}{*}{\makecell[c]{Fold acc. $\uparrow$}}
			 & \multirow{2}{*}{\makecell[c]{Chamfer Dist (CD) $\downarrow$}}                                          \\
			\cmidrule(lr){4-6}
			\cmidrule(lr){7-9}

			 &
			 &
			 & $\varepsilon=0.01$
			 & $\varepsilon=0.03$
			 & $\varepsilon=0.1$
			 & $\varepsilon=0.01$
			 & $\varepsilon=0.03$
			 & $\varepsilon=0.1$
			 &
			 &                                                                                                        \\
			\midrule

			TripoSG
			 & 14.97                                                                 & 9.81
			 & 415.17                                                                & 148.79         & 11.26
			 & 271.31                                                                & 83.34          & 6.10
			 & 78.45\%                                                               & 18.95                          \\

			Hunyuan3D-2.1
			 & 14.19                                                                 & 9.33
			 & 368.38                                                                & 141.91         & 13.33
			 & 242.07                                                                & 82.02          & 7.41
			 & 73.99\%                                                               & 20.32                          \\

			\midrule

			TRELLIS.2
			 & 7.47                                                                  & 4.58
			 & 200.03                                                                & 63.43          & 5.61
			 & 118.42                                                                & 34.09          & 2.74
			 & 83.96\%                                                               & \textbf{13.79}                 \\

			SymTRELLIS
			 & 3.35                                                                  & 2.24
			 & 54.43                                                                 & 26.61          & 4.66
			 & 39.35                                                                 & 17.97          & 2.76
			 & 83.81\%                                                               & 15.33                          \\

			SymTRELLIS with gt-symm
			 & \textbf{2.80}                                                         & \textbf{1.80}
			 & \textbf{45.36}                                                        & \textbf{21.99} & \textbf{3.81}
			 & \textbf{31.63}                                                        & \textbf{13.79} & \textbf{2.17}
			 & \textbf{93.53\%}                                                      & 16.02                          \\

			\bottomrule
		\end{tabular}
	}
\end{table*}

\subsection{Symmetry Dataset and Evaluation}
\paragraph{Symmetry Dataset}
We curate an evaluation dataset of 266 strictly symmetric meshes, covering \(2\)- to \(20\)-fold rotational, reflection, tetrahedral, octahedral, and icosahedral symmetries.
The dataset includes turbine blades, historical relics, electric fans, furniture, cartoon characters, and vehicles, among others.
Although a shape may contain multiple rotation axes or reflection planes, we evaluate only the dominant rotational symmetry, or reflection symmetry when it is the only symmetry relation present.
All meshes are normalized to the \([-0.5,0.5]^3\) cube.
For rotational cases, the major rotation axis passes through the origin.
For reflection-only cases, the reflection plane is aligned with the \(yz\)-plane.
We render 64 views per shape and use them as 2D image conditions for each baseline to generate 3D shapes for evaluation.

\paragraph{Evaluation Metrics}
We evaluate two aspects of 3D generation, reconstruction accuracy and symmetry quality.
Let \(\mathcal{M}\) denote the ground-truth mesh and \(\widehat{\mathcal{M}}\) the generated mesh after rescaling and ICP alignment to \(\mathcal{M}\).
For reconstruction, we compute the Chamfer distance between \(\mathcal{M}\) and \(\widehat{\mathcal{M}}\).

To evaluate symmetry, we measure the self-consistency of \(\widehat{\mathcal{M}}\) under the detected symmetry transformation set \(\mathcal{T}\).
Let \(d(\mathbf{p},\mathcal{M})\) denote the point-to-surface distance from point \(\mathbf{p}\) to mesh \(\mathcal{M}\).
We define the max and average symmetry distance as
\begin{align}
	\mathrm{SD}_{\max}
	 & :=
	\mathbb{E}_{\mathbf{p}\in \widehat{\mathcal{M}}}
	\left[
		\max_{g\in\mathcal{T}}
		d(\mathbf{p}, g\cdot\widehat{\mathcal{M}})
	\right], \\
	\mathrm{SD}_{\mathrm{avg}}
	 & :=
	\mathbb{E}_{\mathbf{p}\in \widehat{\mathcal{M}}}
	\left[
		|\mathcal{T}|^{-1}
		\sum_{g\in\mathcal{T}}
		d(\mathbf{p}, g\cdot\widehat{\mathcal{M}})
		\right].
\end{align}

For rotational symmetry, distance-based errors can become large for points far from the rotation axis.
We therefore additionally report thresholded symmetry errors,
\begin{align}
	\varepsilon\text{-Err}_{\max}
	 & :=
	\mathbb{E}_{\mathbf{p}\in \widehat{\mathcal{M}}}
	\left[
		\max_{g\in\mathcal{T}}
		\mathbbm{1}
		\left[
			d(\mathbf{p}, g\cdot\widehat{\mathcal{M}}) > \varepsilon
			\right]
	\right], \\
	\varepsilon\text{-Err}_{\mathrm{avg}}
	 & :=
	\mathbb{E}_{\mathbf{p}\in \widehat{\mathcal{M}}}
	\left[
		|\mathcal{T}|^{-1}
		\sum_{g\in\mathcal{T}}
		\mathbbm{1}
		\left[
			d(\mathbf{p}, g\cdot\widehat{\mathcal{M}}) > \varepsilon
			\right]
		\right].
\end{align}
We also report predicted fold accuracy $\mathbb{E}[\mathbbm{1}\{\text{fold}_{\text{pred}}=\text{fold}_{\text{gt}}\}]$.

\subsection{Image-to-3D Symmetrization Results}

We use TRELLIS.2~\cite{xiang2025trellis2} as the voxel-based baseline, and include two representative non-voxel-based 3D generation models,
Hunyuan3D-2.1~\cite{yang2024hunyuan3d,hunyuan3d22025tencent,hunyuan3d2025hunyuan3d} and TripoSG~\cite{li2025triposg}.
For our method, we set $\lambda_{\text{ss}}=\lambda_{\text{shape}}=1.0$, $\tau=0.3$,
and evaluate two settings: using folds detected from shapes generated by vanilla TRELLIS.2, or the ground-truth fold.
The quantitative comparison is shown in \Cref{tab:main_results}.
Our method produces shapes with stronger symmetry and more accurate fold prediction than the baselines.
With symmetry enforcement, all symmetry errors are substantially reduced, with only a slight decrease in reconstruction accuracy.
Although using ground-truth folds further improves fold accuracy, perfect fold prediction is still not guaranteed.
We observe that the flow model has a bias toward common folds such as \(4,6,8,12,\) and \(16\).
Despite these biases, our symmetry enforcement enables the model to generate shapes with accurate target folds.

\paragraph{Ablation}
\label{sec:ablation}
\begin{table}[t!]
	\begin{center}
		\caption{Ablation on symmetrization at two stages of TRELLIS.2 generation.}
		\label{tab:ablation}
		\resizebox{1.0\columnwidth}{!}{
			\begin{tabular}{lcccc}
				\toprule

				 & \multicolumn{3}{c}{$\varepsilon\text{-Err}_{\max/\mathrm{avg}}\downarrow$ ($\times 10^3$)}
				 & \multirow{2}{*}{\makecell[c]{CD $\downarrow$                                               \\ ($\times 10^3$)}} \\
				\cmidrule(lr){2-4}
				 & $\varepsilon=0.003$
				 & $\varepsilon=0.01$
				 & $\varepsilon=0.03$
				 &                                                                                            \\
				\midrule

				Vanilla
				 & 396.89/272.08
				 & 208.36/124.46
				 & 68.24/37.09
				 & \textbf{13.90}                                                                             \\

				Sparse structure
				 & 247.51/143.44
				 & 71.55/42.98
				 & 23.61/14.53
				 & 16.19                                                                                      \\

				Full
				 & \textbf{110.76/68.52}
				 & \textbf{45.29/31.54}
				 & \textbf{22.11/13.80}
				 & 16.30                                                                                      \\

				\bottomrule
			\end{tabular}
		}
	\end{center}

\end{table}

Figure~\ref{fig:ablation} provides a qualitative comparison of symmetry enforcement at different stages of generation.
Enforcing symmetry only during sparse-structure generation already substantially improves the global symmetric layout of the generated shape.
Applying symmetry enforcement to both sparse-structure and shape-latent generation further improves reconstruction quality and local geometric consistency,
which is also reflected in the quantitative results shown in Table~\ref{tab:ablation}.

\section{Discussions}
\label{sec:discussions}
\paragraph{Complex symmetry.}
Although optimization-based symmetrization methods before the deep-learning era can handle complex symmetry~\cite{wu2014real}, to our knowledge,
our method is the first deep learning framework to enforce finite point-group symmetry beyond reflection symmetry.
Our method supports arbitrary finite point groups in 3D, including cyclic and dihedral groups with rotations and reflections, as well as polyhedral groups such as tetrahedral, octahedral, and icosahedral symmetries.

\begin{figure*}[t!]
	\centering
	\small
	\setlength{\tabcolsep}{2pt}
	\renewcommand{\arraystretch}{0.8}

	\newlength{\visualresultsize}
	\setlength{\visualresultsize}{0.18\textwidth}

	\newcommand{\placeholder}[1]{%
		\begingroup
		\setlength{\fboxsep}{0pt}%
		\fbox{%
			\begin{minipage}[c][\visualresultsize][c]{\visualresultsize}
				\centering
				\vspace*{\fill}
				{\small #1}
				\vspace*{\fill}
			\end{minipage}%
		}%
		\endgroup
	}

	\newcommand{\zoomin}[2]{%
		\begin{tikzpicture}
			\node[inner sep=0, anchor=south west] (base) at (0,0) {%
				\includegraphics[
					width=\visualresultsize,
					height=\visualresultsize
				]{#1}%
			};
			\begin{scope}[
					x={(base.south east)},
					y={(base.north west)}
				]
				\node[inner sep=0, anchor=south west] at (0.70,0.70) {%
					\includegraphics[
						width=0.37\visualresultsize,
						height=0.37\visualresultsize
					]{#2}%
				};
			\end{scope}
		\end{tikzpicture}%
	}

	\newcommand{\zoominii}[2]{%
		\begin{tikzpicture}
			\node[inner sep=0, anchor=south west] (base) at (0,0) {%
				\includegraphics[
					width=\visualresultsize,
					height=\visualresultsize
				]{#1}%
			};
			\begin{scope}[
					x={(base.south east)},
					y={(base.north west)}
				]
				\node[inner sep=0, anchor=south west] at (0.73,0.02) {%
					\includegraphics[
						width=0.40\visualresultsize,
						height=0.40\visualresultsize
					]{#2}%
				};
			\end{scope}
		\end{tikzpicture}%
	}

	\begin{tabular}{ccccc}
		Input image                                                                                                                               &
		TripoSG                                                                                                                                   &
		Hunyuan3D-2.1                                                                                                                             &
		TRELLIS.2                                                                                                                                 &
		SymTRELLIS (Ours)                                                                                                                           \\[6pt]

		\includegraphics[width=\visualresultsize,height=\visualresultsize]{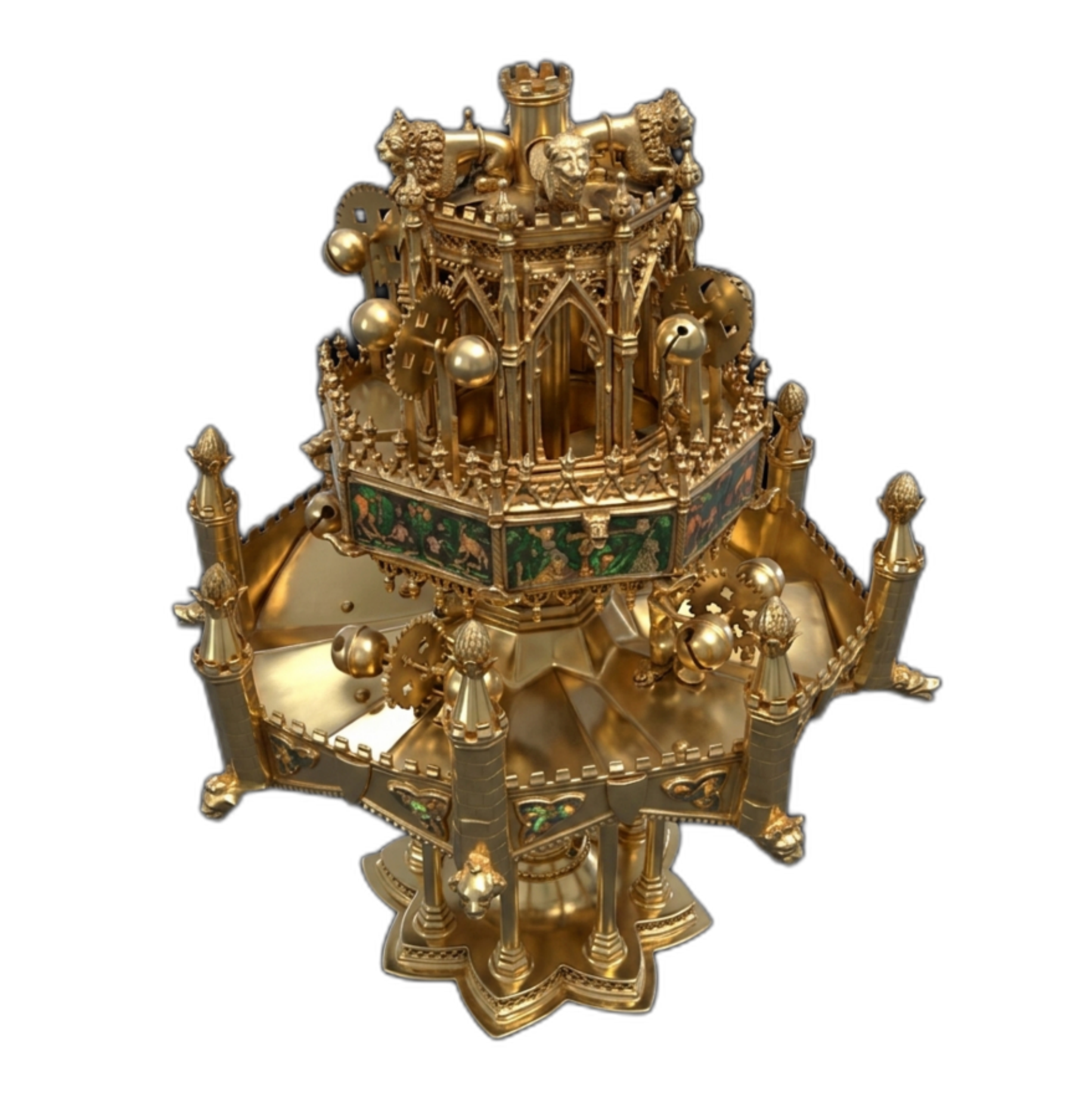}                     &
		\zoomin {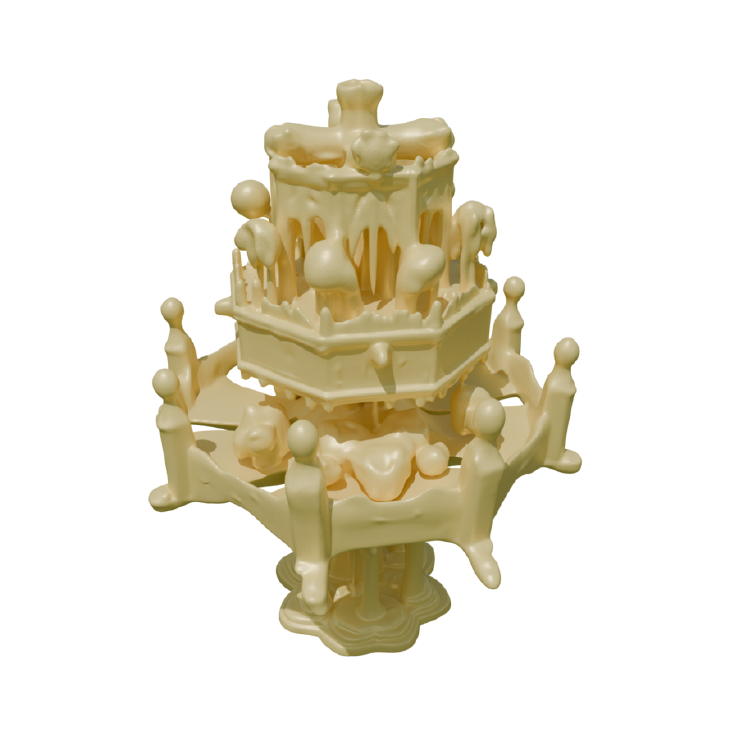} {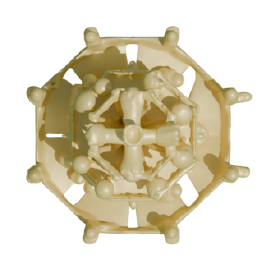}                         &
		\zoomin {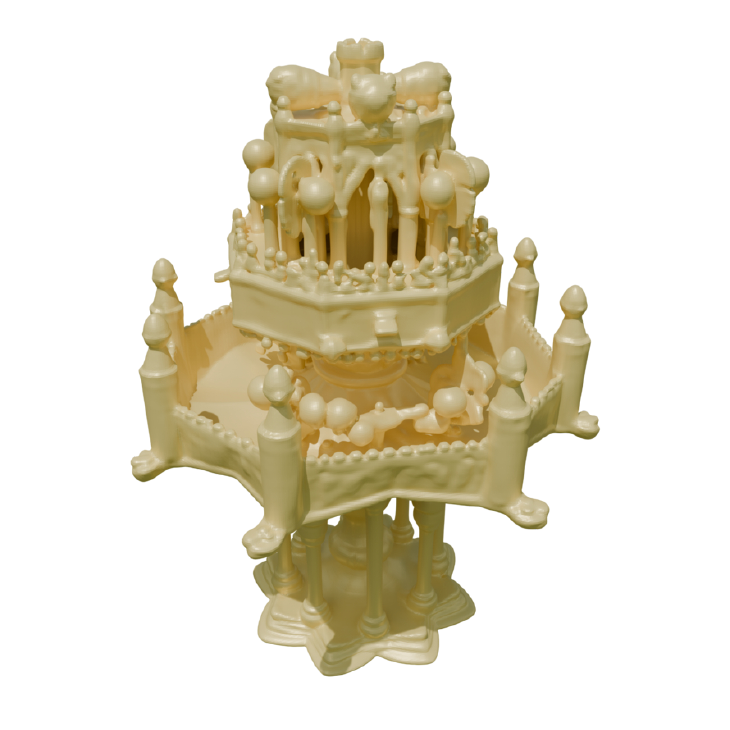} {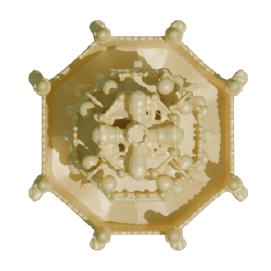}             &
		\zoomin {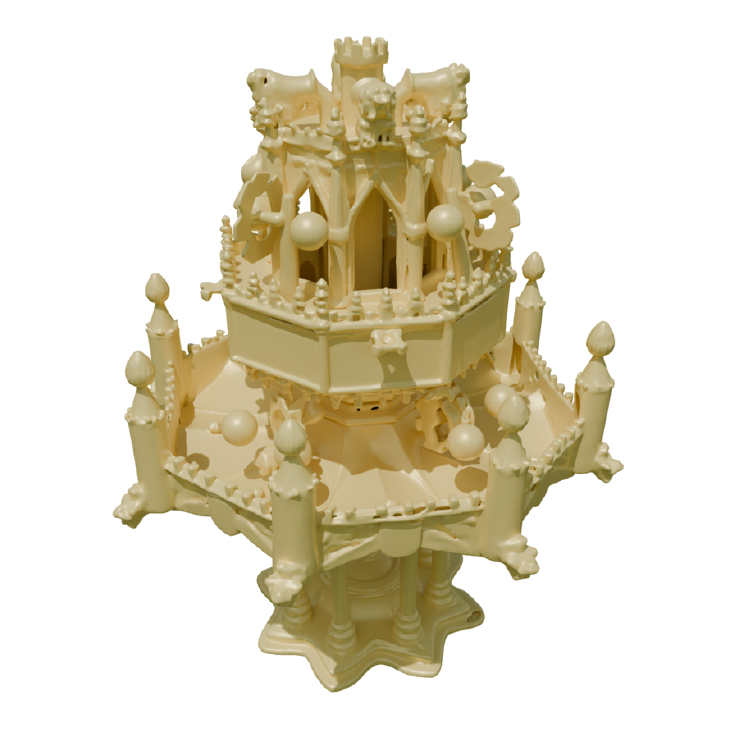} {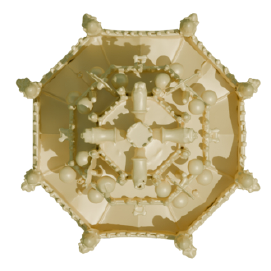}       &
		\zoomin {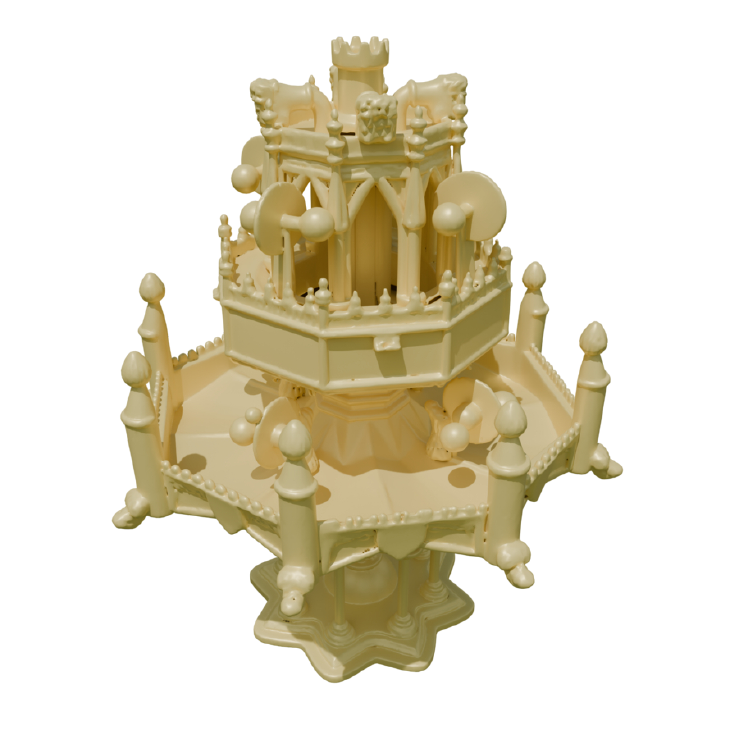} {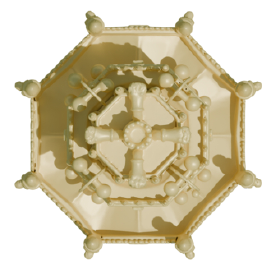}                     \\[4pt]

		\includegraphics[width=\visualresultsize,height=\visualresultsize]{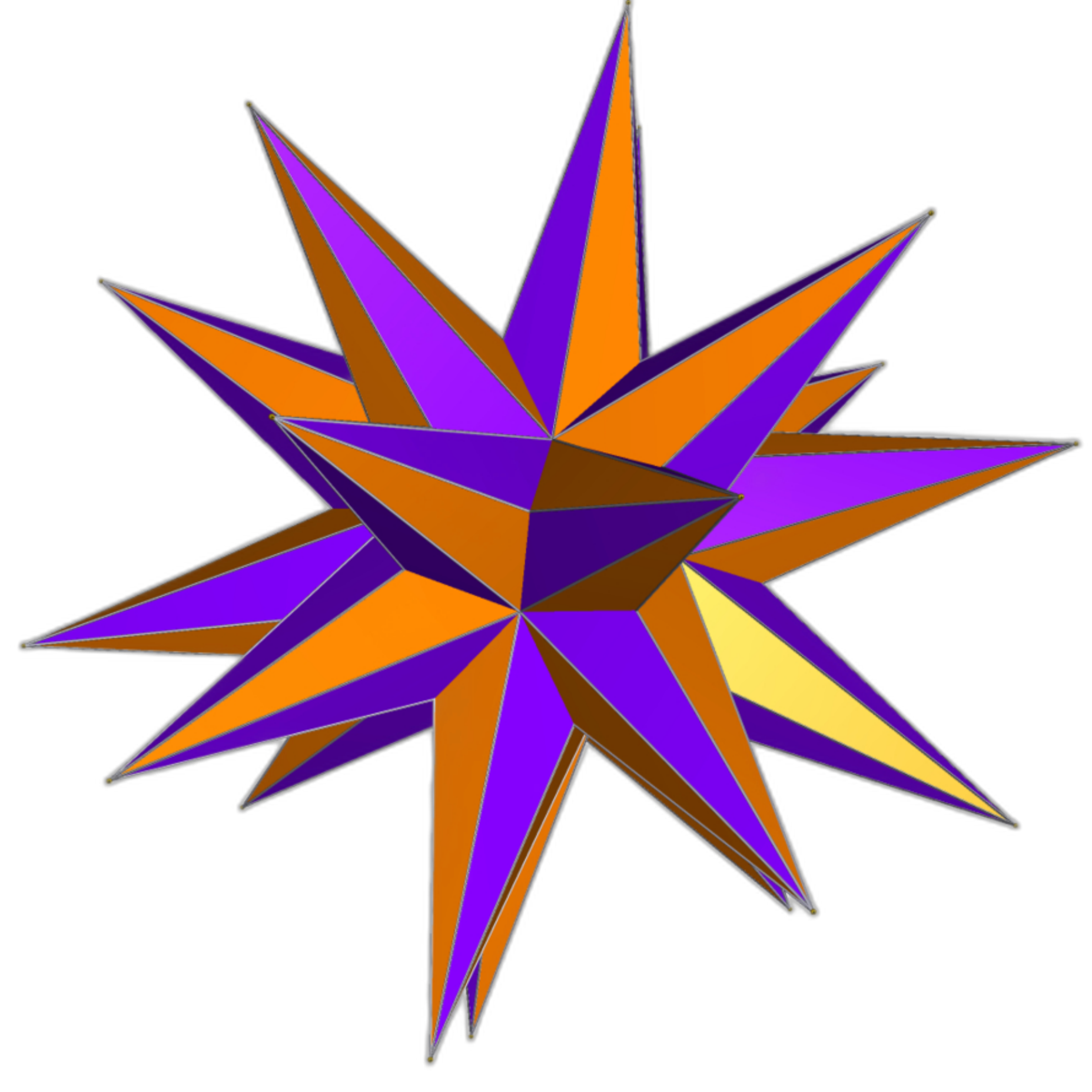}                  &
		\zoomin {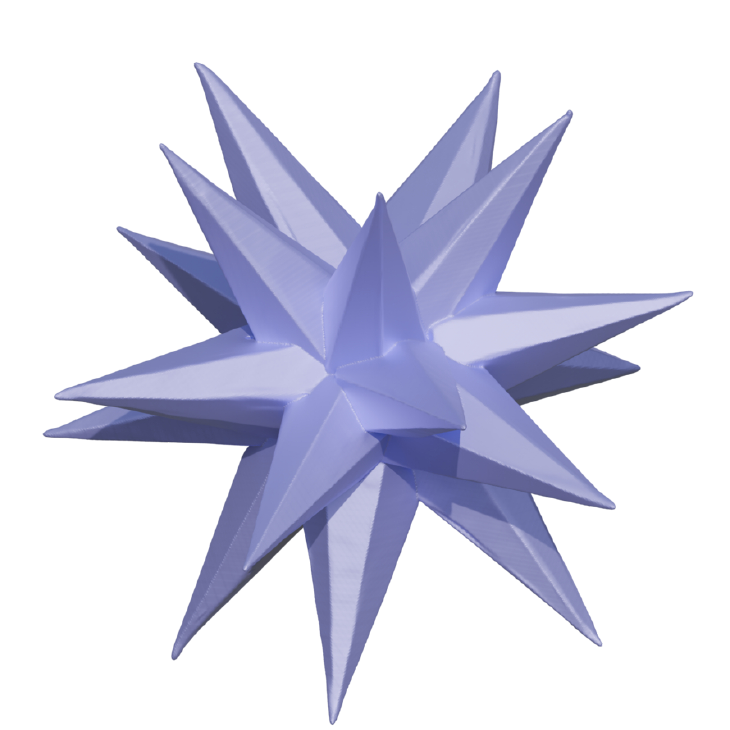} {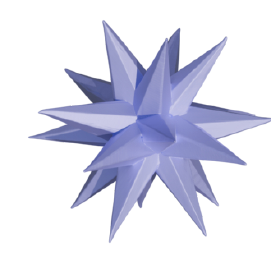}                      &
		\zoomin {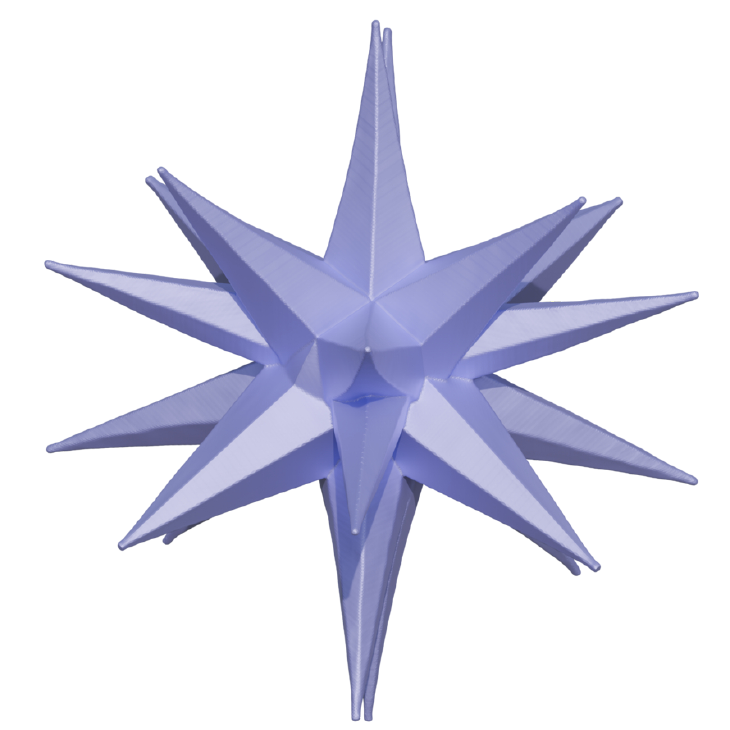} {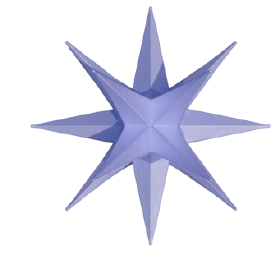}                      &
		\zoomin {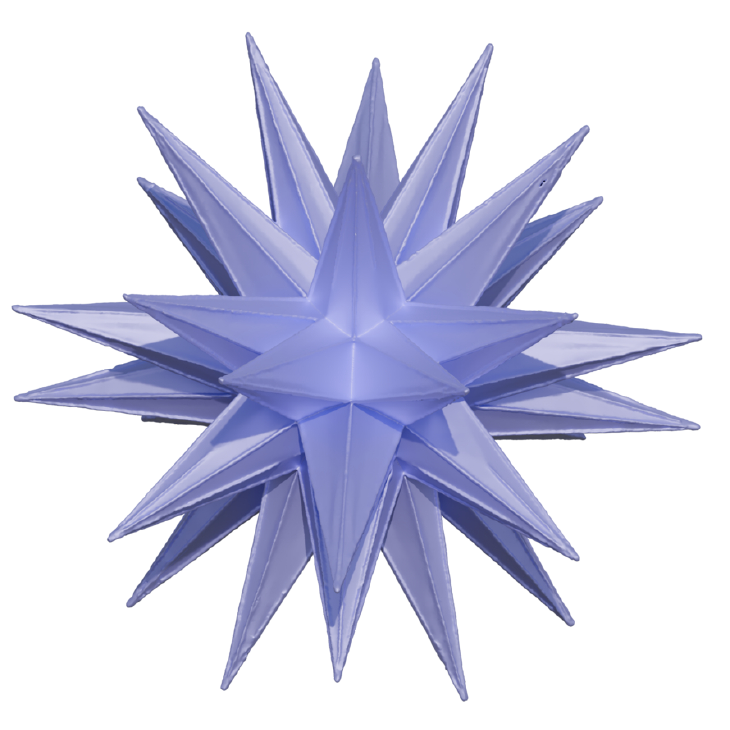} {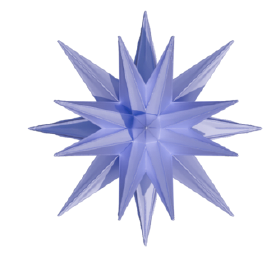}                    &
		\zoomin {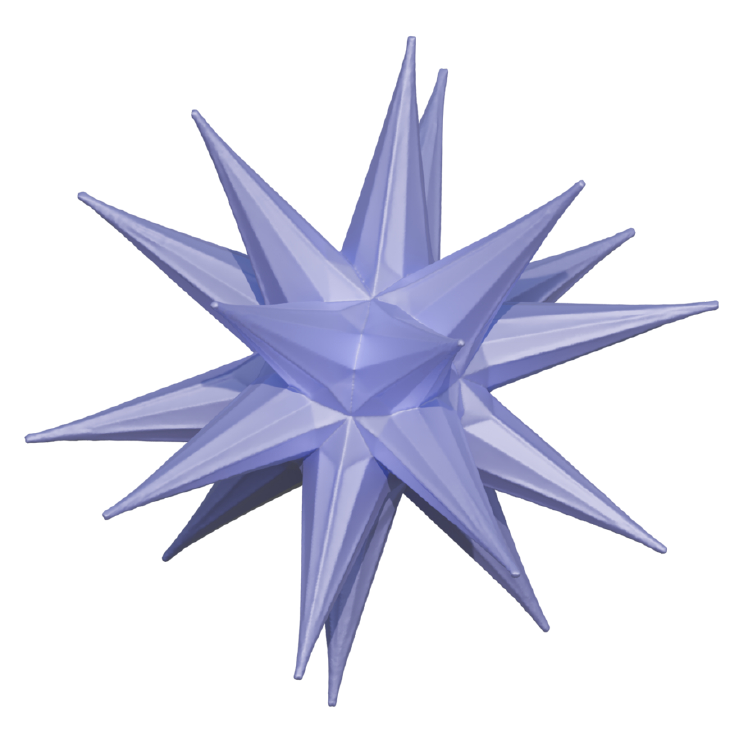} {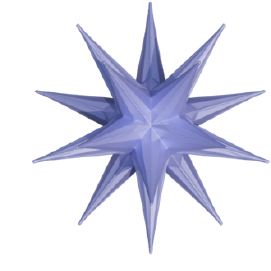}                  \\[4pt]

		\includegraphics[width=\visualresultsize,height=\visualresultsize]{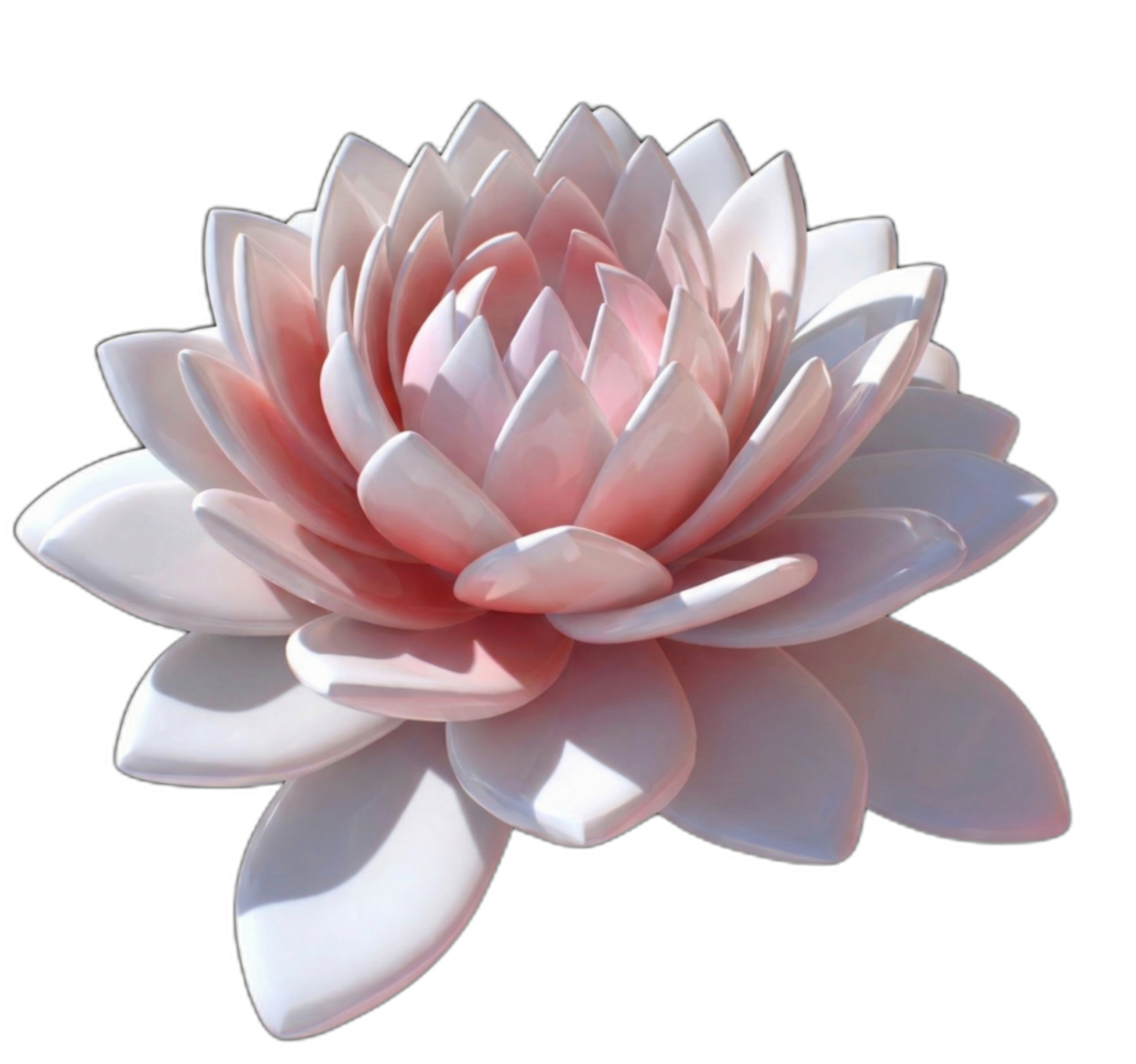}                        &
		\zoomin {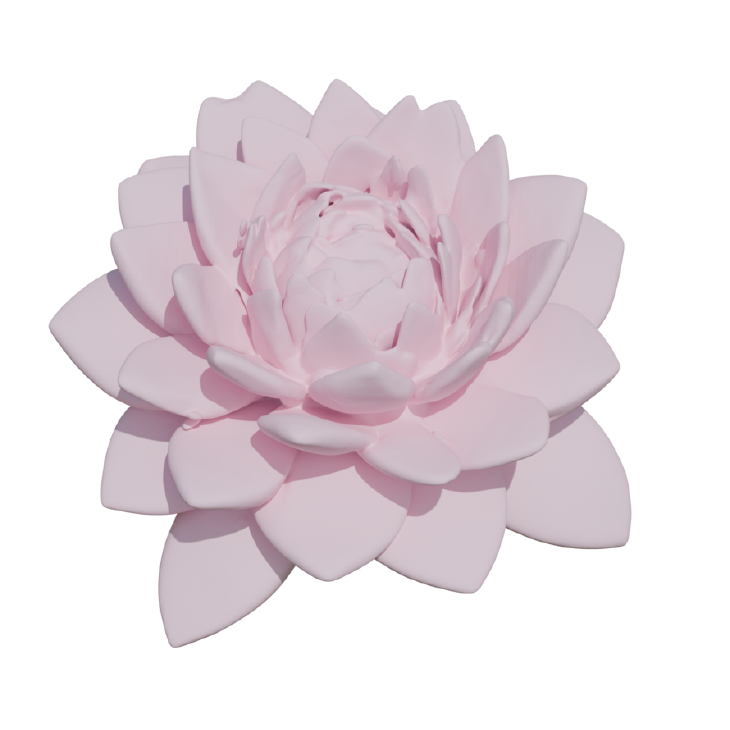} {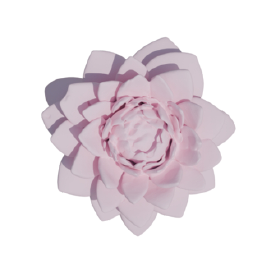}                                  &
		\zoomin {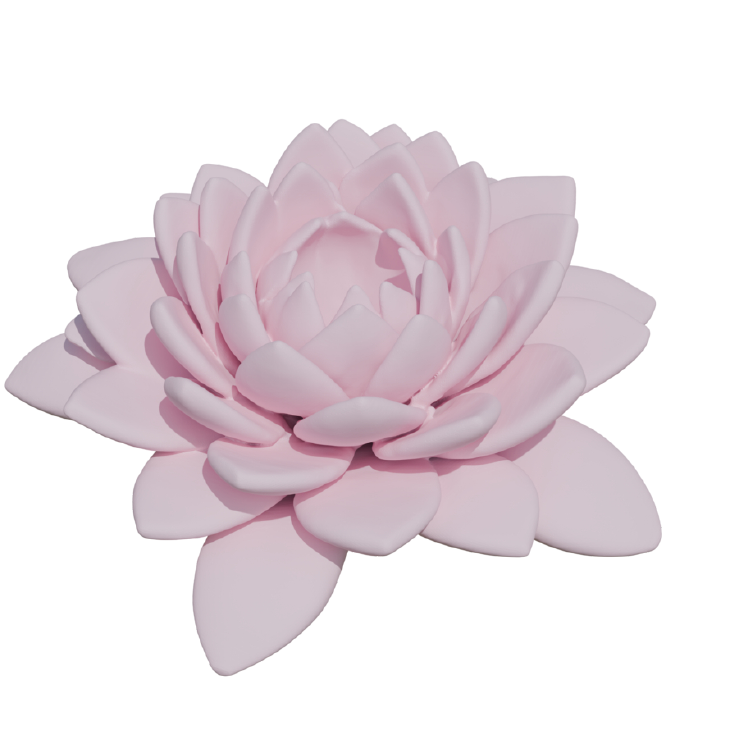} {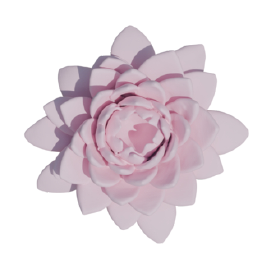}                                  &
		\zoomin {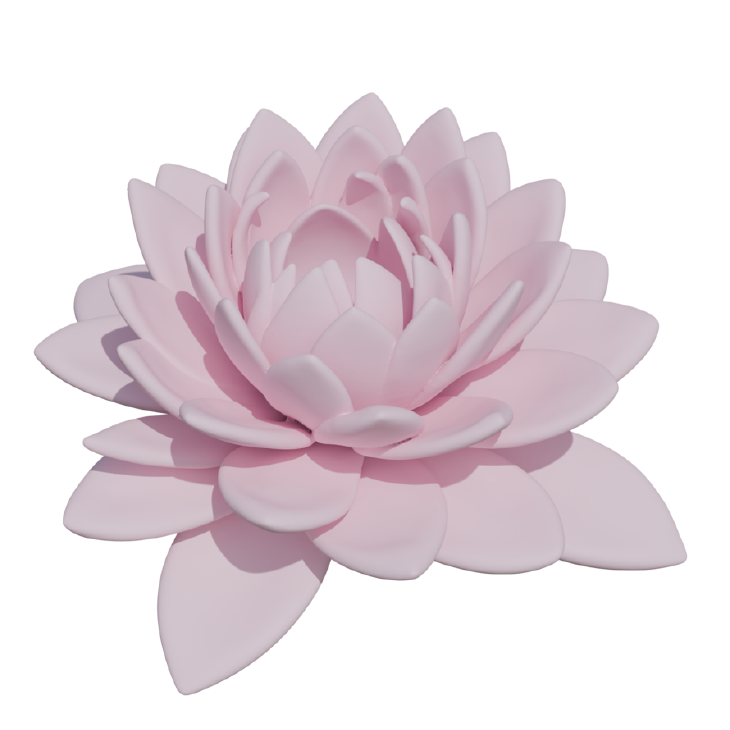} {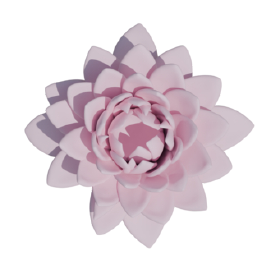}                                &
		\zoomin {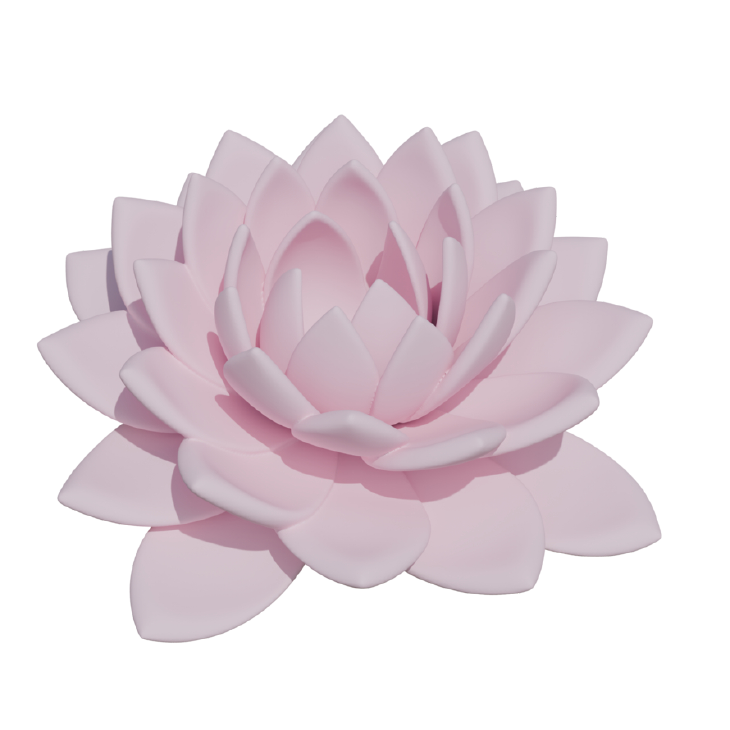} {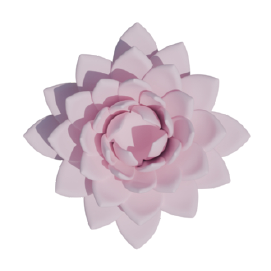}                              \\[4pt]

		\includegraphics[width=\visualresultsize,height=\visualresultsize]{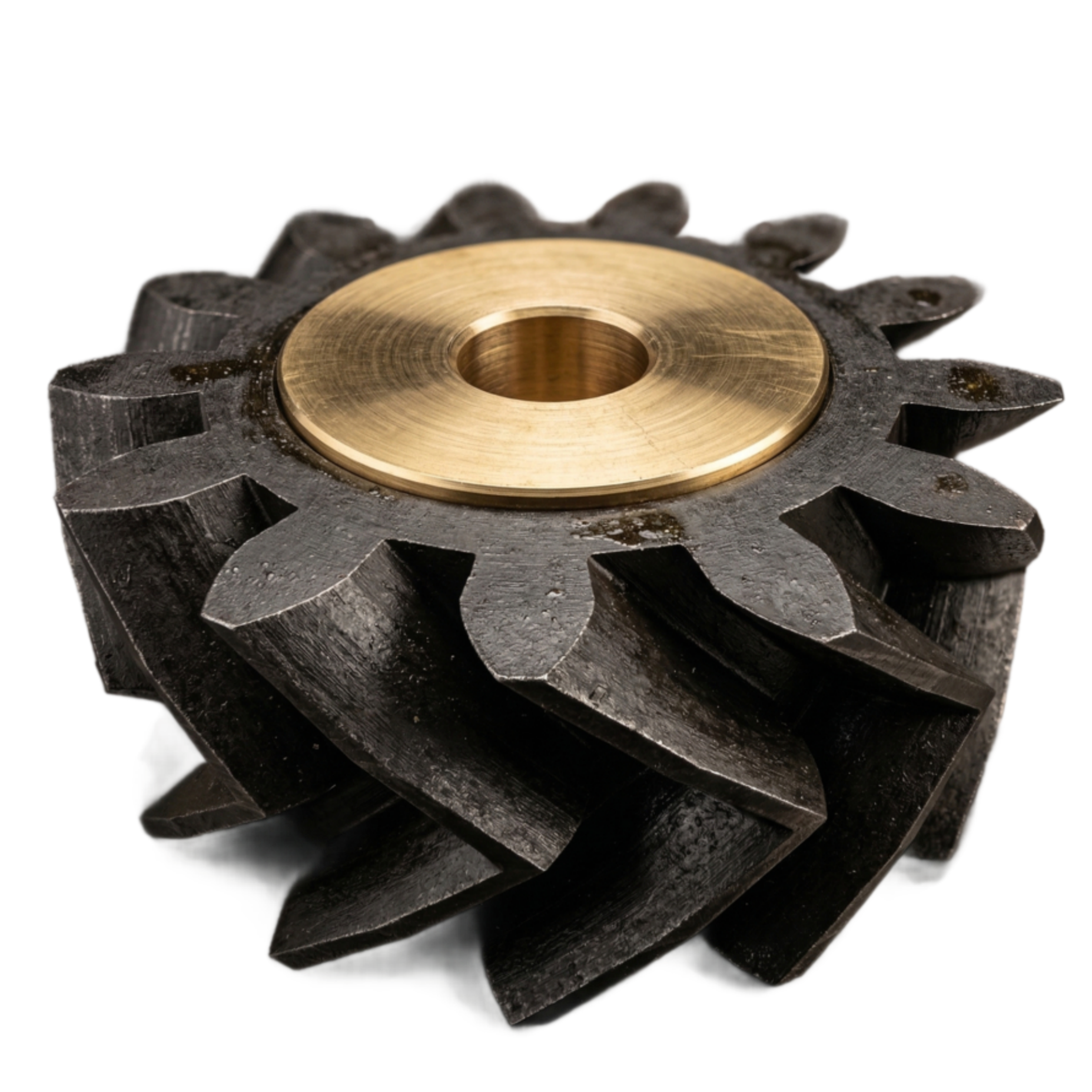}                         &
		\zoomin {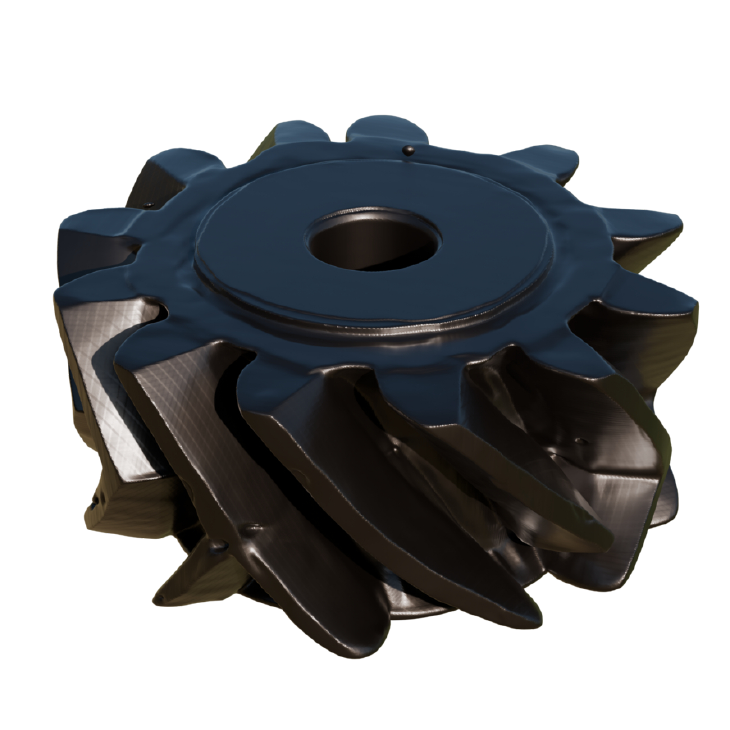} {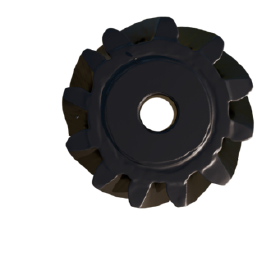}                                    &
		\zoomin {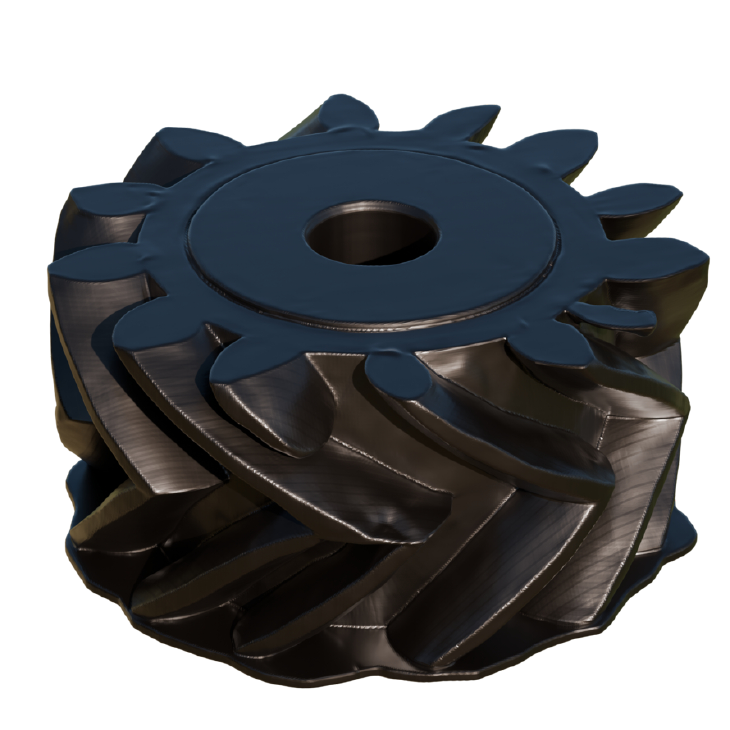} {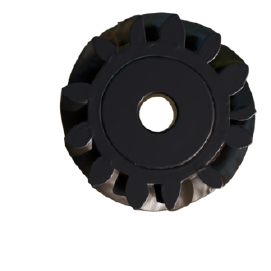}                                    &
		\zoomin {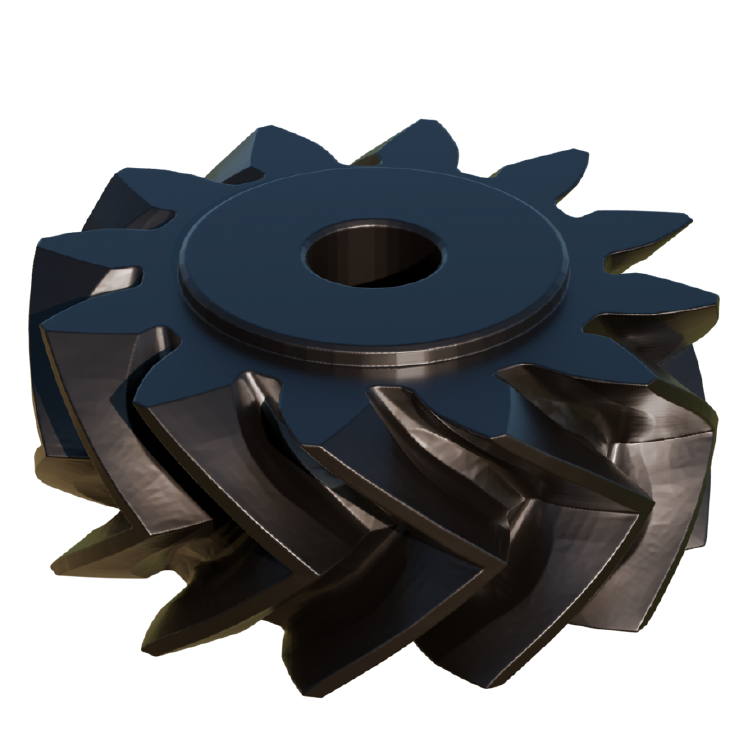} {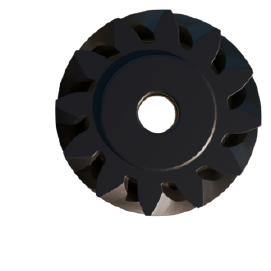}                                  &
		\zoomin {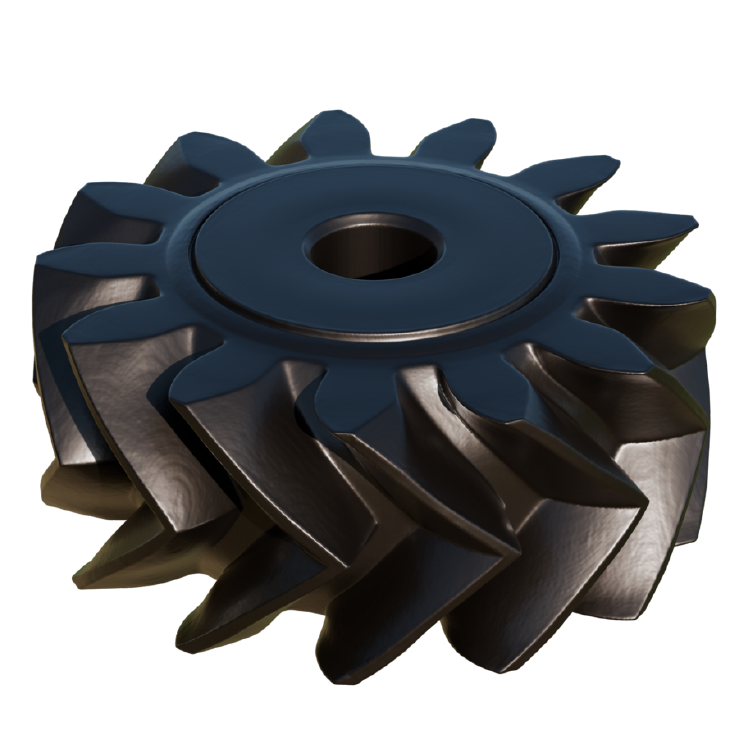} {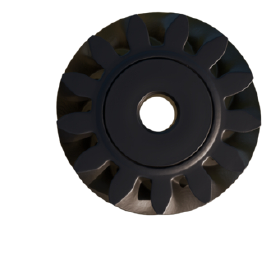}                                \\[4pt]

		\includegraphics[width=\visualresultsize,height=\visualresultsize]{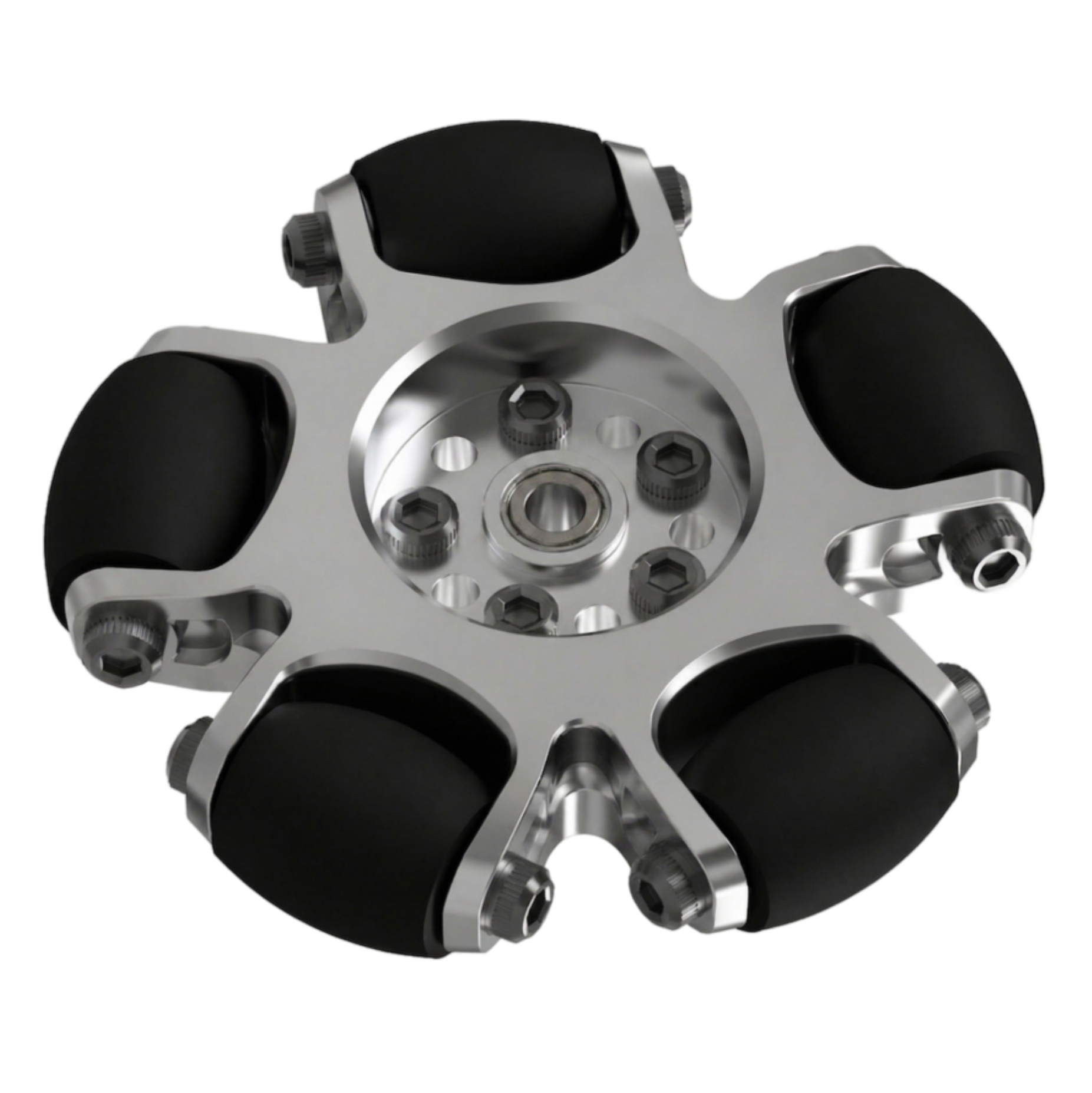}                    &
		\zoomin {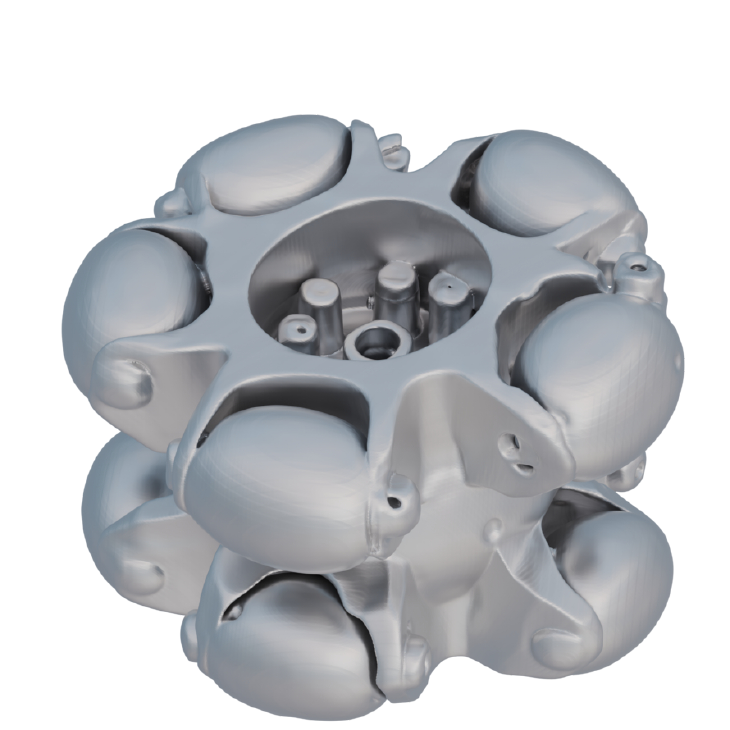} {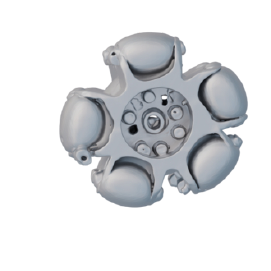}                          &
		\zoomin {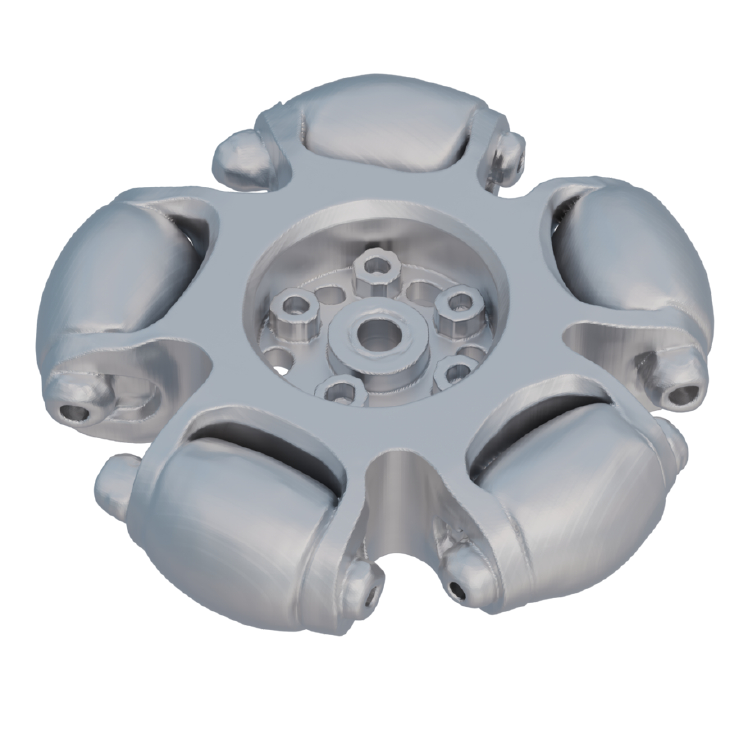} {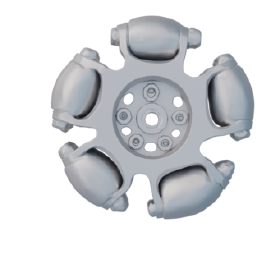}                          &
		\zoomin {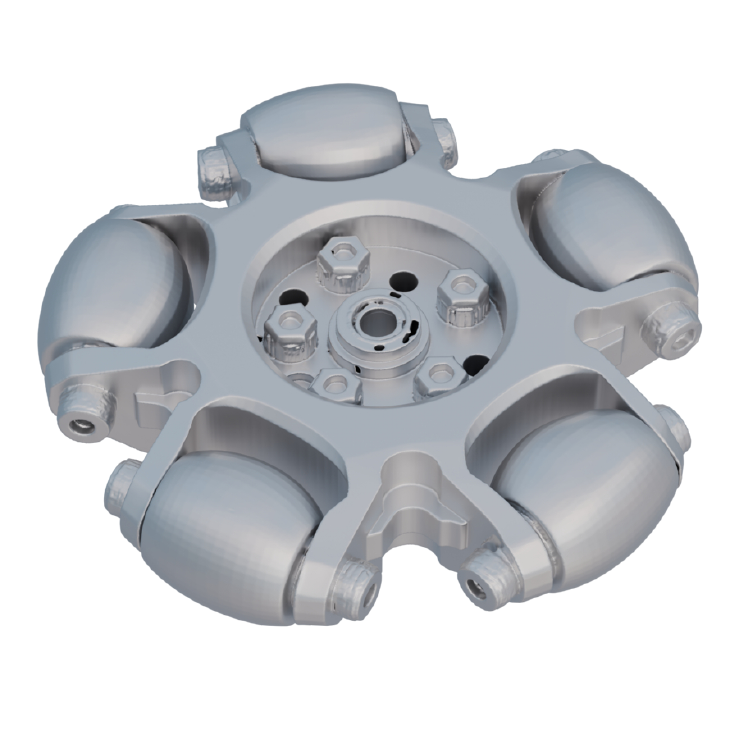} {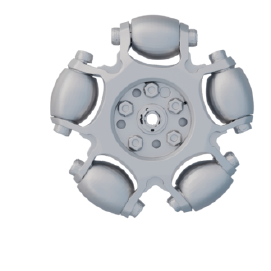}                        &
		\zoomin {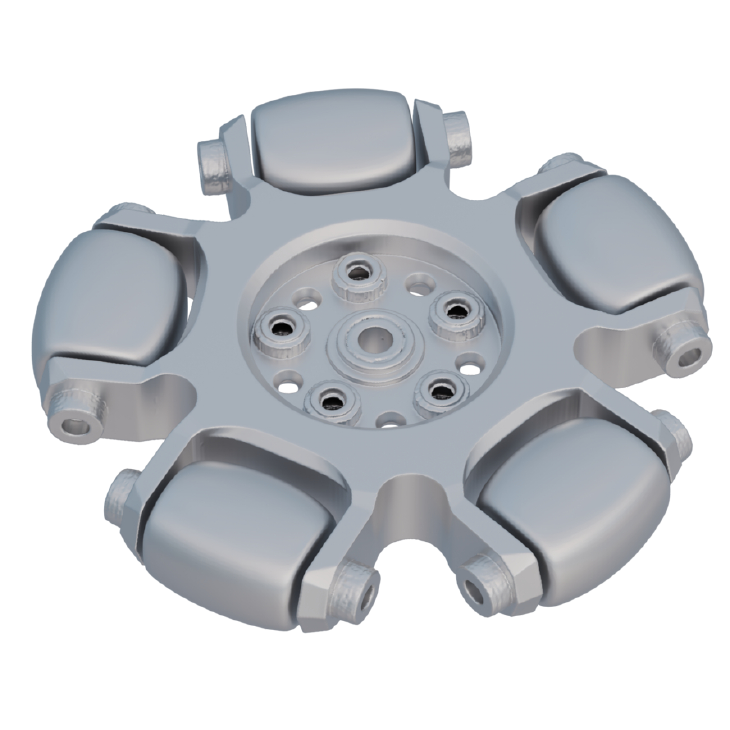} {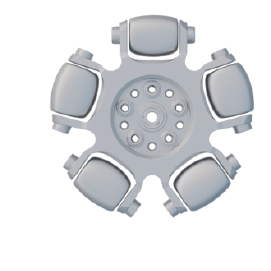}                      \\[4pt]

		\includegraphics[width=\visualresultsize,height=\visualresultsize]{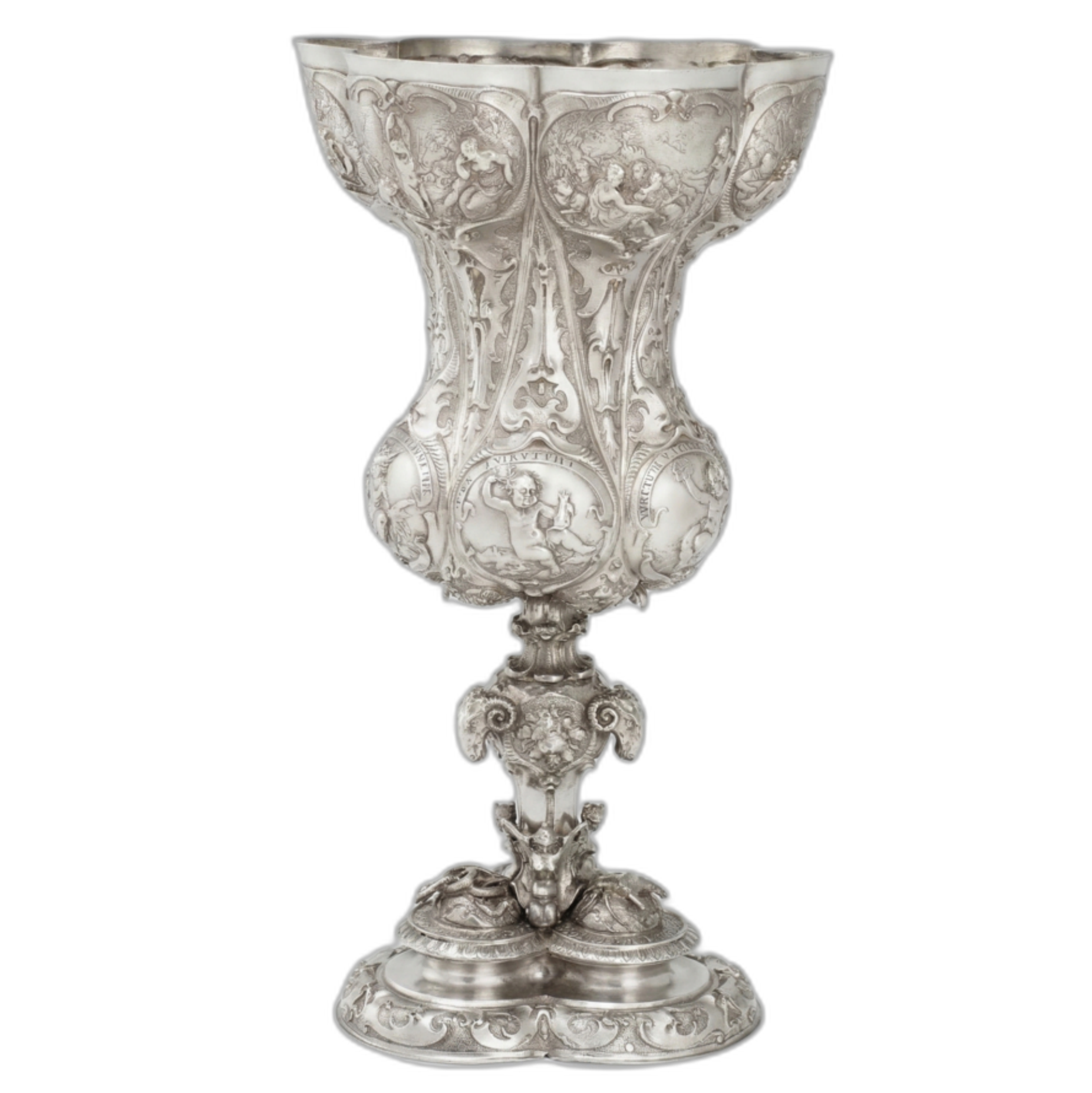}                   &
		\zoominii {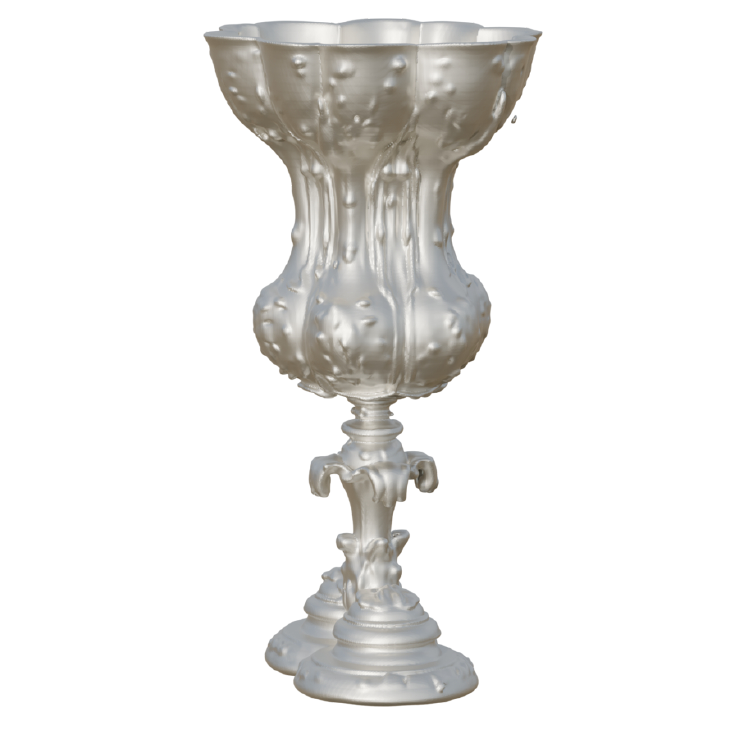} {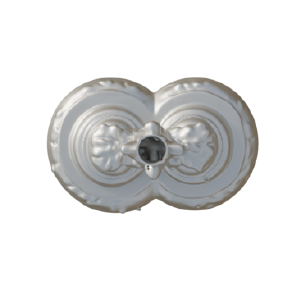}                   &
		\zoominii {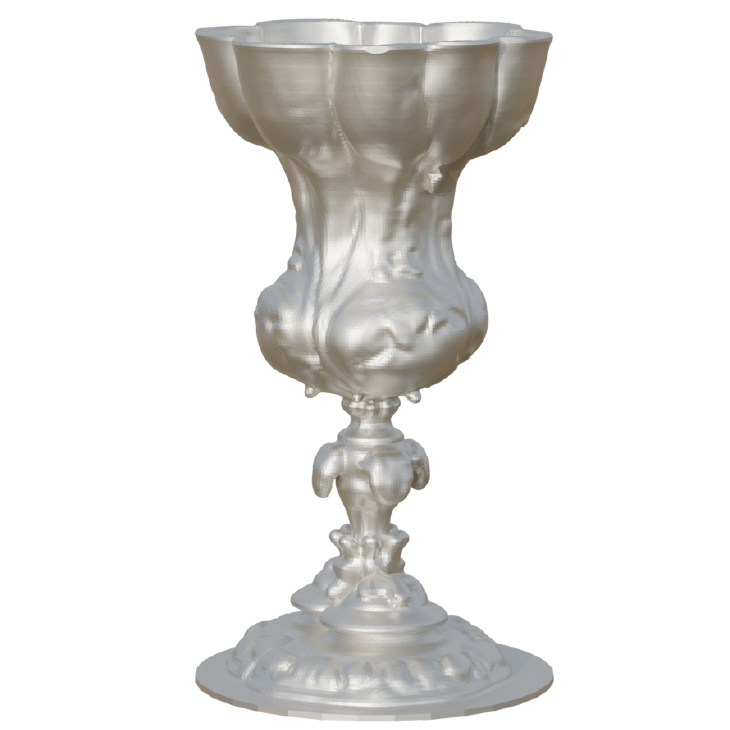} {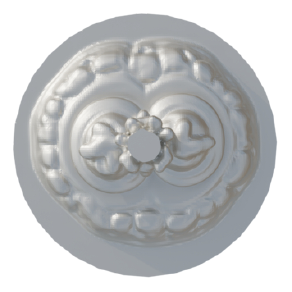}       &
		\zoominii {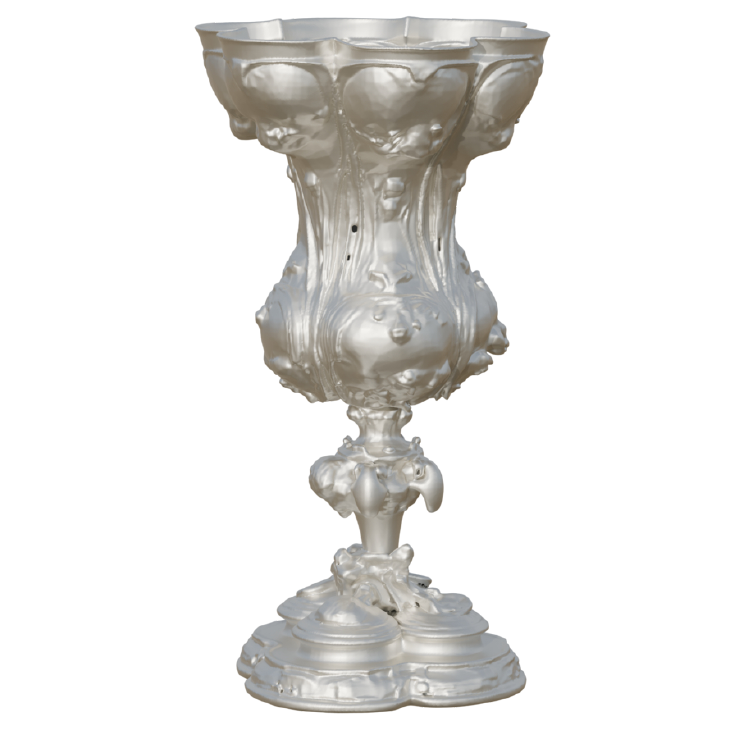} {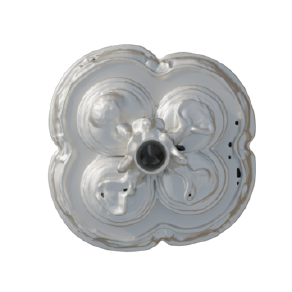} &
		\zoominii {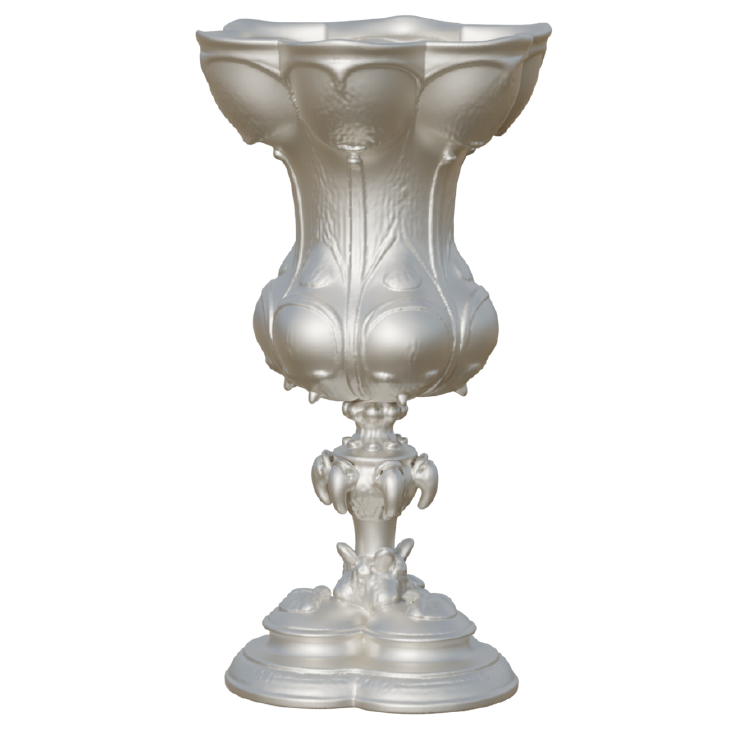} {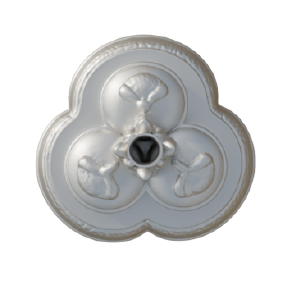}
	\end{tabular}
	\caption{
		\label{fig:visuals}
		Visuals of:
		(1) \href{https://www.clevelandart.org/art/1924.859}{Table Fountain},
		(2) \href{https://polytope.miraheze.org/wiki/Great_disdyakis_triacontahedron}{Great disdyakis triacontahedron} with full icosahedral symmetry,
		(3) lotus,
		(4) gear of 15 fold,
		(5) asymmetric omniwheel generated by ChatGPT for the attempt of changing fold from 4 to 5.
		(6) \href{https://www.britishmuseum.org/collection/object/H_-103?selectedImageId=1595205001}{Silver cup}. Refer to the video for visualizing the results in 3D.}
\end{figure*}

In \Cref{fig:visuals}, we show that our method can enforce the symmetries of a fountain with a 4-fold rotational symmetry and four reflection planes, producing a cleaner shape while recovering missing symmetric sectors.
This is challenging for post-hoc optimization-based symmetrization, which can regularize existing sectors but cannot easily recover missing ones.
\Cref{fig:sym_failure} demonstrates our handling of icosahedral symmetries using a \href{https://en.wikipedia.org/wiki/Dodecadodecahedron}{dodecadodecahedron}, which contains six 5-fold axes, ten 3-fold axes, and fifteen 2-fold axes.

\begin{table*}[t!]
	\centering
	\caption{
		\label{tab:mapper_quantitative}
		Cosine similarity of mapped latents and their ground truth for single-transform, identity, inverse, composition, and closure tests, evaluated on Toys4K.
		Further evaluation includes IoU for the decoded sparse structure and four cascaded shape-subdivision stages, and \(L_2\) error on decoded O-Voxel features.
	}
	\resizebox{\textwidth}{!}{%
		\begin{tabular}{lccccccc}
			\toprule
			Mapper
			 & \makecell[c]{Cosine Similarity (\%) $\uparrow$                    \\
				$M(g)X$ vs.\ $X_{\mathrm{gt}}'$}
			 & \makecell[c]{Identity (\%) $\uparrow$                             \\
				$M(e)X$ vs.\ $X$}
			 & \makecell[c]{Inverse  (\%)$\uparrow$                              \\
				$M(g)M(g^{-1})X$ vs.\ $X$}
			 & \makecell[c]{Composition (\%) $\uparrow$                          \\
				$M(g_1 g_2)X$ vs.\ $M(g_1)M(g_2)X$}
			 & \makecell[c]{Closure (\%) $\uparrow$                              \\
				$M(r_{60^\circ})^6X$ vs.\ $X$}
			 & Decoded IoU (\%) $\uparrow$
			 & Decoded L2 $\downarrow$
			\\

			\midrule

			Sparse structure
			 & 58.82
			 & 95.76
			 & 73.93
			 & 77.47
			 & 57.16
			 & 74.51
			 & --
			\\

			\midrule

			4M shape
			 & 74.27
			 & 98.03
			 & 79.99
			 & 89.92
			 & 76.53
			 & 84.30 / 72.29 / 44.61 / 47.56
			 & 30.63
			\\

			12M shape
			 & \textbf{88.15}
			 & \textbf{98.11}
			 & \textbf{89.55}
			 & \textbf{94.27}
			 & \textbf{81.22}
			 & \textbf{92.68} / \textbf{86.48} / \textbf{71.69} / \textbf{66.64}
			 & \textbf{17.48}
			\\

			\bottomrule
		\end{tabular}%
	}
\end{table*}

\paragraph{Fold manipulation.}
SymTRELLIS can manipulate fold counts.
In \Cref{fig:fold_manipulation}, given an input image of a 12-fold gear, our method can generate shapes with folds ranging from 9 to 18 proper symmetries, demonstrating robust and controllable symmetry manipulation.


\begin{figure}[t!]
	\centering
	\small
	\setlength{\tabcolsep}{2pt}
	\renewcommand{\arraystretch}{0.8}

	\newlength{\multisymresultsize}
	\setlength{\multisymresultsize}{0.31\columnwidth}

	\begin{tabular}{ccc}
		Input image                                                                                      &
		Space segmentation                                                                               &
		SymTRELLIS                                                                                         \\[4pt]

		\includegraphics[width=\multisymresultsize]{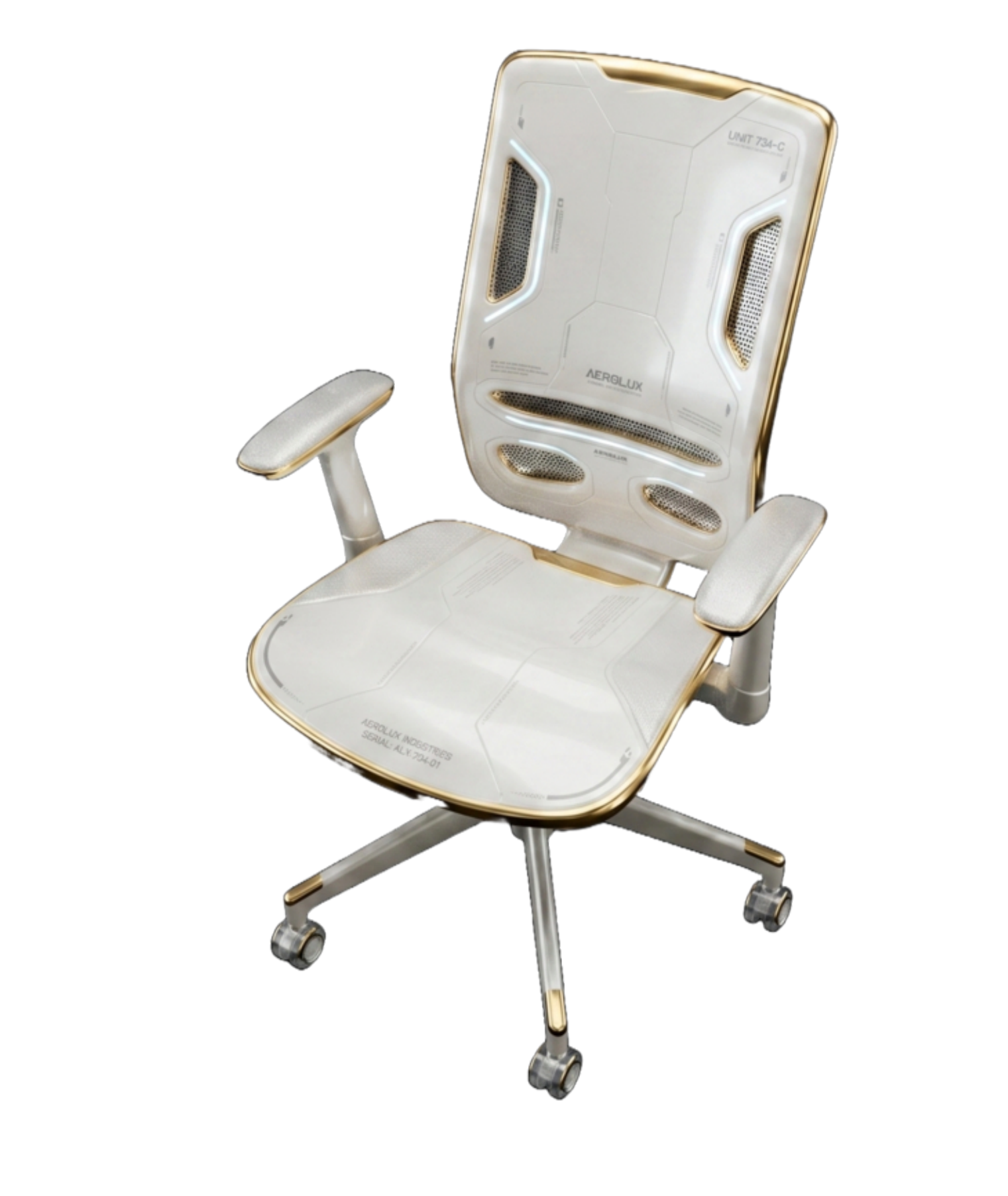}        &
		\includegraphics[width=\multisymresultsize]{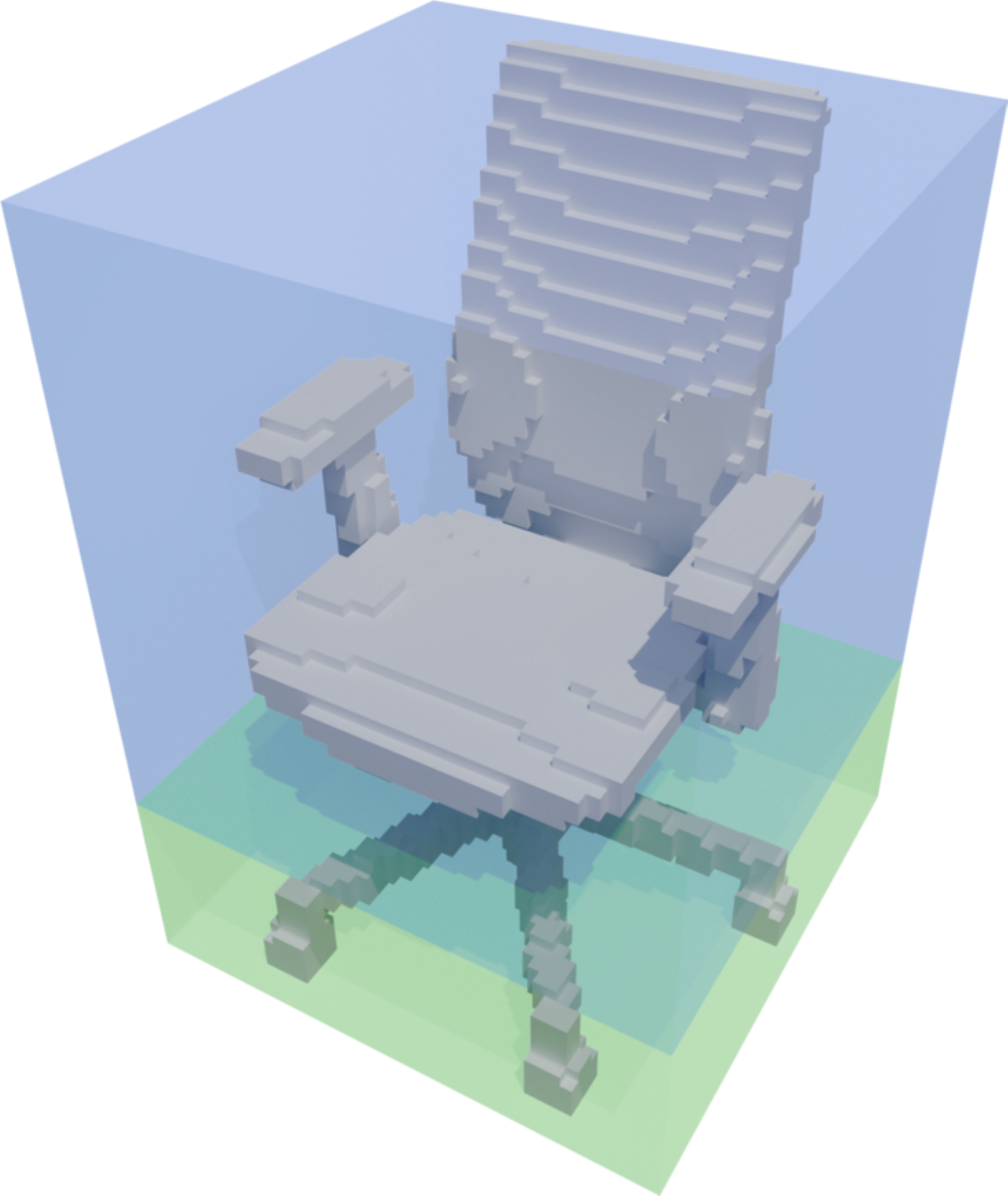} &
		\includegraphics[width=\multisymresultsize]{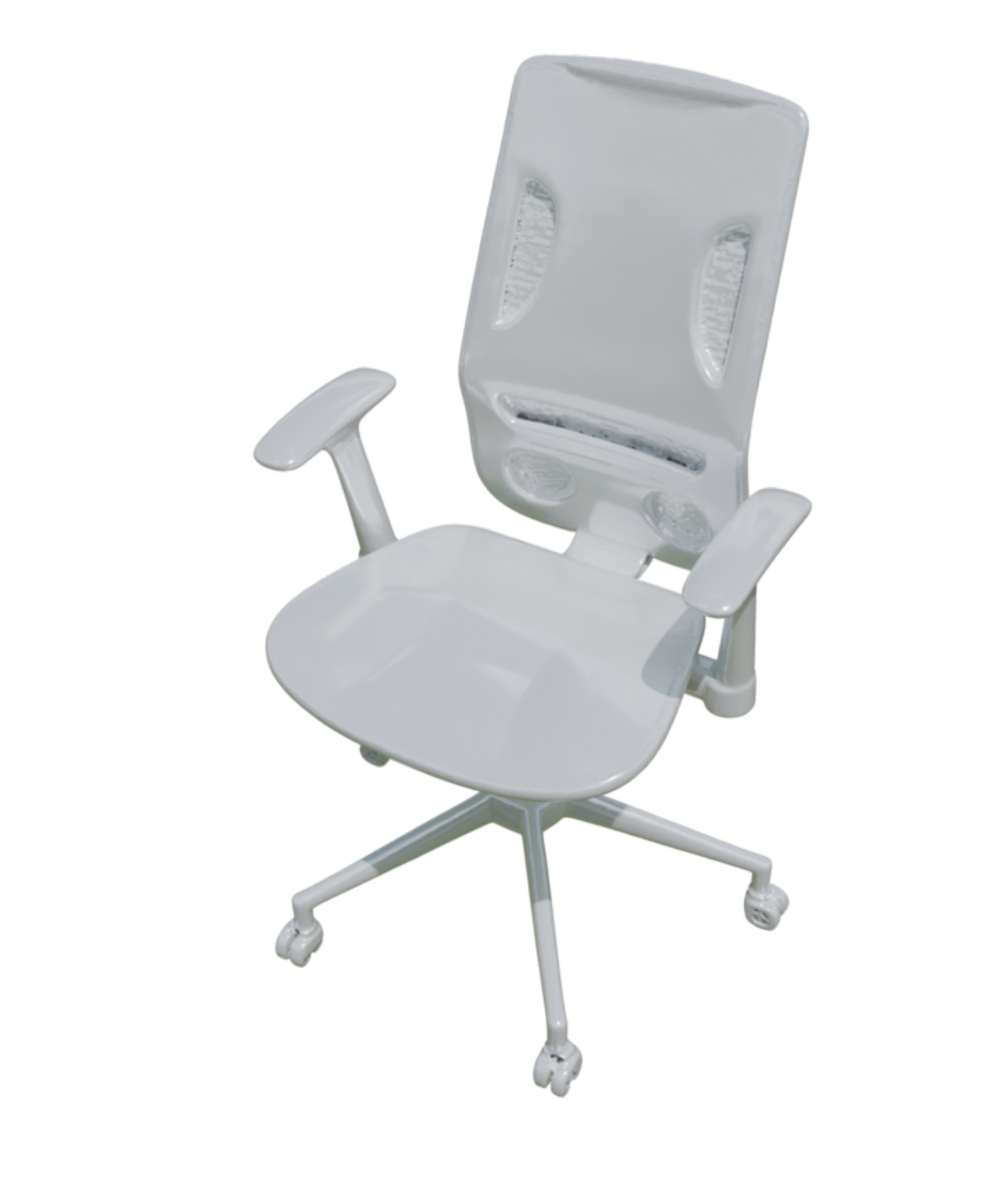}
	\end{tabular}

	\caption{
		\label{fig:multi_symmetry}
		Our method is not limited to a single global symmetry relation.
		Given an input image of a swivel chair, we partition the voxel space into regions with reflectional symmetry for the upper chair structure and 5-fold rotational symmetry for the wheeled base.
		SymTRELLIS applies the spatial-transform latent mapper separately within each segmented region.
	}
\end{figure}
\paragraph{Multi-symmetry enforcement.}
Our method is not limited to a single symmetry relation.
In \Cref{fig:multi_symmetry}, the swivel chair has a reflection symmetry on the upper part and a 5-fold rotational symmetry on the base.
We enforce multiple symmetry relations by assigning voxel regions to different symmetry specifications and computing spatial-transform mapper coefficients separately for each region.
When multiple symmetry relations overlap, a voxel may belong to multiple regions.
Since our focus is on symmetry enforcement rather than multi-symmetry detection, the main experiments use one symmetry relation per generation.
Thus, for the swivel chair, the voxel regions and symmetry relations are specified manually.

\paragraph{Speed}
We test inference speed on an RTX~5090 to quantify the overhead introduced by symmetry enforcement.
Mapper inference happens once per generation and takes \(65~\mathrm{ms}\) for sparse structure and \(77~\mathrm{ms}\) for shape.
The noise symmetrization step takes \(36~\mathrm{ms}\) for sparse structure and \(43~\mathrm{ms}\) for shape.
The coefficient application step happens at each flow step and takes \(0.7~\mathrm{ms}\) for sparse structure and \(0.9~\mathrm{ms}\) for shape.
As a comparison, a vanilla flow step takes \(175~\mathrm{ms}\) for sparse structure and \(104~\mathrm{ms}\) for shape and these per step values show that coefficient application incurs negligible overhead relative to the vanilla flow computation.

\paragraph{Failure Cases}
Besides the smoothing effects arising from velocity symmetrization, we have observed two additional types of failure cases.
The first is attributed to the existence of empty active voxel sets, in which case the averaged latents cancel each other.
This occurs with only a 0.17\%  chance over our evaluation set.
The second failure type is due to our method failing to detect appropriate shape symmetries,
which accounts for about 0.85\% of the cases over our entire evaluation of SymTRELLIS.

On the ground truth meshes we collected, we did not observe any failures by our symmetry detection module.
For generated meshes with imperfect symmetries,
we observe that the erroneously detected symmetry fold counts tend to be multiples or factors of the true fold counts, e.g., predicting 16 or 4 folds when the ground truth is 8, with the mesh nearing the threshold of fold ambiguity.

\begin{figure}[t!]
	\centering
	\small
	\setlength{\tabcolsep}{2pt}
	\renewcommand{\arraystretch}{0.8}

	\newlength{\limitationresultsize}
	\setlength{\limitationresultsize}{0.31\columnwidth}

	\newcommand{\zoomin}[2]{%
		\begin{tikzpicture}
			\node[inner sep=0, anchor=south west] (base) at (0,0) {%
				\includegraphics[
					width=\limitationresultsize,
					height=\limitationresultsize
				]{#1}%
			};
			\begin{scope}[
					x={(base.south east)},
					y={(base.north west)}
				]
				\node[inner sep=0, anchor=south west] at (0.68,0.68) {%
					\includegraphics[
						width=0.40\limitationresultsize,
						height=0.40\limitationresultsize
					]{#2}%
				};
			\end{scope}
		\end{tikzpicture}%
	}

	\begin{tabular}{ccc}
		Input image                                                                                                          &
		TRELLIS.2                                                                                                            &
		SymTRELLIS                                                                                                             \\[6pt]

		\includegraphics[width=\limitationresultsize,height=\limitationresultsize]{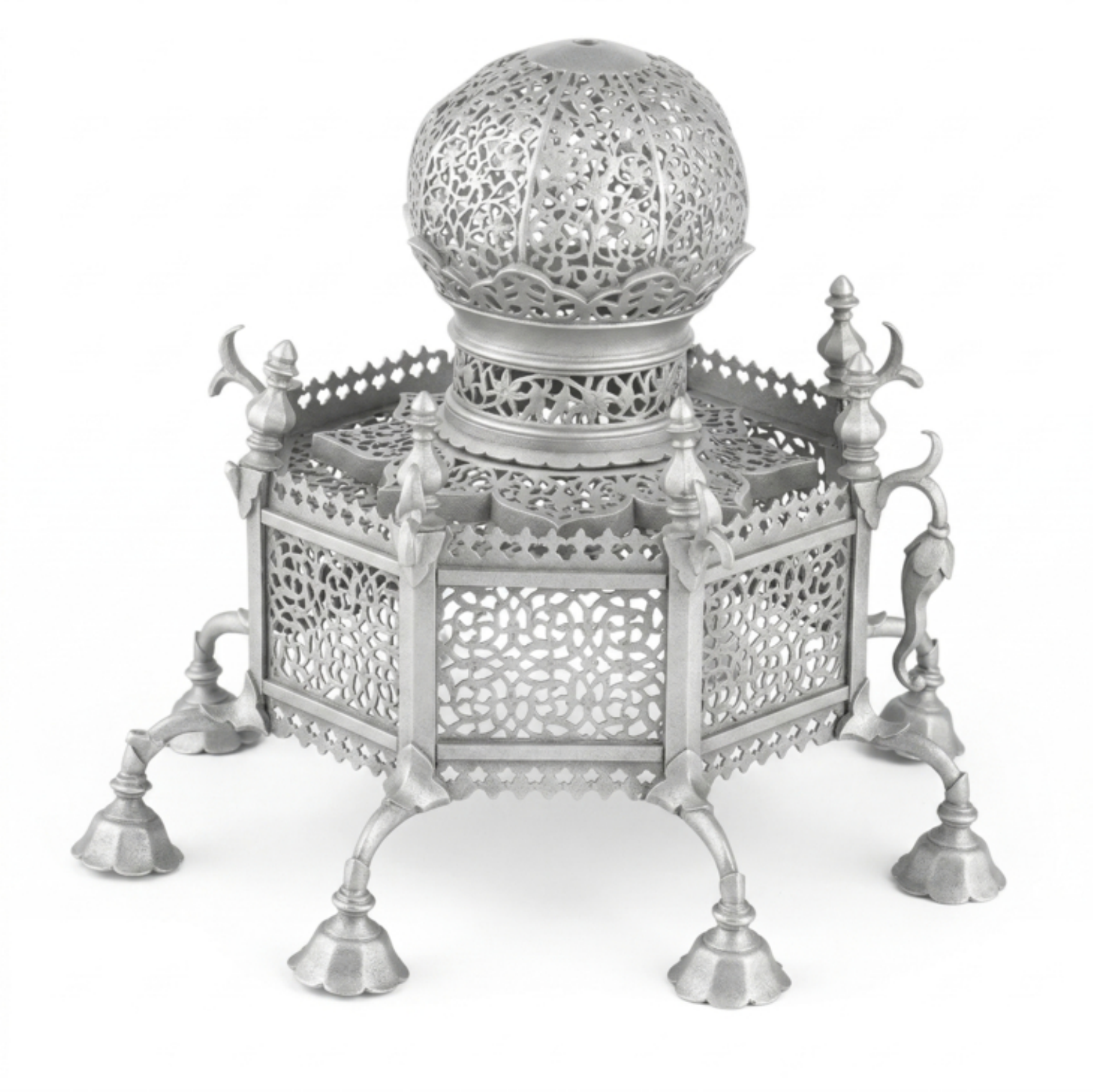} &
		\zoomin{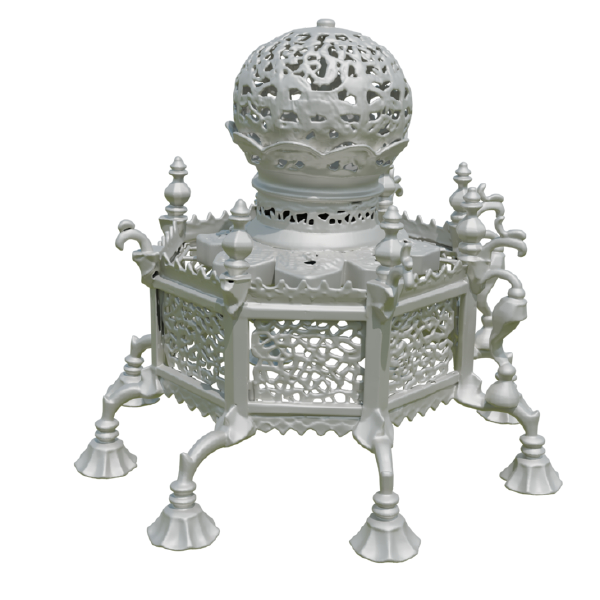}{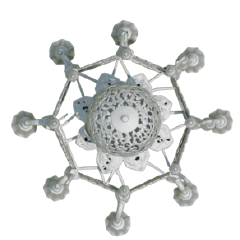}      &
		\zoomin{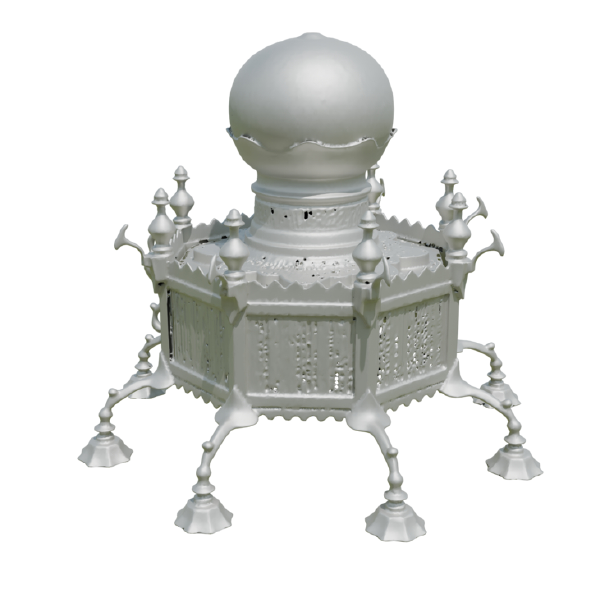}{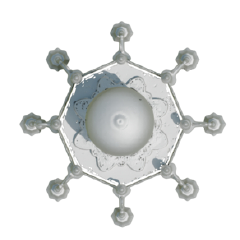}        \\[4pt]
	\end{tabular}

	\caption{
		\label{fig:smoothing}
		For objects with dense fine-scale details, such as the perforated dome and ornamental structures in the input image, averaging latents mapped from multiple symmetry-related spatial transforms can smooth high-frequency geometry.
	}
\end{figure}
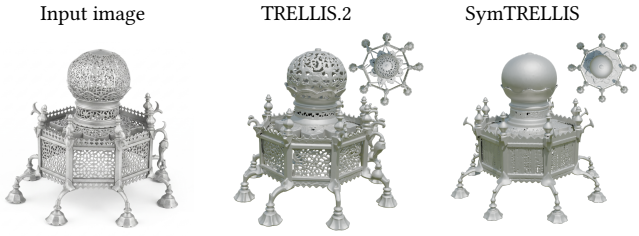
\paragraph{Limitation.}

The main limitation of SymTRELLIS is that it is designed to only enforce \emph{global extrinsic}, not partial~\cite{niloy2006partial}
or intrinsic \cite{xu_siga12}, symmetries.
Also, our method averages latents from multiple spatial transforms, which can smooth fine details, as shown in~\Cref{fig:smoothing}.
We attribute this to imperfect consistency among latents mapped from different poses.
This effect can be alleviated by shortening the symmetry-guidance duration in the shape-latent stage, allowing later flow-matching steps to recover the smoothed details.
In some cases, the method may produce empty active voxels or only a very small active voxel set.
This issue can often be resolved by changing the random seed.

A key factor behind the effectiveness of our mapper is that voxel-based latents make spatial information explicit by attaching features to fixed grid coordinates.
Given a spatial transform,
the correspondence between coordinates is directly known,
allowing the mapper to learn feature-space transport on top of this explicit geometric relation.
The same assumption applies to voxel-based latent generation models such as TRELLIS, TRELLIS.2, and SAM3D~\cite{chen2025sam}.
In contrast, ShapeVAE-based models such as TripoSG or Hunyuan3D-2.1 represent both geometry and token positions implicitly, making it substantially more difficult to apply our method in their latent space for symmetry enforcement.
Enforcing symmetries on such models may require training a separate layer for decoupling position features from texture features.

\section{Conclusion and future work}

We have presented SymTRELLIS, a method to enforce arbitrary finite point group symmetry during flow-based 3D generation of TRELLIS.2.
Given a symmetry specification, either automatically detected or user-supplied,
SymTRELLIS learns a latent-space action of spatial transformations and uses the resulting group-averaged operator to symmetrize flow predictions at each ODE step.
Applying this guidance in both the sparse-structure and shape latent stages enforces symmetry before mesh decoding,
allowing symmetric regions to be generated jointly rather than corrected in a post-processing step.
On a curated benchmark of symmetric objects, SymTRELLIS substantially reduces symmetry errors while maintaining reconstruction accuracy comparable to the base model,
suggesting that learned latent-space transform operators provide a practical interface for imposing geometric constraints on voxel-based 3D generative models without retraining the flow model.

Several future directions emerge from this work.
The smoothing effect from velocity symmetrization suggests that a more expressive latent mapper, e.g., one that better preserves high-frequency features across spatial transformations,
could recover finer geometric detail while maintaining symmetry.
The occasional empty-generation failure points toward a more principled guidance schedule that adapts the symmetry strength dynamically across ODE steps rather than applying a fixed $\lambda$ and $\tau$.
More broadly, our method is currently tied to voxel-based latents,
hence extending the spatial-transform mapper to implicit or token-based latent spaces would broaden its applicability.
Another promising future direction is to integrate part generation with partial symmetry enforcement and detection.
Beyond these direct extensions, it would be interesting to explore learning the symmetry specification end-to-end jointly with generation, rather than relying on a separate detection stage,
and to investigate whether velocity symmetrization generalizes to other functional requirements beyond symmetry, such as co-planarity, connectivity, and mechanical fit,
opening a broader path toward functionally valid 3D generation.

\begin{figure*}[t!]
	\centering
	\small
	\setlength{\tabcolsep}{1.5pt}
	\renewcommand{\arraystretch}{0.8}

	\newlength{\sixsize}
	\newlength{\eightsize}
	\setlength{\sixsize}{0.155\textwidth}
	\setlength{\eightsize}{0.112\textwidth}

	\newcommand{\phsquare}[2]{%
		\begin{tikzpicture}[x=#1,y=#1,baseline=(current bounding box.center)]
			\draw (0,0) rectangle (1,1);
			\node[align=center,font=\scriptsize] at (0.5,0.5) {#2};
		\end{tikzpicture}%
	}

	\newcommand{\phzoomsquare}[3]{%
		\begin{tikzpicture}[x=#1,y=#1,baseline=(current bounding box.center)]
			\draw (0,0) rectangle (1,1);
			\node[align=center,font=\scriptsize] at (0.5,0.5) {#2};
			\draw (0.60,0.60) rectangle (0.98,0.98);
			\node[align=center,font=\tiny] at (0.79,0.79) {#3};
		\end{tikzpicture}%
	}

	\newcommand{\imgcell}[3]{%
		\begin{tabular}{@{}c@{}}
			#1 \\[-1pt]
			\parbox[t]{#2}{\centering\scriptsize #3}
		\end{tabular}%
	}

	\newcommand{\phcell}[3]{%
		\imgcell{\phsquare{#1}{#2}}{#1}{#3}%
	}

	\newcommand{\phzoomcell}[4]{%
		\imgcell{\phzoomsquare{#1}{#2}{zoom}}{#1}{#4}%
	}

	\newcommand{\realcell}[3]{%
		\imgcell{%
			\includegraphics[width=#1,height=#1]{#2}%
		}{#1}{#3}%
	}

	\newcommand{\zoominreal}[3]{%
		\begin{tikzpicture}
			\node[inner sep=0, anchor=south west] (base) at (0,0) {%
				\includegraphics[
					width=#1,
					height=#1
				]{#2}%
			};
			\begin{scope}[
					x={(base.south east)},
					y={(base.north west)}
				]
				\node[inner sep=0, anchor=south west] at (0.60,0.60) {%
					\includegraphics[
						width=0.38#1,
						height=0.38#1
					]{#3}%
				};
			\end{scope}
		\end{tikzpicture}%
	}

	\newcommand{\realzoomcell}[4]{%
		\imgcell{%
			\zoominreal{#1}{#2}{#3}%
		}{#1}{#4}%
	}

	\begin{tabular}{@{}*{6}{c}@{}}
		\realcell{\sixsize}{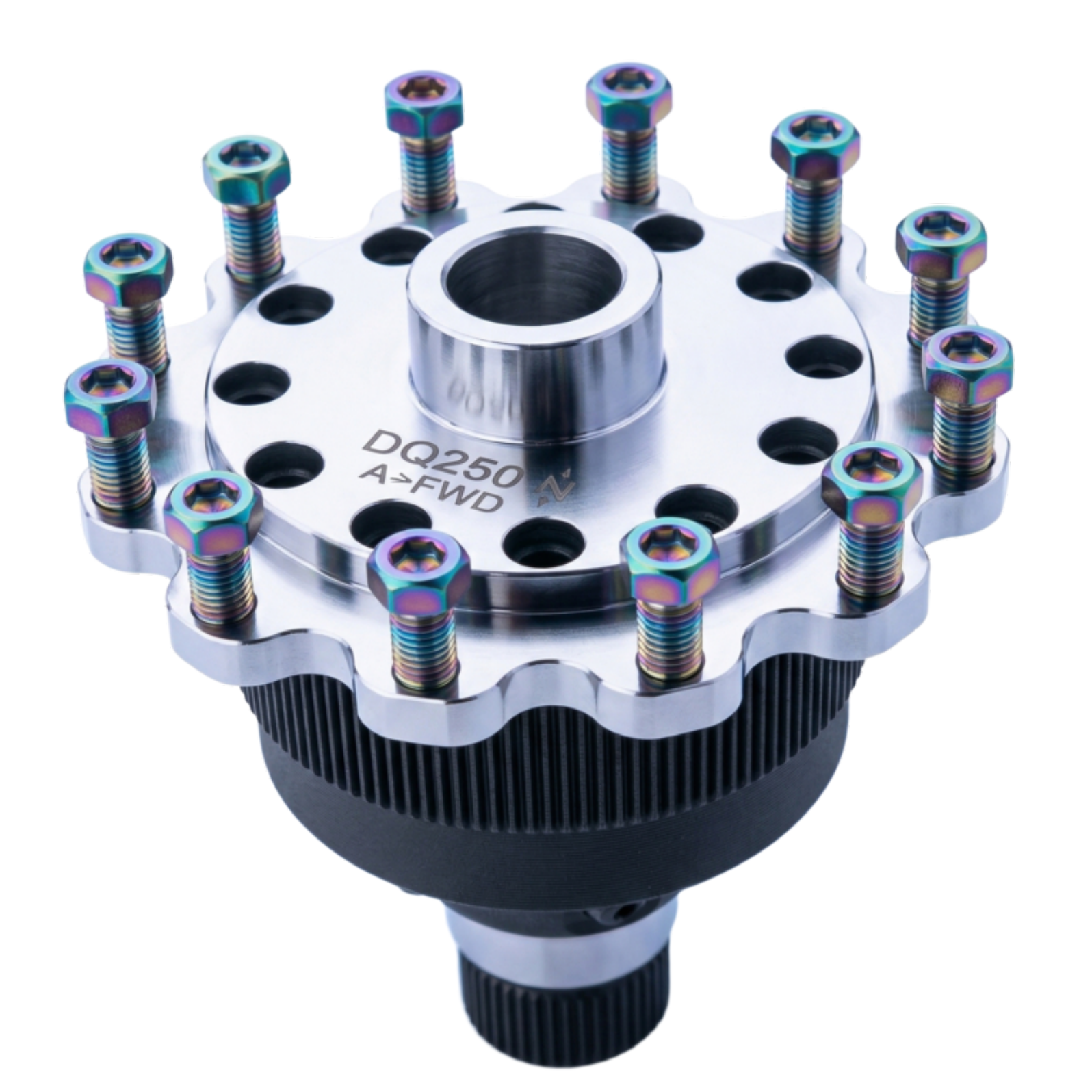}{Input Image}    &
		\realcell{\sixsize}{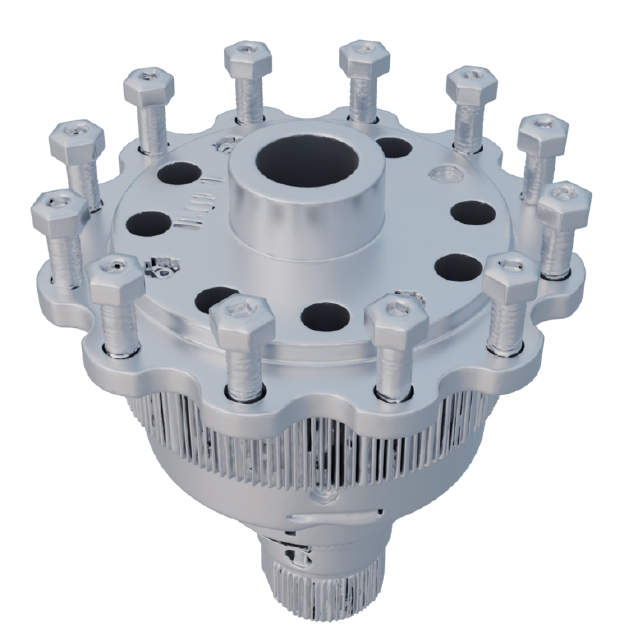}{TRELLIS.2} &
		\realcell{\sixsize}{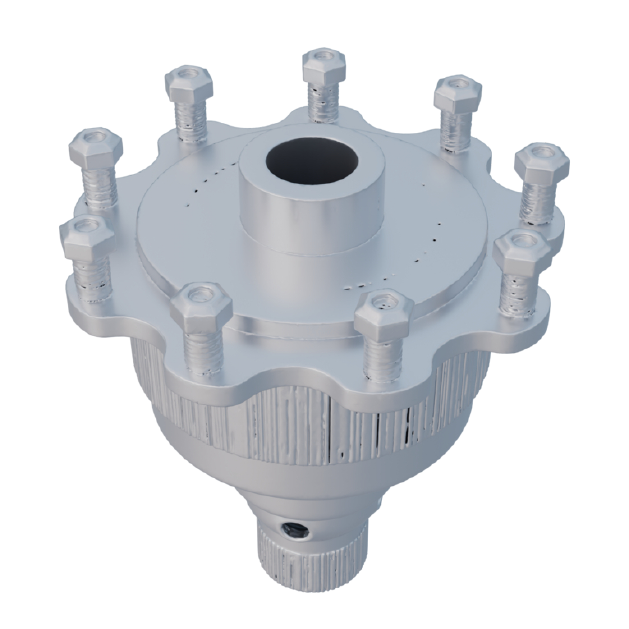}{9 Fold}             &
		\realcell{\sixsize}{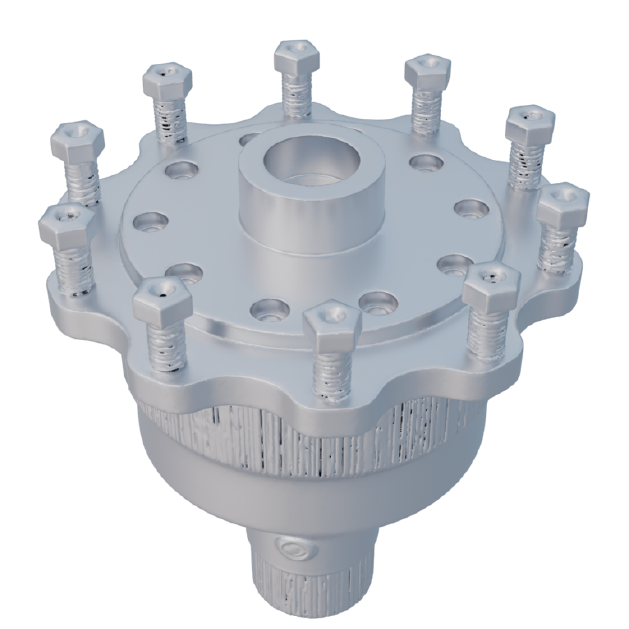}{10 Fold}            &
		\realcell{\sixsize}{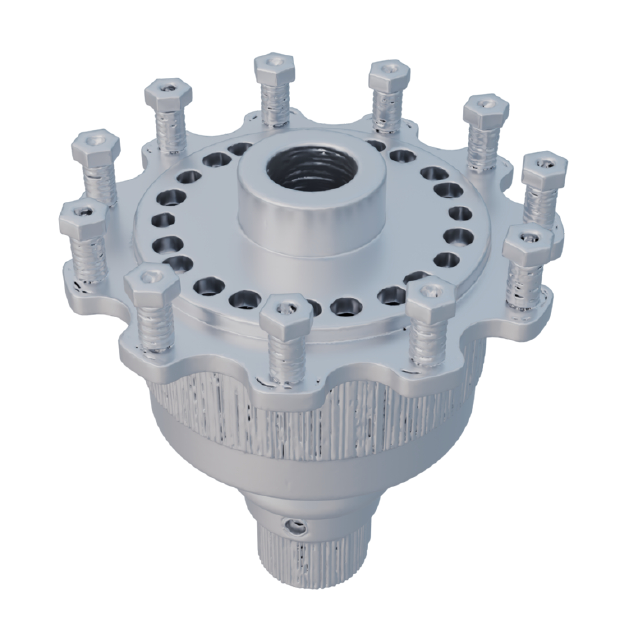}{11 Fold}            &
		\realcell{\sixsize}{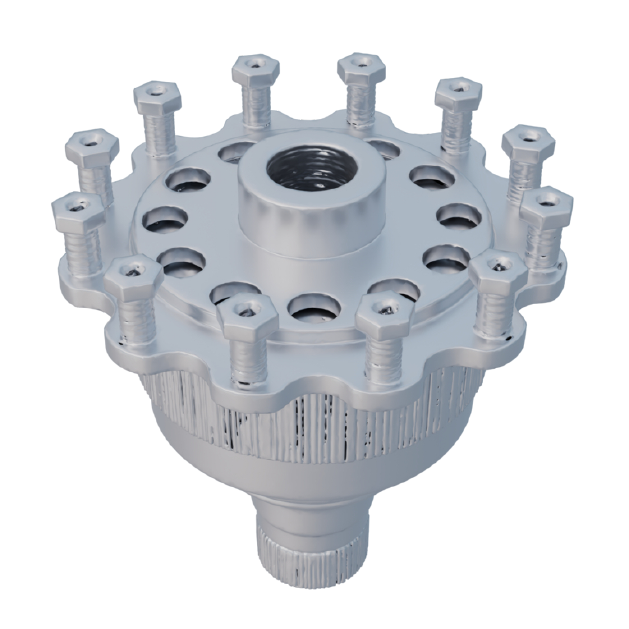}{12 Fold}
	\end{tabular}

	\vspace{4pt}

	\begin{tabular}{@{}*{6}{c}@{}}
		\realcell{\sixsize}{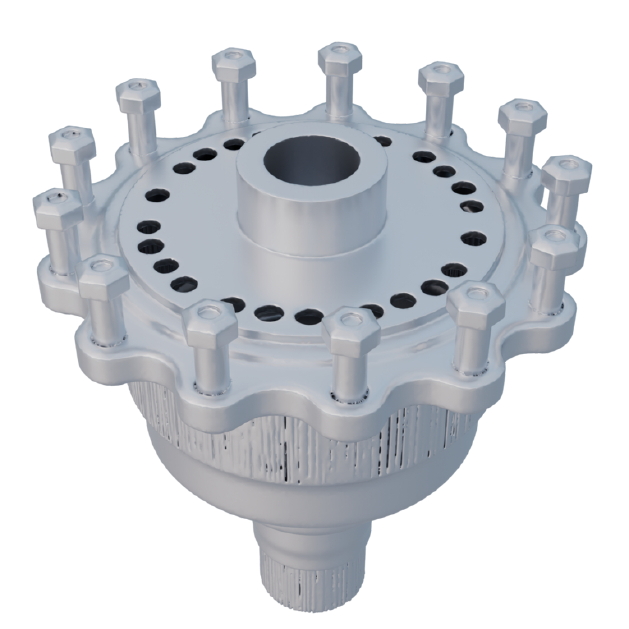}{13 Fold} &
		\realcell{\sixsize}{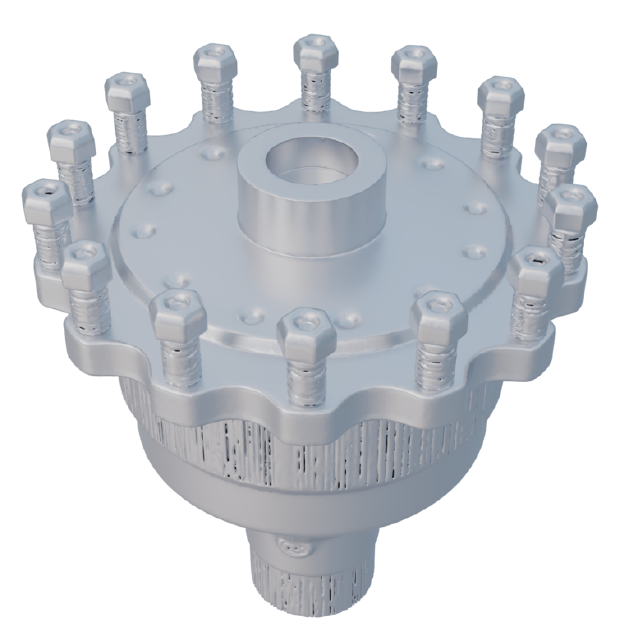}{14 Fold} &
		\realcell{\sixsize}{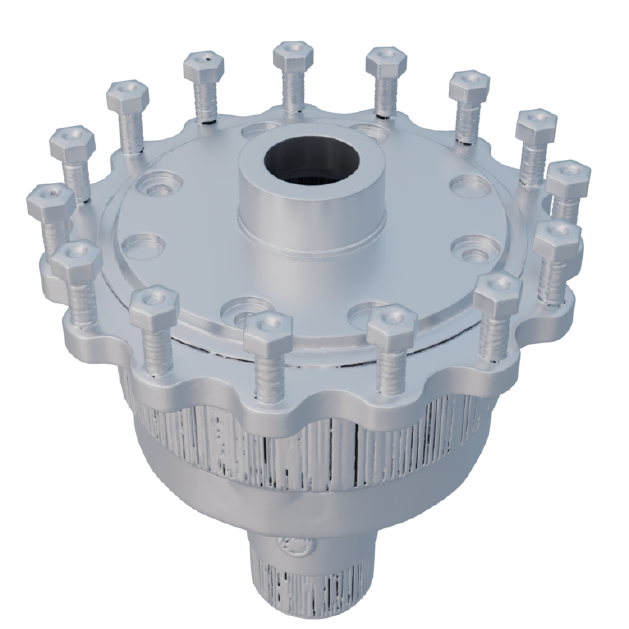}{15 Fold} &
		\realcell{\sixsize}{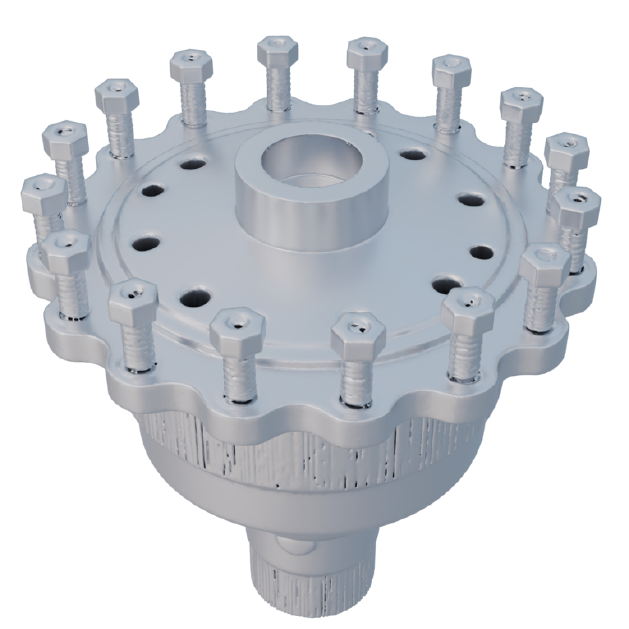}{16 Fold} &
		\realcell{\sixsize}{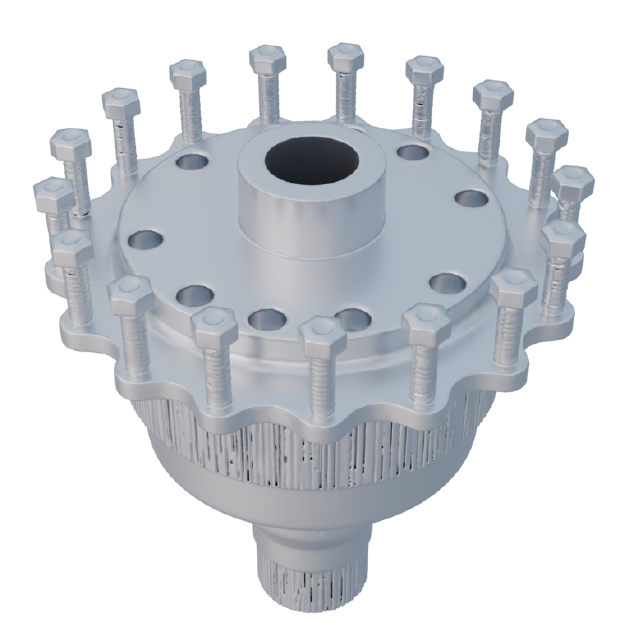}{17 Fold} &
		\realcell{\sixsize}{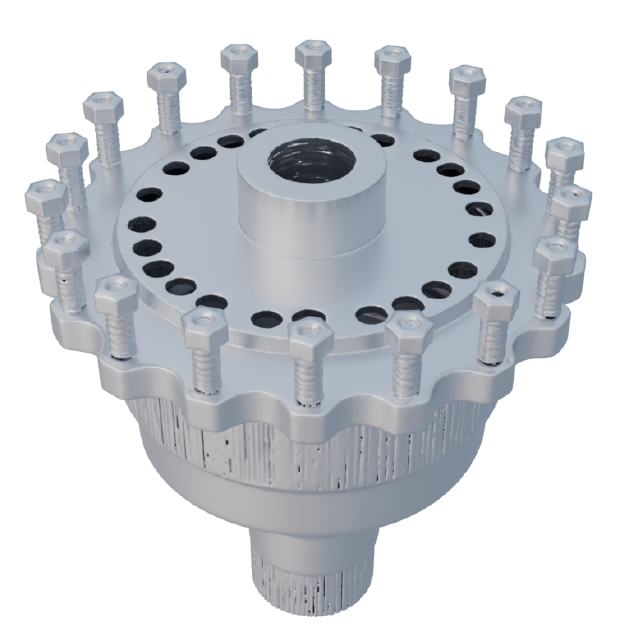}{18 Fold}
	\end{tabular}

	\vspace{4pt}

	\begin{tabular}{@{}*{8}{c}@{}}
		\realcell{\eightsize}{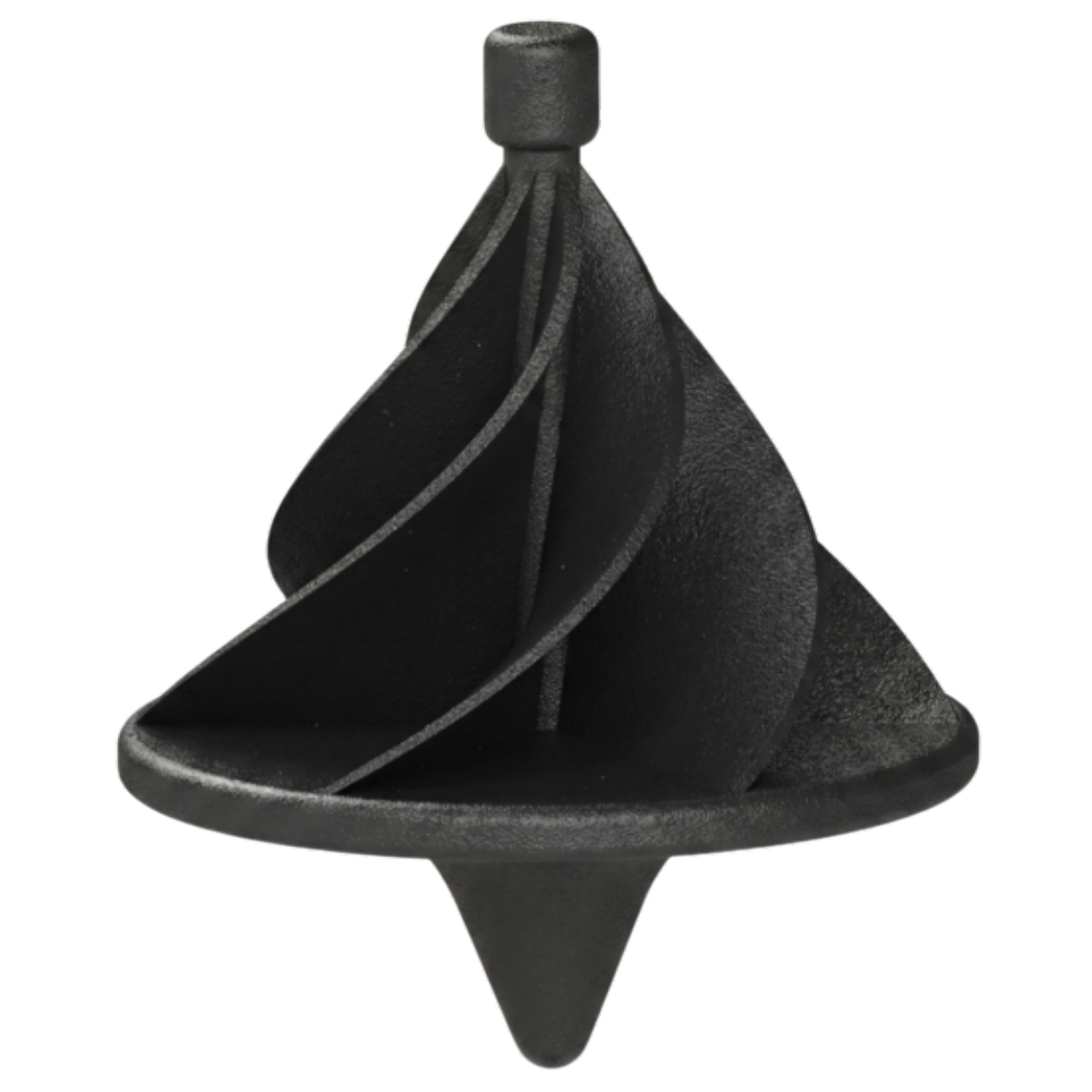}{Input Image} &
		\realzoomcell{\eightsize}
		{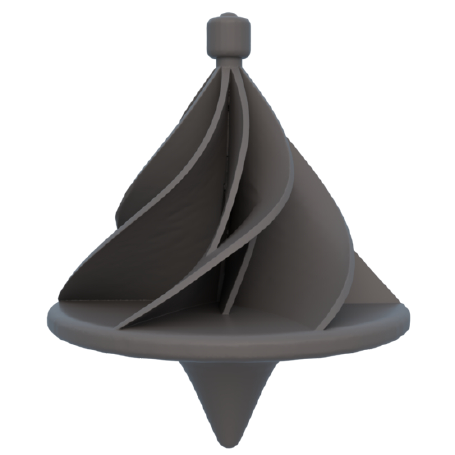}
		{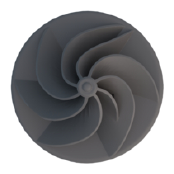}
		{TRELLIS.2}                                                                                 &
		\realzoomcell{\eightsize}
		{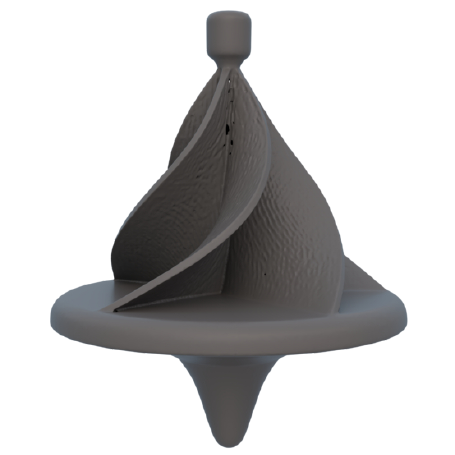}
		{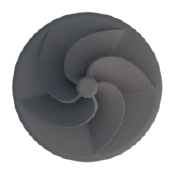}
		{5 Fold}                                                                                    &
		\realzoomcell{\eightsize}
		{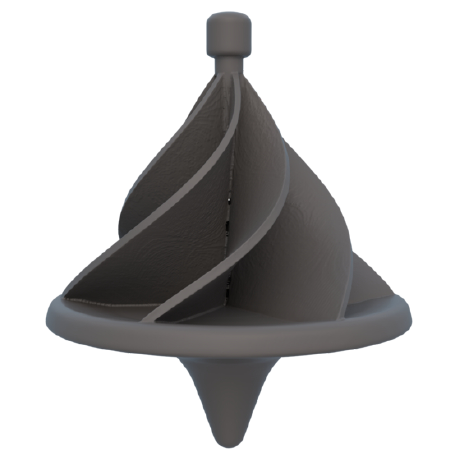}
		{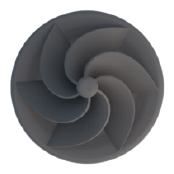}
		{6 Fold}                                                                                    &
		\realzoomcell{\eightsize}
		{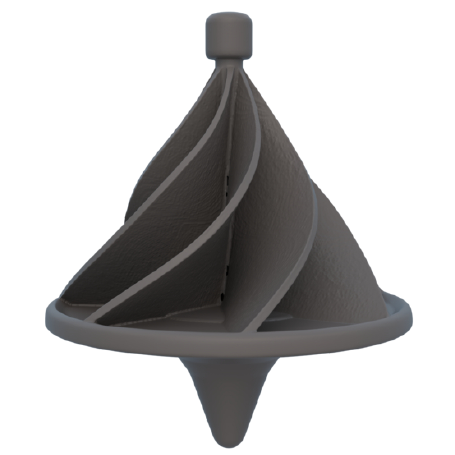}
		{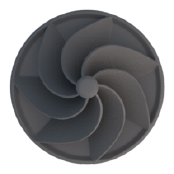}
		{7 Fold}                                                                                    &
		\realzoomcell{\eightsize}
		{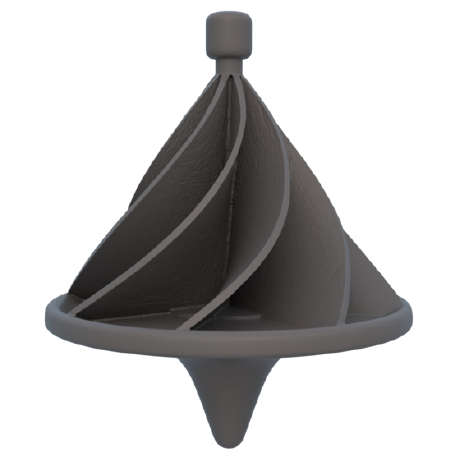}
		{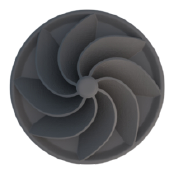}
		{8 Fold}                                                                                    &
		\realzoomcell{\eightsize}
		{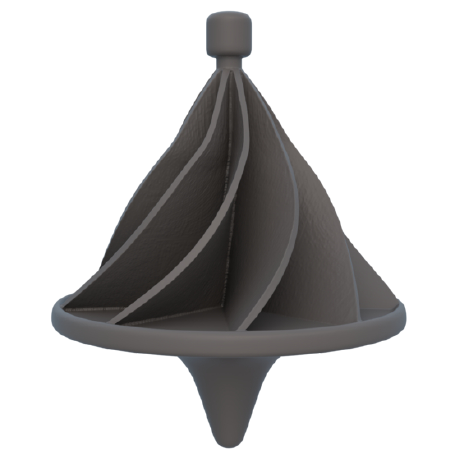}
		{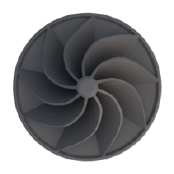}
		{9 Fold}                                                                                    &
		\realzoomcell{\eightsize}
		{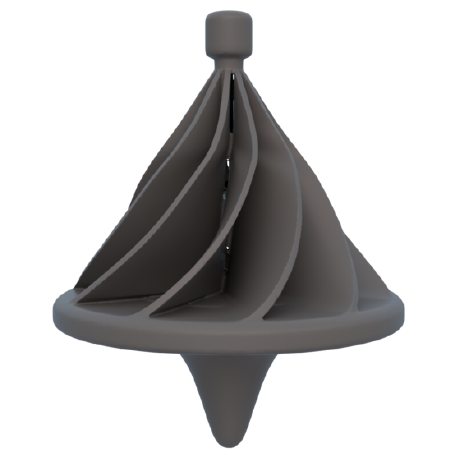}
		{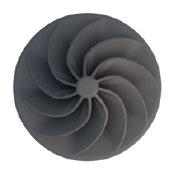}
		{10 Fold}
	\end{tabular}

	\caption{
		\label{fig:fold_manipulation}
		We explicitly control the rotational fold by injecting different symmetry specifications during generation.
		For the gear example, we generate 9 to 18-fold symmetric variants from a 12-fold input. For the spinning top example, we generate 5 to 10-fold symmetric variants.
	}
\end{figure*}
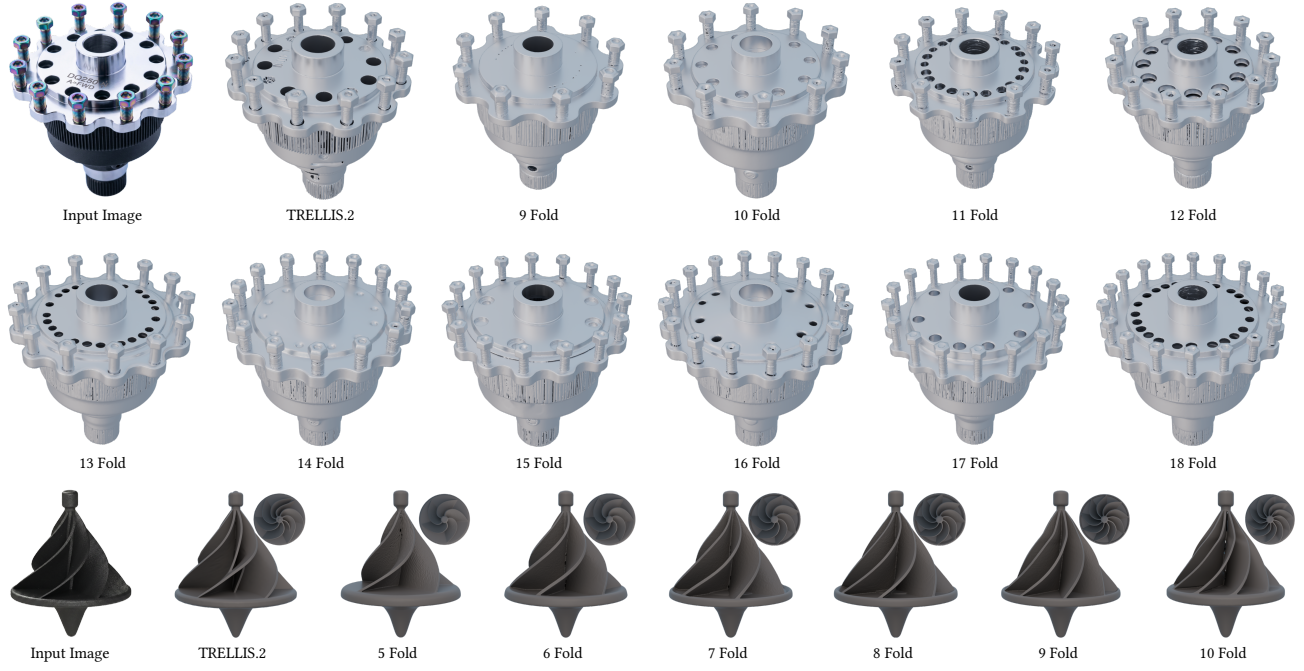
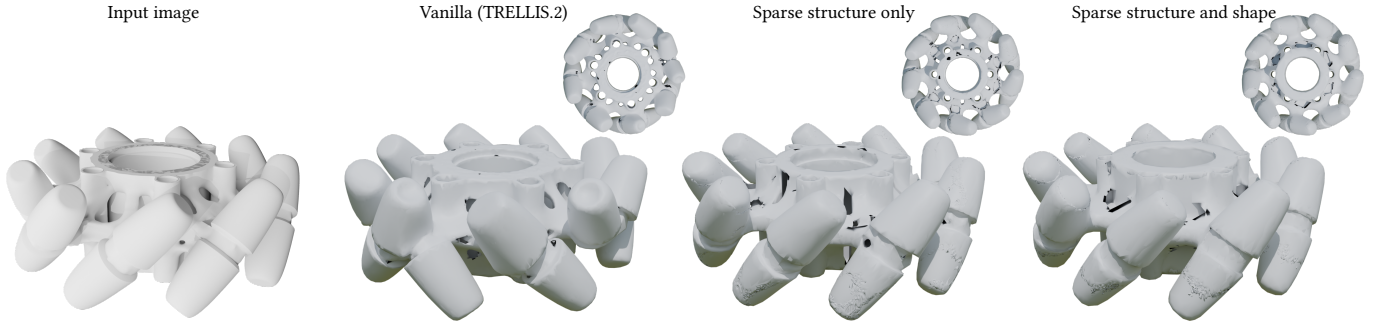
\begin{figure*}[p]
	\centering

	{%
		\small

		\setlength{\tabcolsep}{1pt}
		\renewcommand{\arraystretch}{0.8}

		\def\ablationresultsize{\dimexpr(\textwidth-6\tabcolsep)/4\relax}
		\def\ablationzoomsize{\dimexpr(\textwidth-6\tabcolsep)/10\relax}

		\def\ablhead#1{%
			\makebox[\ablationresultsize][c]{%
				\begingroup
				\setbox0=\hbox{\footnotesize\strut #1}%
				\ifdim\wd0>\ablationresultsize
					\resizebox{\ablationresultsize}{!}{\footnotesize\strut #1}%
				\else
					\footnotesize\strut #1%
				\fi
				\endgroup
			}%
		}

		\def\ablationzoomin#1#2{%
			\begin{tikzpicture}
				\node[inner sep=0, anchor=south west] (base) at (0,0) {%
					\includegraphics[
						width=\ablationresultsize,
						height=\ablationresultsize
					]{#1}%
				};
				\begin{scope}[
						x={(base.south east)},
						y={(base.north west)}
					]
					\node[inner sep=0, anchor=south west] at (0.68,0.68) {%
						\includegraphics[
							width=\ablationzoomsize,
							height=\ablationzoomsize
						]{#2}%
					};
				\end{scope}
			\end{tikzpicture}%
		}

		\begin{tabular}{@{}*{4}{>{\centering\arraybackslash}m{\ablationresultsize}}@{}}
			\ablhead{Input image}                                                                                                   &
			\ablhead{Vanilla (TRELLIS.2)}                                                                                           &
			\ablhead{Sparse structure only}                                                                                         &
			\ablhead{Sparse structure and shape}                                                                                      \\[4pt]

			\includegraphics[
				width=\ablationresultsize,
				height=\ablationresultsize
			]{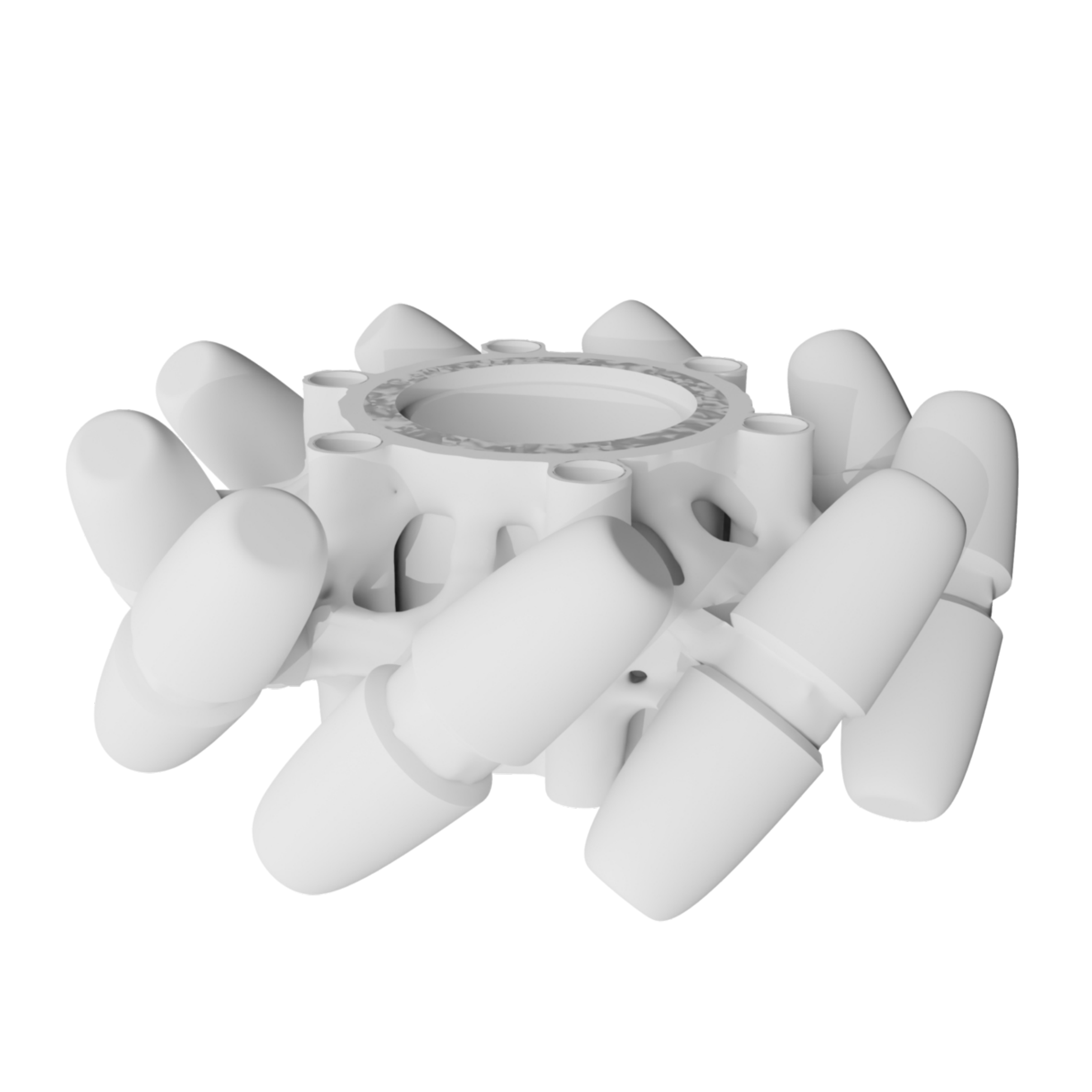}                                                                               &
			\ablationzoomin{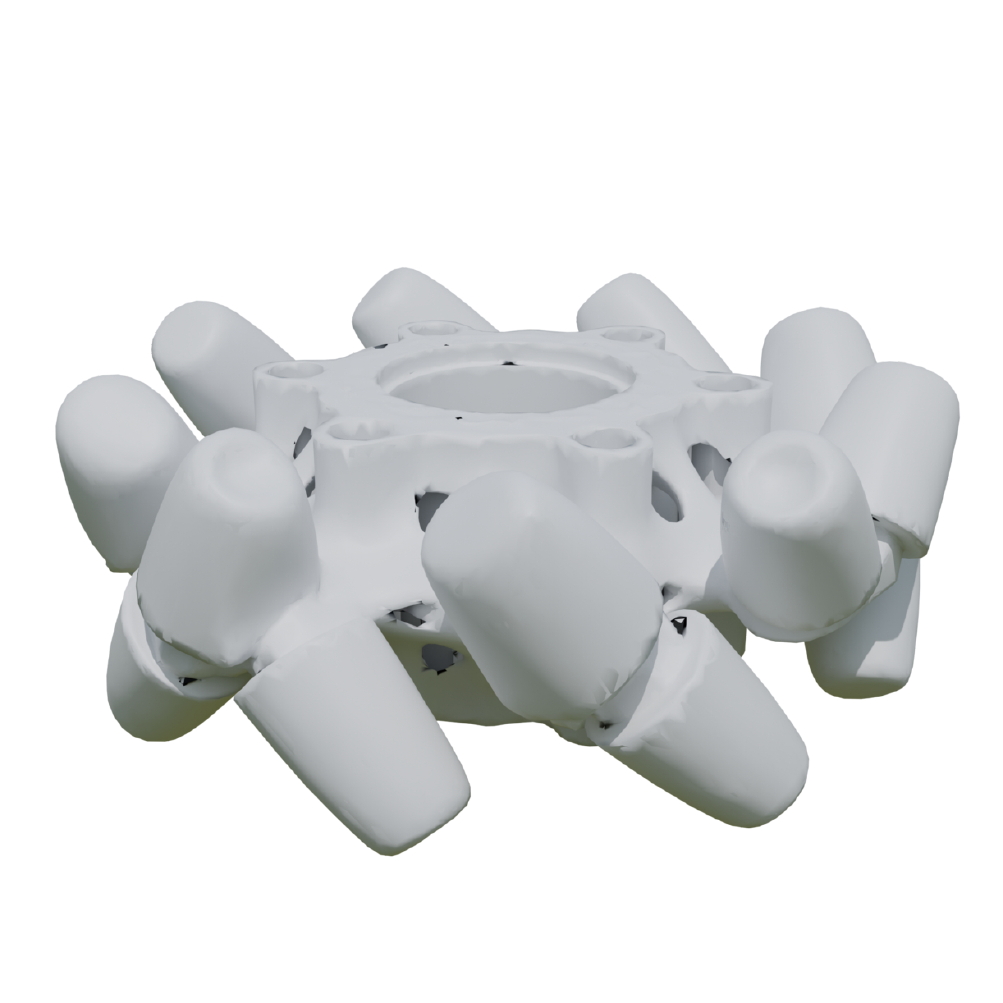}{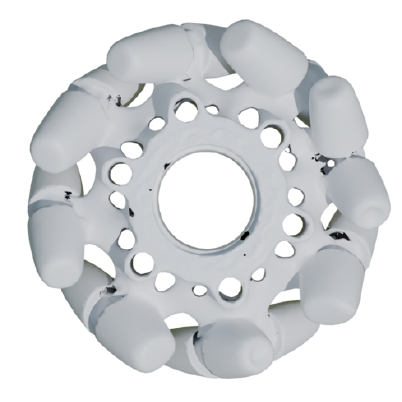}     &
			\ablationzoomin{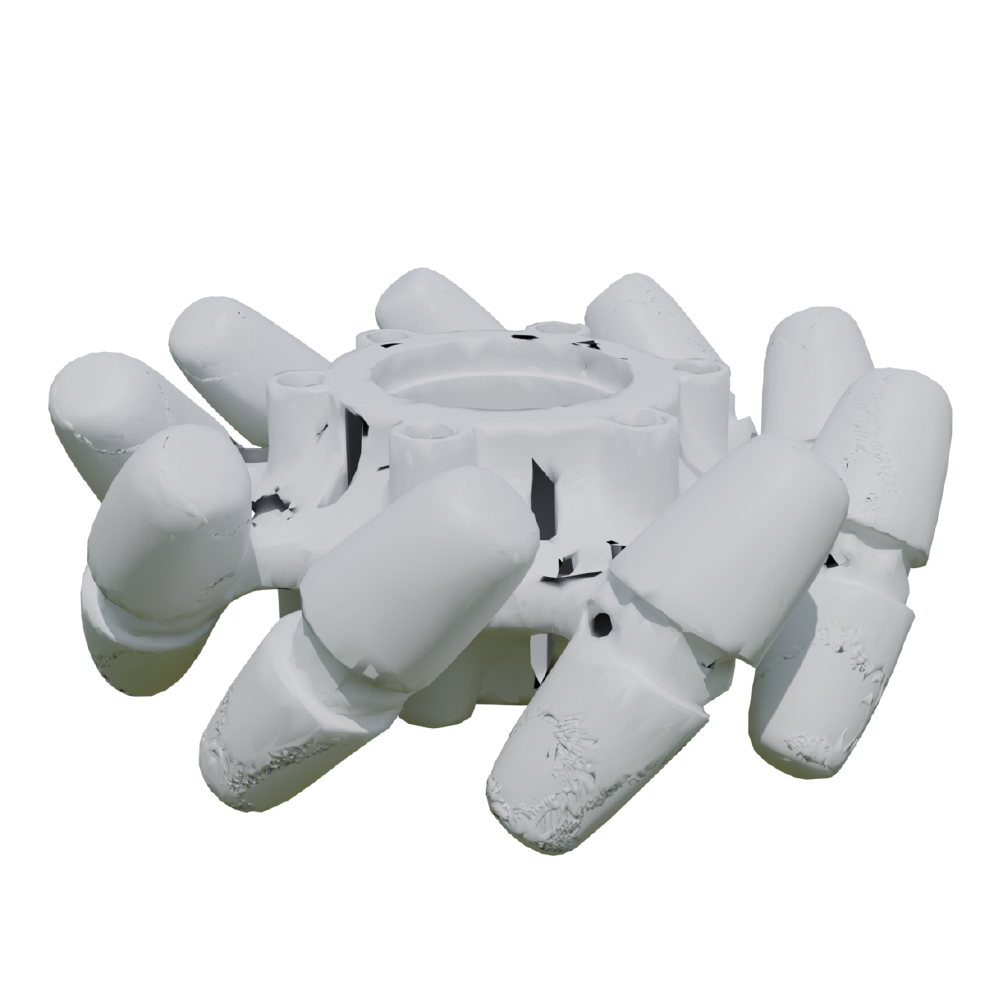}{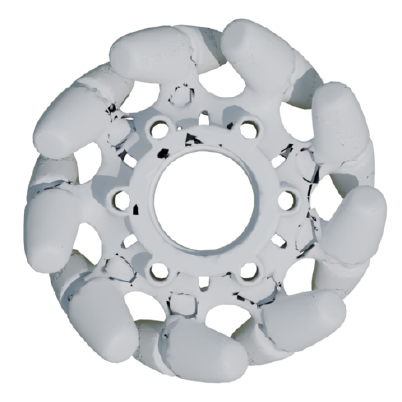} &
			\ablationzoomin{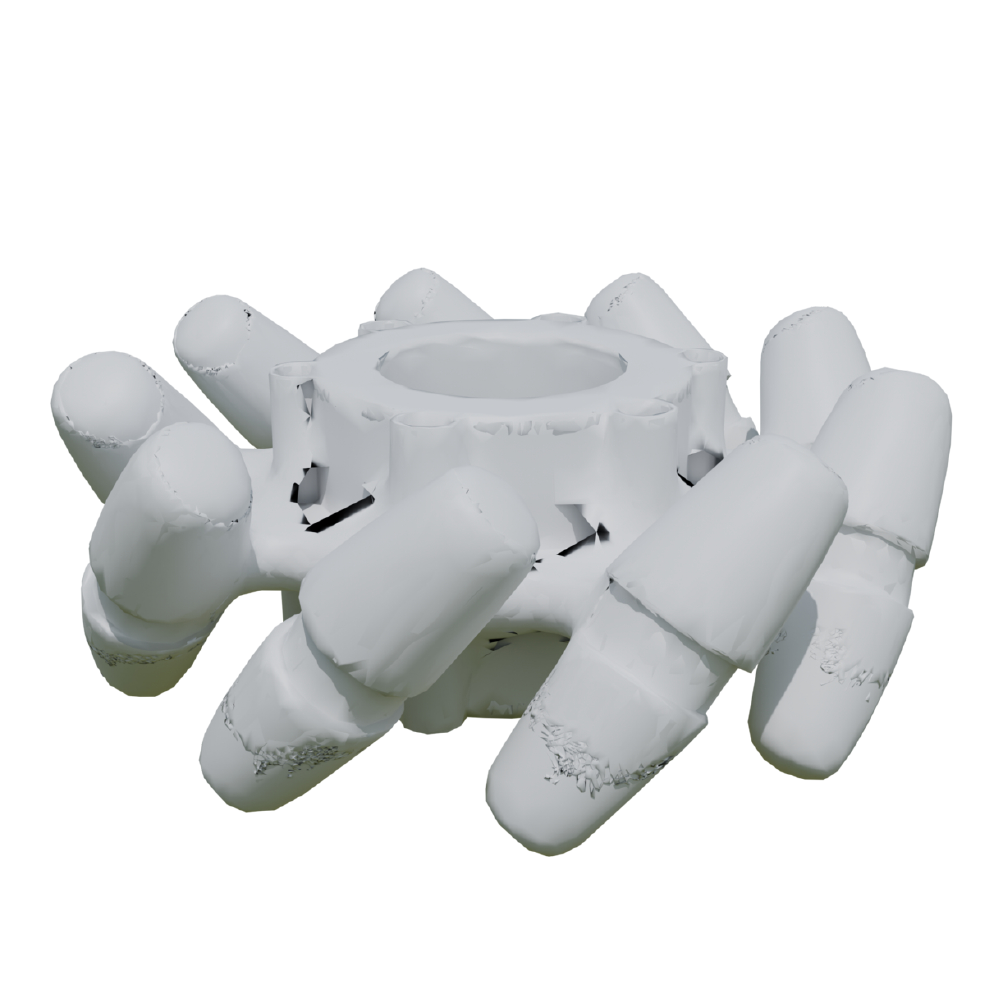}{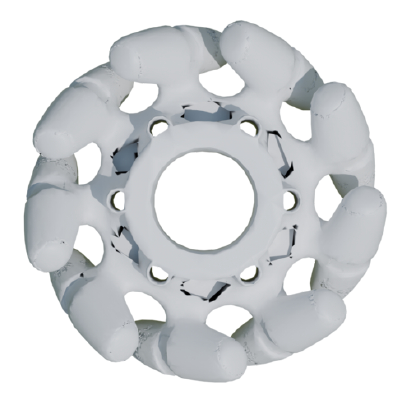}
		\end{tabular}

		\caption{
			\label{fig:ablation}
			We compare the TRELLIS.2 baseline, SymTRELLIS on sparse structure latent only, and full SymTRELLIS.
			The corresponding $\varepsilon\text{-Err}_{\max}$ values at $\varepsilon=0.03$, reported in units of $10^{-3}$, are $34.28$, $2.73$, and $1.57$, respectively.
		}
	}
\end{figure*}

\begin{acks}
	The authors thank the anonymous reviewers for their valuable comments and feedback.
	We also thank Pengshuai Wang from Peking University for helpful discussions and feedback.
	This work is supported in part by an NSERC Discovery Grant (RGPIN-2026-07641).
\end{acks}

\clearpage

\bibliographystyle{ACM-Reference-Format}
\bibliography{references}

\appendix
\numberwithin{figure}{section}
\numberwithin{table}{section}

\renewcommand{\thefigure}{\thesection\arabic{figure}}
\renewcommand{\thetable}{\thesection\arabic{table}}

\section{Spatial-Transform Latent Mapper}
\label{app:mapper}

We provide implementation details of the spatial-transform latent mapper introduced in \optionalCref{sec:3_2_mapper}{Section 4.2}.
The mapper constructs a linear operator on latent feature values.
Its coefficients are predicted only from sparse support geometry and the spatial transform.

\paragraph{Inputs and coordinate convention.}
A mapper input consists of a source support \(\mathcal{C}^{\mathrm{src}}\), a target support \(\mathcal{C}^{\mathrm{tgt}}\), source features \(X^{\mathrm{src}}\), and a known source-to-target Euclidean transformation \(g=(R,t)\).
We write the action of \(g\) on a continuous point \(\mathbf{p}\in\mathbb{R}^3\) as
\begin{equation}
	g(\mathbf{p}) = R\mathbf{p}+t .
\end{equation}
Voxel coordinates are integer grid coordinates in \(\{0,\ldots,G-1\}^3\).
We convert a voxel coordinate \(\mathbf{c}\) to a centered continuous coordinate by
\begin{equation}
	p(\mathbf{c}) = G^{-1}(\mathbf{c}+0.5)-0.5 .
\end{equation}
For each target coordinate \(\mathbf{c}^{\mathrm{tgt}}_i\), we map it back to the source frame:
\begin{equation}
	\mathbf{q}_i
	=
	p^{-1}\!\left(g^{-1}\!\left(p(\mathbf{c}^{\mathrm{tgt}}_i)\right)\right),
\end{equation}
where \(\mathbf{q}_i\) is a continuous source-grid coordinate.
The mapper predicts the feature at \(\mathbf{c}^{\mathrm{tgt}}_i\) by transporting features from source coordinates near \(\mathbf{q}_i\).

\paragraph{Local transport graph.}
For each target coordinate, we construct a radius graph
\begin{equation}
	\mathcal{N}_i
	=
	\left\{
	j:
	\left\|
	\mathbf{c}^{\mathrm{src}}_j-\mathbf{q}_i
	\right\|_2
	\le r
	\right\}.
\end{equation}
This handles the sparse-grid misalignment caused by rotations and reflections.
In our implementation, \(r=2\) grid cells.
Target coordinates whose back-transformed locations leave the valid source grid are masked out.

\paragraph{Coordinate and edge embeddings.}
The mapper first computes coordinate embeddings for source coordinates \(\mathbf{c}^{\mathrm{src}}_j\) and back-transformed target queries \(\mathbf{q}_i\).
We use Fourier positional features followed by a sparse transformer.
The transformer has a source branch that performs sparse self-attention over source coordinates,
and a target branch that attends to the source branch and then performs sparse self-attention over target queries.
The target branch is conditioned on the spatial transform \(g\).

The transform condition contains a 6D representation of \(R\) from its first two columns, Fourier features of \(t\), and a learned token for the orientation sign \(\det(R)\).
The orientation token allows the same mapper to handle both rotations and reflections.

For each local edge \((i,j)\), an edge MLP receives the target embedding, the source embedding, their difference, the relative displacement \(\mathbf{c}^{\mathrm{src}}_j-\mathbf{q}_i\), its squared distance, and the transform condition.
It outputs an aggregation logit \(\ell_{ij}\) and a low-rank transport scale vector \(s_{ij}\).

\paragraph{Linear feature transport.}
The aggregation weights are normalized over the source neighbors of each target
:
\begin{equation}
	\alpha_{ij}
	=
	\frac{\exp(\ell_{ij})}
	{\sum_{k\in\mathcal{N}_i}\exp(\ell_{ik})}.
\end{equation}
The feature-space transport matrix is parameterized as a residual low-rank map, with \(U\) and \(V\) as shared learned bases:
\begin{equation}
	W_{ij}
	=
	I + V\operatorname{diag}(s_{ij})U^\top ,
\end{equation}
The mapped target feature is
\begin{equation}
	\widehat{X}^{\mathrm{tgt}}_i
	=
	\sum_{j\in\mathcal{N}_i}
	\alpha_{ij}
	W_{ij}
	X^{\mathrm{src}}_j .
\end{equation}
For fixed supports and transform, all \(\alpha_{ij}\) and \(W_{ij}\) are fixed.
Therefore, the mapper is nonlinear as a function of support geometry and transform, but exactly linear as a function of \(X^{\mathrm{src}}\).
This is the property used by the velocity symmetrization operator in~\optionalCref{sec:3_1_vsymm}{Section~4.1}.

\paragraph{Training data.}
We train the mapper from paired TRELLIS.2 latents of the same object under known spatial transformations.
For each object, we encode multiple transformed instances using sampled rotations, rotation perturbations, translations, and optional reflections.
During training, two encoded instances are sampled as source and target.
Their known relative transform gives \(g\), and the mapper is trained to predict \(X^{\mathrm{tgt}}\) from \(X^{\mathrm{src}}\).
The training objects do not need to be symmetric.

\paragraph{Loss.}
The main objective is feature regression on valid target coordinates:
\begin{equation}
	\mathcal{L}_{\mathrm{feat}}
	=
	\frac{1}{|\Omega|}
	\sum_{i\in\Omega}
	\left\|
	\widehat{X}^{\mathrm{tgt}}_i
	-
	X^{\mathrm{tgt}}_i
	\right\|_2^2 ,
\end{equation}
where \(\Omega\) is the set of valid target coordinates with nonempty source neighbourhoods.
For the sparse-structure latent, we use a relative-MSE variant and ignore low-norm target features.
We also experimented with an auxiliary frozen-decoder loss, where the predicted sparse-structure latent is decoded and compared in occupancy space.

\paragraph{Configurations.}
The sparse-structure mapper uses a grid size of \(G=16\) and feature dimension \(d=8\), respectively, while for the shape mapper, we set \(G=32\) and \(d=32\).
Both mappers adopt a radius of \(r=2\), window size of \(4\), shifted-window offsets of \((0,0,0)\) and \((2,2,2)\),
Fourier coordinate features, sparse transformer blocks, and with residual low-rank transport.
The trained checkpoints use a width of \(192\), depth of \(8\), \(6\) attention heads, edge dimension of \(96\), and finally, a low-rank transport rank of \(64\).

\section{Comparison to Post-Processing Symmetrization}
\begin{figure}[t!]
	\centering
	\small
	\setlength{\tabcolsep}{2pt}
	\renewcommand{\arraystretch}{0.8}

	\newlength{\postprocessfigresultsize}
	\setlength{\postprocessfigresultsize}{0.23\textwidth}
	\resizebox{\columnwidth}{!}{%
		\begin{tabular}{cccc}
			\textbf{TRELLIS.2}                                                                                                                                   &
			\textbf{Sector Replication}                                                                                                                          &
			\textbf{Occupancy Average}                                                                                                                           &
			\textbf{Closest Vertex Average}                                                                                                                        \\[3pt]

			\includegraphics[width=\postprocessfigresultsize,height=\postprocessfigresultsize]{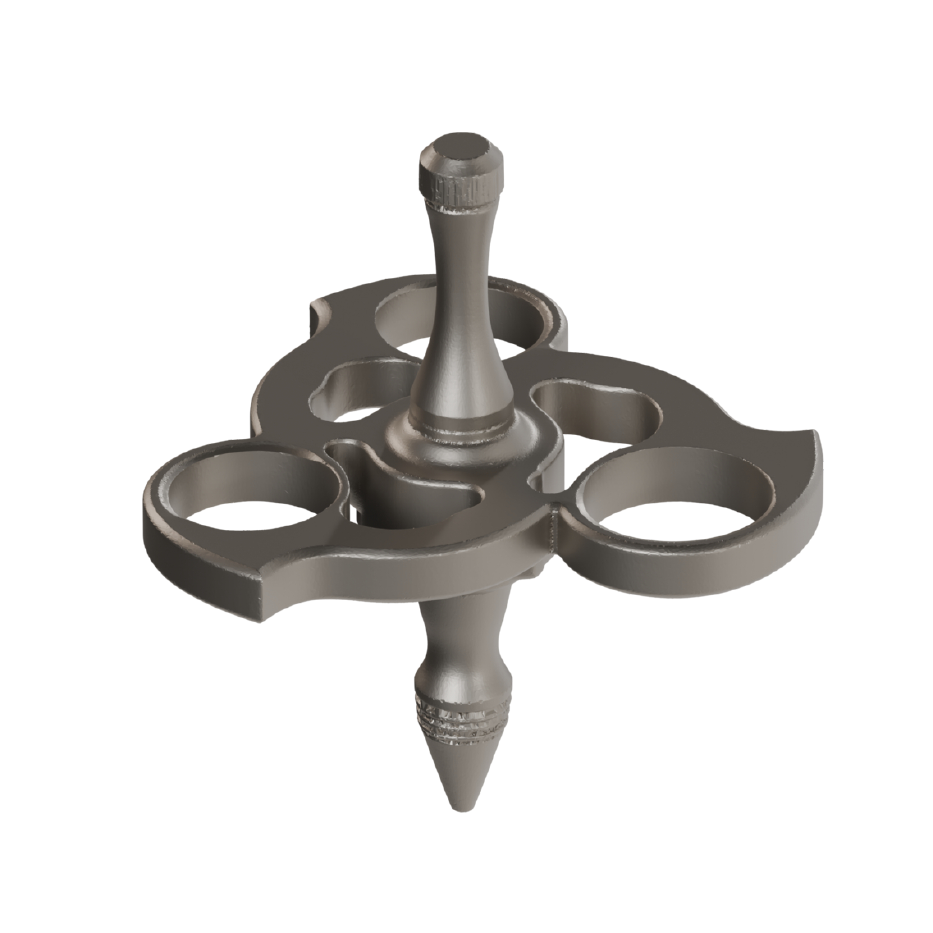}    &
			\includegraphics[width=\postprocessfigresultsize,height=\postprocessfigresultsize]{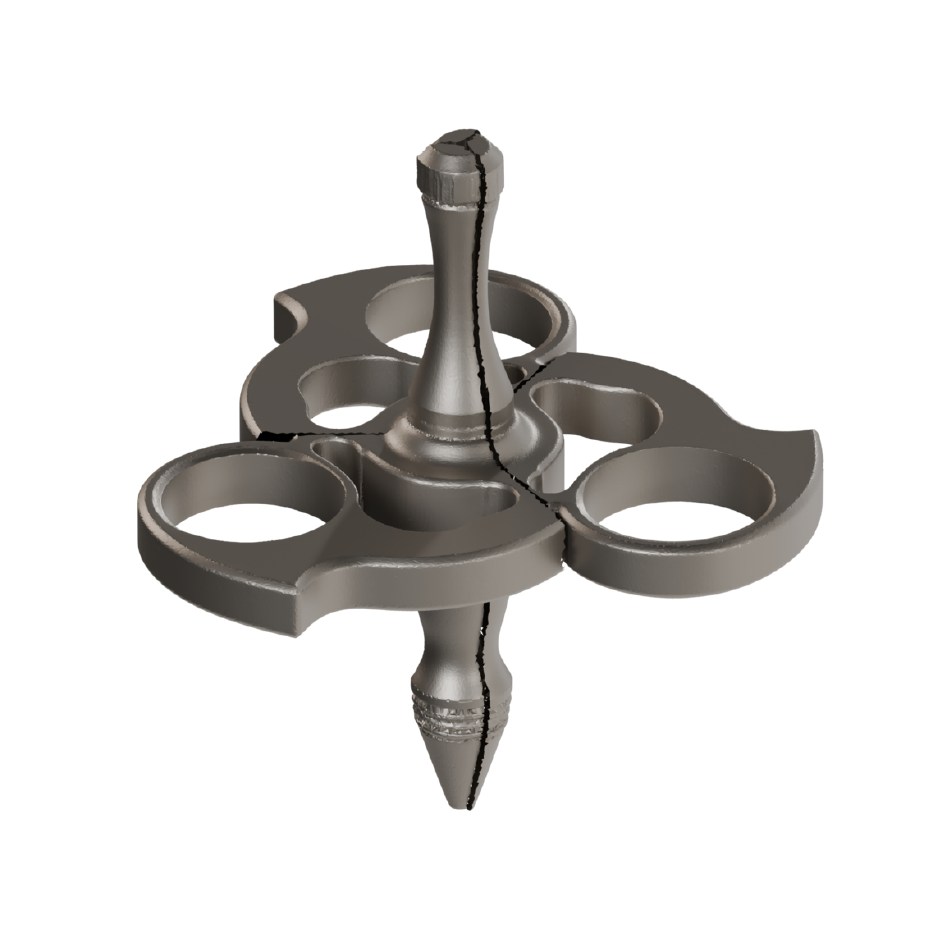} &
			\includegraphics[width=\postprocessfigresultsize,height=\postprocessfigresultsize]{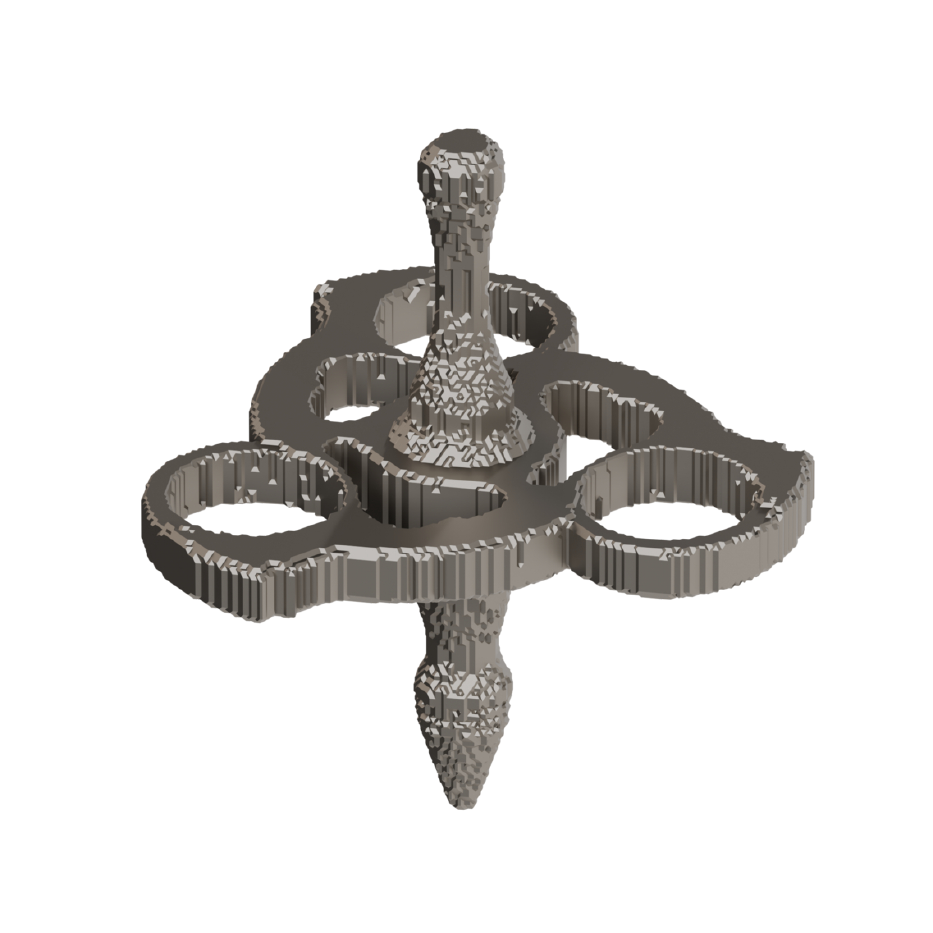}          &
			\includegraphics[width=\postprocessfigresultsize,height=\postprocessfigresultsize]{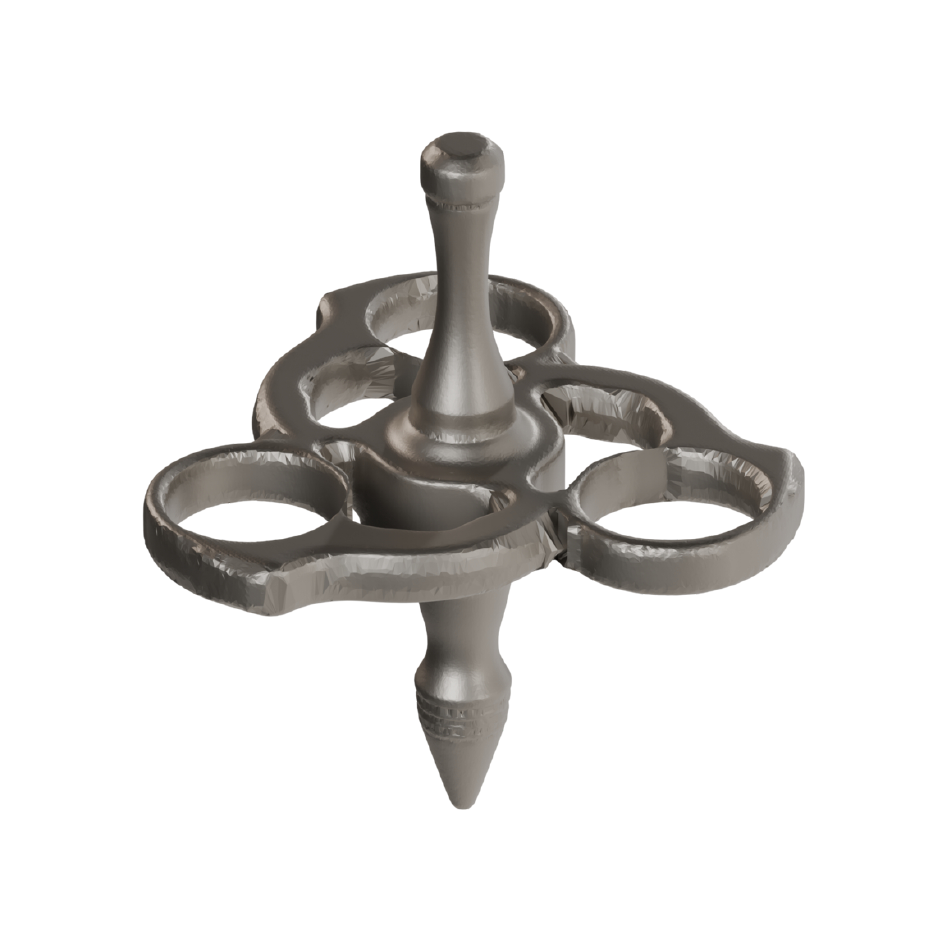}   \\[4pt]

			\includegraphics[width=\postprocessfigresultsize,height=\postprocessfigresultsize]{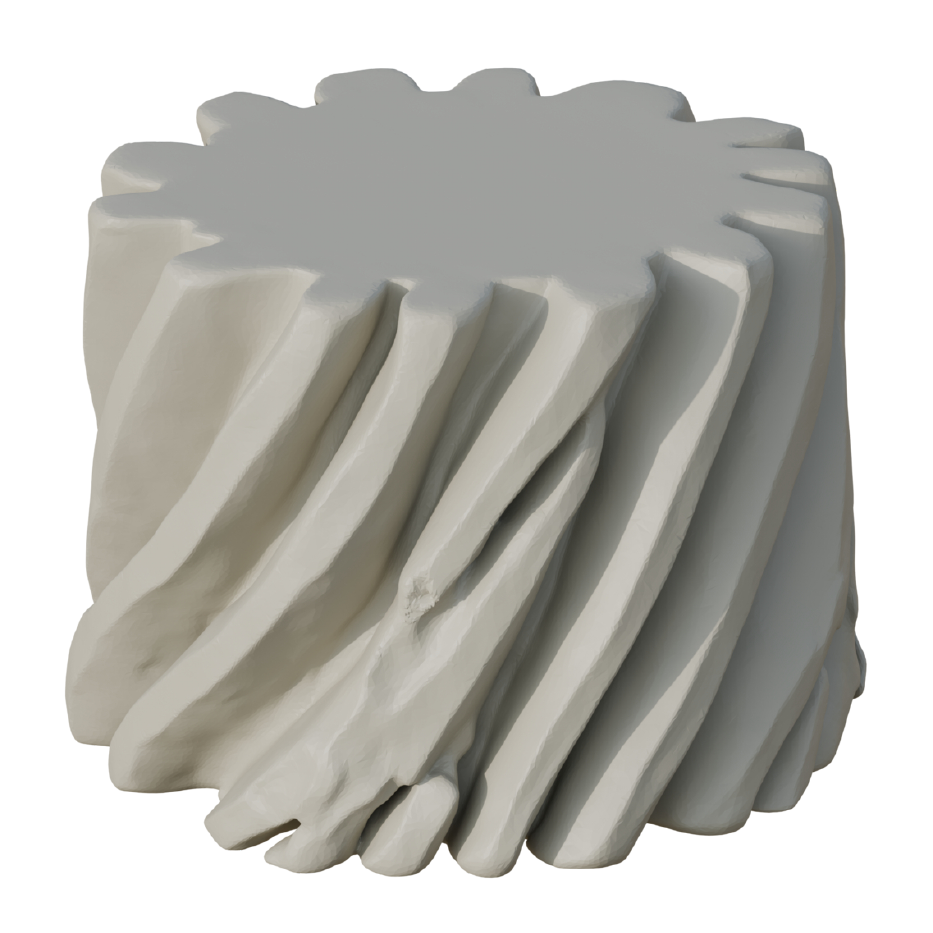}            &
			\includegraphics[width=\postprocessfigresultsize,height=\postprocessfigresultsize]{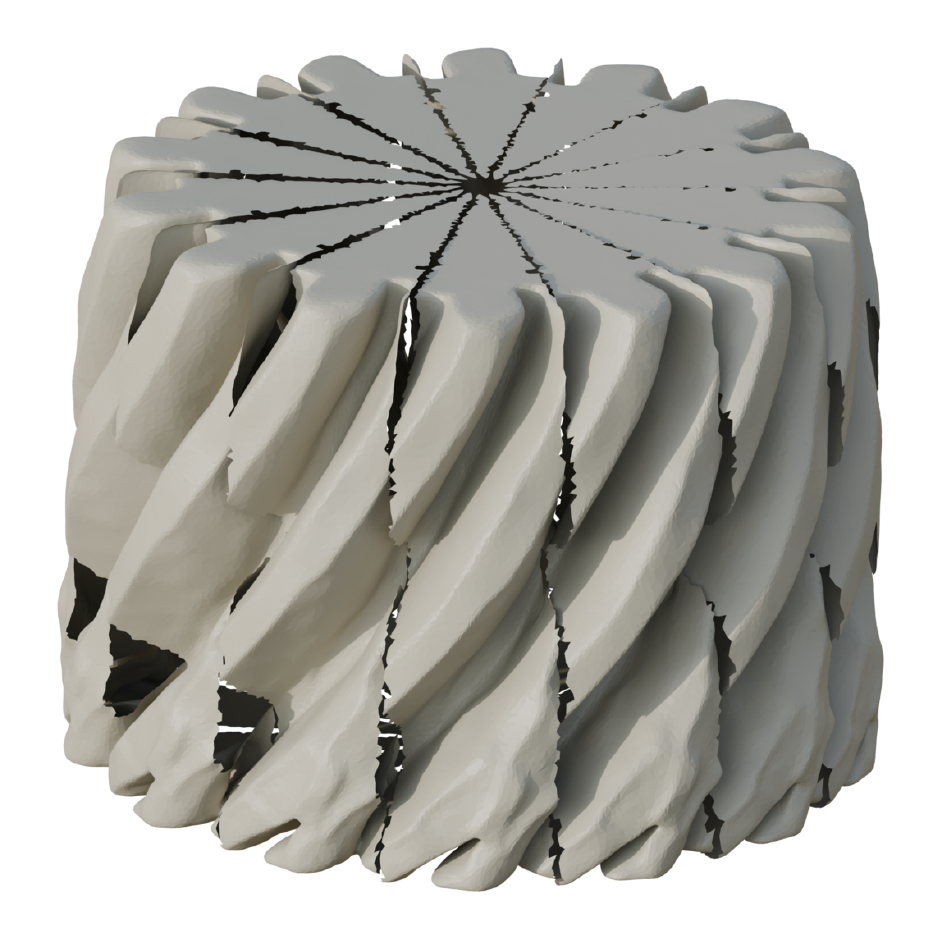}         &
			\includegraphics[width=\postprocessfigresultsize,height=\postprocessfigresultsize]{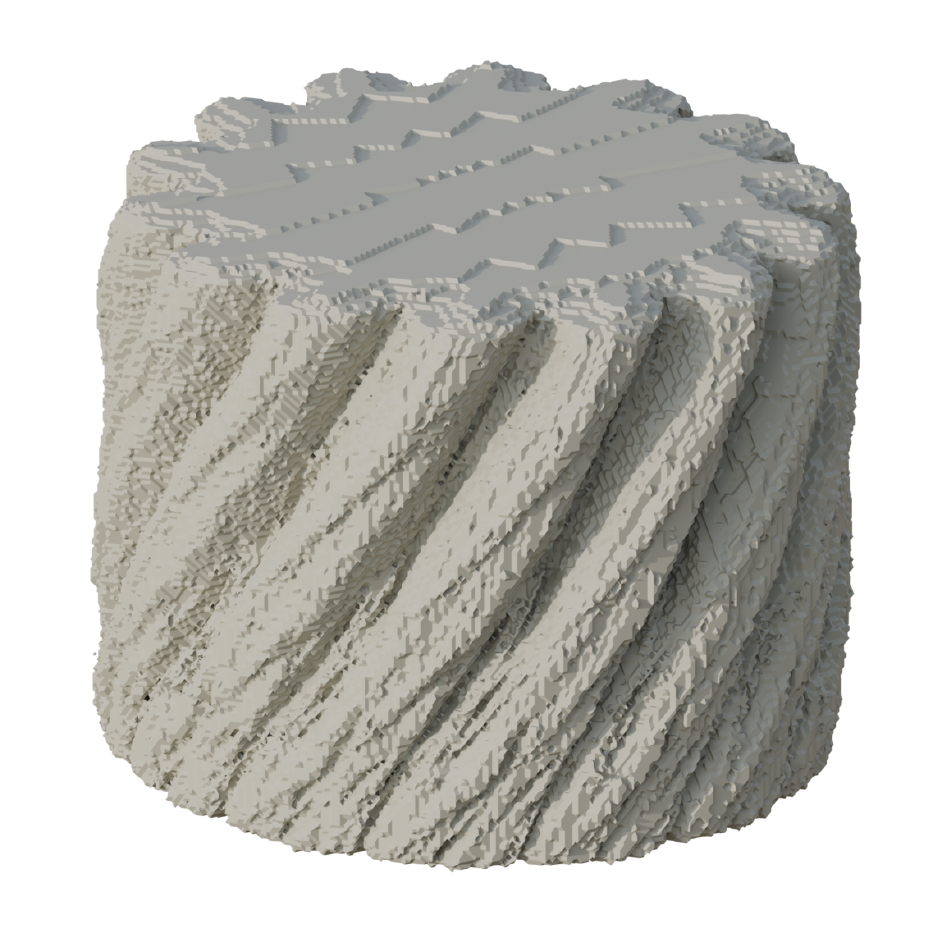}                  &
			\includegraphics[width=\postprocessfigresultsize,height=\postprocessfigresultsize]{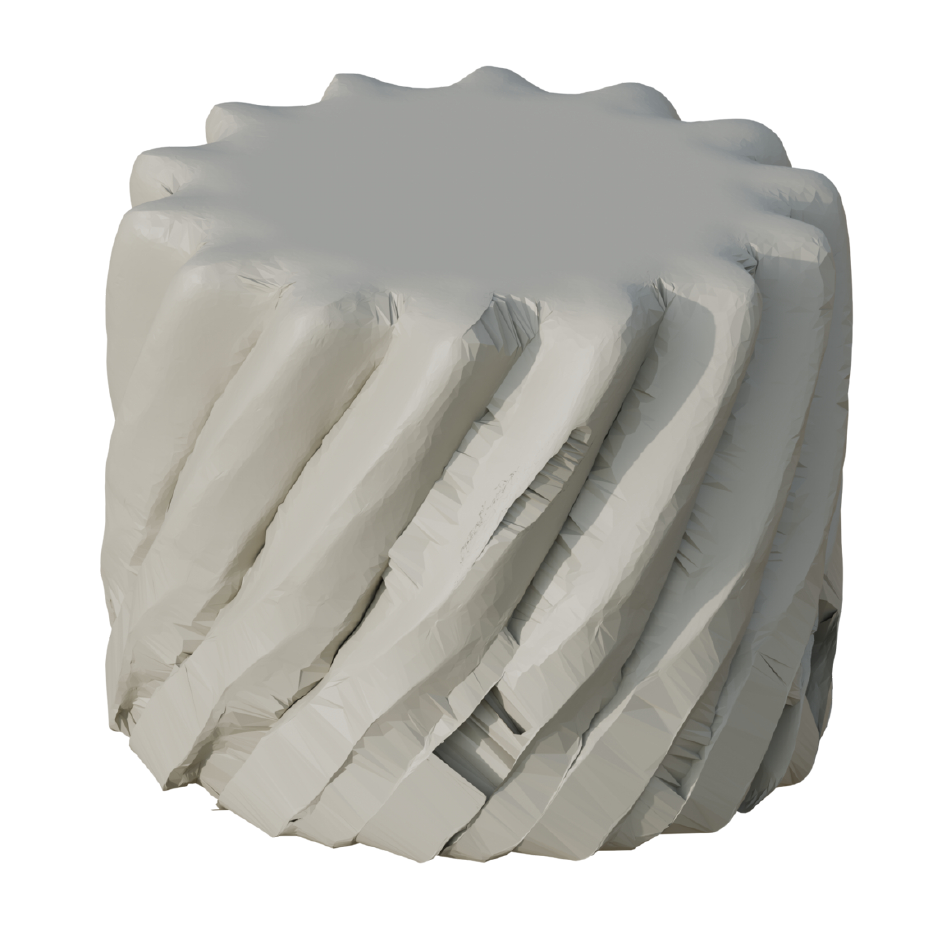}           \\[4pt]

			\includegraphics[width=\postprocessfigresultsize,height=\postprocessfigresultsize]{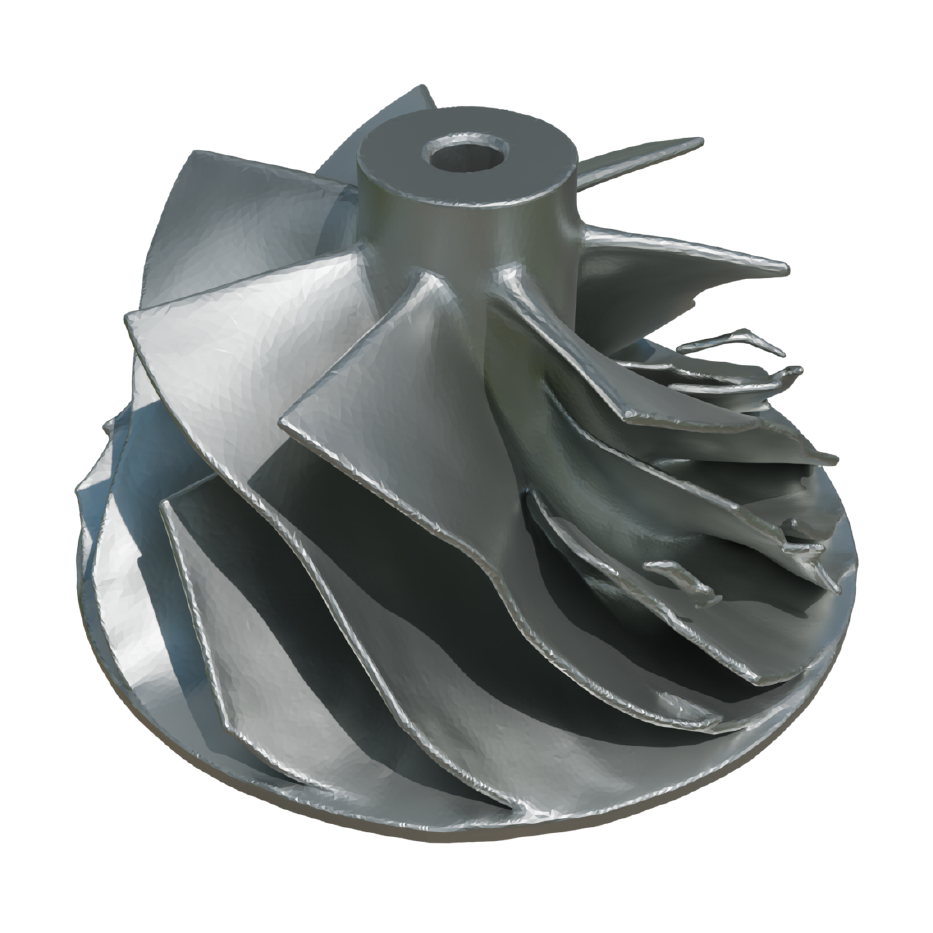}     &
			\includegraphics[width=\postprocessfigresultsize,height=\postprocessfigresultsize]{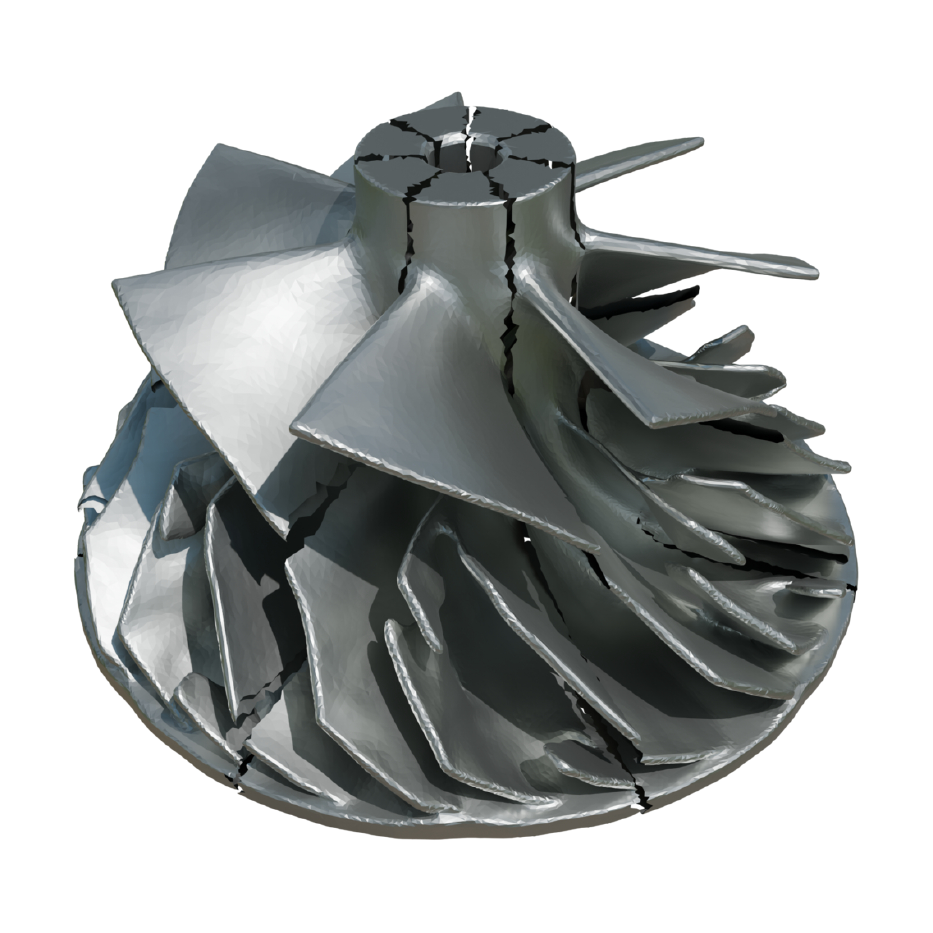}  &
			\includegraphics[width=\postprocessfigresultsize,height=\postprocessfigresultsize]{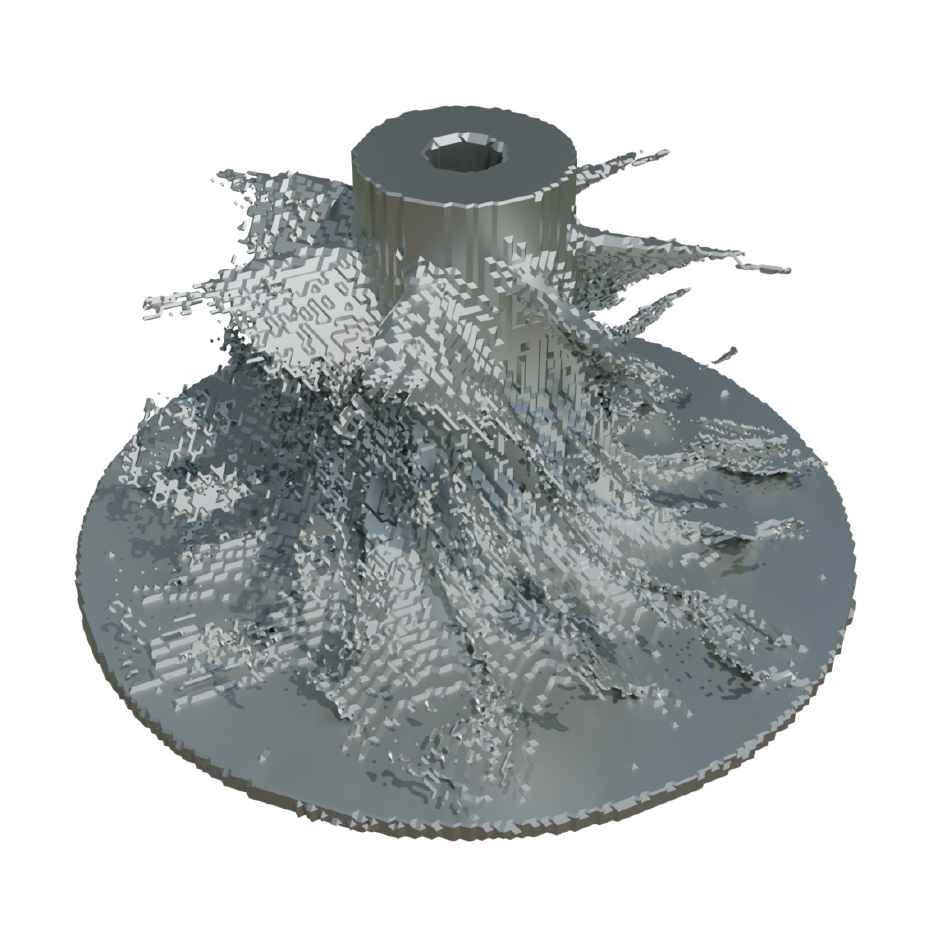}           &
			\includegraphics[width=\postprocessfigresultsize,height=\postprocessfigresultsize]{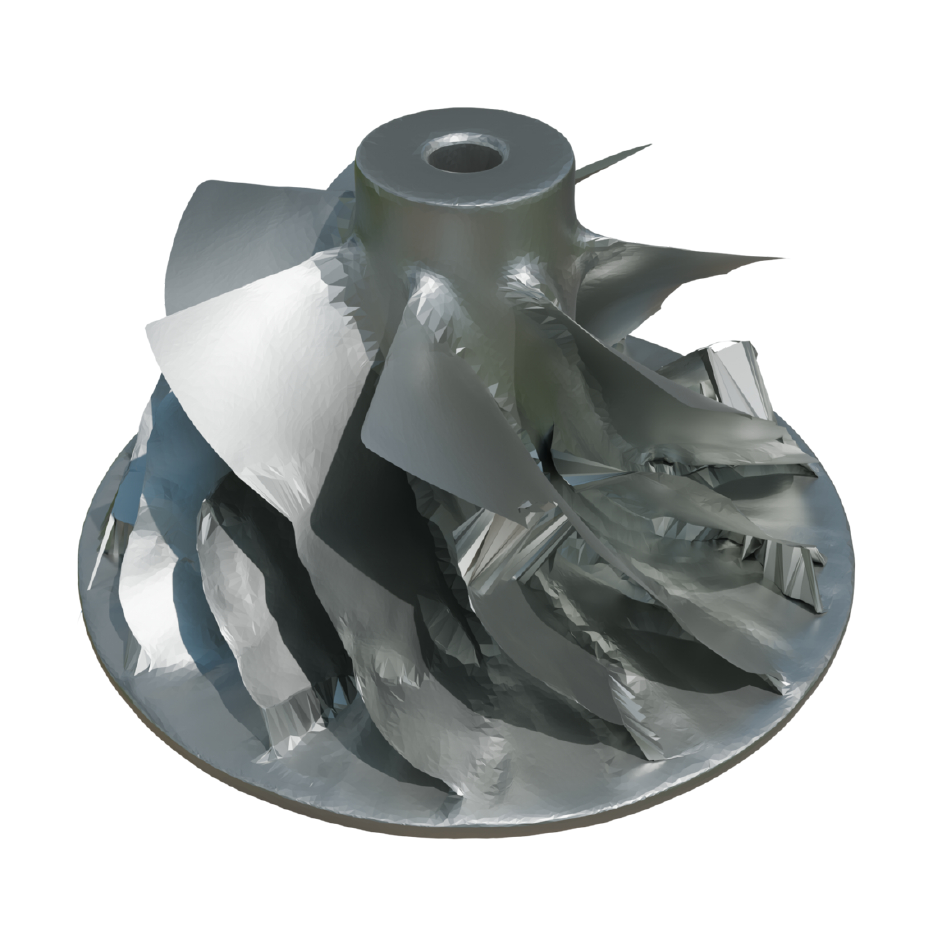}    \\[4pt]
		\end{tabular}%
	}
	\caption{
		\label{fig:postprocess}
		Visuals of simple post-processing baselines show that these methods are not suitable for symmetry enforcement.
		Sector replication creates cracks and requires the method to intelligently segment out the most nearly perfect sector.
		Occupancy averaging cancels occupancy if the mesh has thin structures and is not perfectly symmetric.
		Closest vertex averaging degrades details on the mesh.
	}
\end{figure}

\begin{table*}[t!]
	\centering
	\caption{Post-processing results.
		$\mathrm{SD}$, $\varepsilon$-Err, and CD are reported in units of $10^{-3}$.
		Fold accuracies are reported in percentages.}
	\label{tab:post_processing}

	\resizebox{1.0\textwidth}{!}{
		\begin{tabular}{lcccccccccc}
			\toprule
			 & \multicolumn{9}{c}{Symmetry}
			 & \multicolumn{1}{c}{Reconstruction}                                                                     \\
			\cmidrule(lr){2-10} \cmidrule(lr){11-11}

			\multirow{2}{*}{Method}
			 & \multirow{2}{*}{\makecell[c]{$\mathrm{SD}_{\max}\downarrow$}}
			 & \multirow{2}{*}{\makecell[c]{$\mathrm{SD}_{\mathrm{avg}}\downarrow$}}
			 & \multicolumn{3}{c}{$\varepsilon\text{-Err}_{\max}\downarrow$}
			 & \multicolumn{3}{c}{$\varepsilon\text{-Err}_{\mathrm{avg}}\downarrow$}
			 & \multirow{2}{*}{\makecell[c]{Fold acc. $\uparrow$}}
			 & \multirow{2}{*}{\makecell[c]{Chamfer Dist $\downarrow$}}                                               \\
			\cmidrule(lr){4-6}
			\cmidrule(lr){7-9}

			 &
			 &
			 & $\varepsilon=0.01$
			 & $\varepsilon=0.03$
			 & $\varepsilon=0.1$
			 & $\varepsilon=0.01$
			 & $\varepsilon=0.03$
			 & $\varepsilon=0.1$
			 &
			 &                                                                                                        \\
			\midrule

			TRELLIS.2
			 & 7.47                                                                  & 4.58
			 & 200.03                                                                & 63.43          & 5.61
			 & 118.42                                                                & 34.09          & 2.74
			 & 83.96\%                                                               & \textbf{13.79}                 \\

			\quad + Sector Replication
			 & \textbf{0.19}                                                         & \textbf{0.11}
			 & \textbf{2.21}                                                         & \textbf{0.47}  & \textbf{0.00}
			 & \textbf{0.96}                                                         & \textbf{0.16}  & \textbf{0.00}
			 & 99.35\%                                                               & 18.08                          \\

			\quad + Occupancy Average
			 & 6.01                                                                  & 3.58
			 & 125.32                                                                & 23.84          & 4.42
			 & 63.30                                                                 & 12.24          & 2.32
			 & \textbf{99.60\%}                                                      & 42.89                          \\

			\quad + Closest Vertex Average
			 & 41.01                                                                 & 23.25
			 & 112.02                                                                & 42.50          & 9.87
			 & 70.06                                                                 & 26.59          & 6.42
			 & 92.78\%                                                               & 18.50                          \\
			\bottomrule
		\end{tabular}
	}

\end{table*}

\Cref{tab:post_processing} and \Cref{fig:postprocess} evaluate three post-processing schemes to symmetrize meshes generated by the vanilla TRELLIS.2 model:
(1) \textit{Sector Replication}: segmentation followed by picking one sector and replicating it across folds;
(2) \textit{Occupancy Averaging}: averaging the group transformed occupancies; and
(3) \textit{Closest Vertex Averaging}: iteratively updating mesh vertices towards their group-averaged closest points.
All these post-processing schemes are explicitly designed to produce symmetric results, while Section Replication is particularly effective in forcing high degrees of symmetry in the final results,
even outperforming SymTRELLIS.
However, symmetry is only one desirable outcome and it can be trivially achieved, but at the expense of reconstruction quality.
Quantitatively, as measured by Chamfer Distance, the three post-processing options yielded worse results than both TRELLIS.2 and SymTRELLIS.
What is most telling though is the visual quality, as shown in ~\Cref{fig:postprocess} with significant artifacts produced by naive shape symmetrization during post-processing.
Closest-point averaging and voxel averaging degrade details, while sector replication inevitably creates unfixable cracks, among other issues.
In contrast, SymTRELLIS bypasses these limitations and avoids such artifacts.

\section{Noise Symmetrization, Gaussian Rescaling, and Side Effects}

We clarify that we symmetrize the flow matching noise in our SymTRELLIS pipeline.
However, if the symmetrized noise was not normalized to be in the form of standard Gaussians, it would cause some failure cases,
more frequently when the symmetry groups have high orders, leading to the generation of empty active voxel sets.
We can fix this issue by rescaling the noise back to standard Gaussians, effectively eliminating most cases of empty active voxel sets and the corresponding failure cases.
But a side effect of such a rescaling is the occasional (around 5\% over the SymTRELLIS evaluation) creation of abnormal active voxels around the symmetry origin, thus degrading reconstruction quality.

\section{Subset Selection Rules}

The velocity symmetrization operator in \optionalCref{sec:3_1_vsymm}{Section~4.1} averages mapped flow predictions over a transformation subset \(\mathcal{T}\) of the full target point group \(\mathcal{G}\).
We construct \(\mathcal{T}\) by including all rotations around the group's symmetry axes, all reflections across its symmetry planes, and inversion when present, while excluding additional transformations obtained by composing them.
Under this rule, \(\mathcal{T}=\mathcal{G}\) for pure rotational groups such as \(C_n,D_n,T,O,I\).
For other groups, \(\mathcal{T}\) typically contains about 63\%--88\% of the elements in \(\mathcal{G}\).

This choice balances two competing effects.
Too many transformed copies can cancel occupancy or over-smooth shape latents,
while using only a minimal set of group generators would require multiple projection steps for information to reach all equivalent folds.
Our heuristic subset rule reduces the number of averaged elements while allowing information to reach all equivalent folds with at most one additional message-passing step.

\end{document}